\documentclass[a4paper,12pt,english,twoside]{scrreprt}

\usepackage{babel}
\usepackage{graphicx}
\usepackage[T1]{fontenc}
\usepackage[ansinew]{inputenc}
\usepackage{amsmath}
\usepackage{indentfirst}
\usepackage{fancyhdr}
\usepackage{feynmf,multicol}
\usepackage{amssymb}
\usepackage{multicol}
\usepackage{pstricks} 
\usepackage{dsfont}

\newcommand{\orch}[1]{#1\rput(-0.1,.35){\leftrightarrow}}

\begin{document}

%\review[Fundamentals and recent developments in non-perturbative canonical Quantum Gravity \LaTeXe]{Fundamentals and recent developments in non-perturbative canonical Quantum Gravity}

%\author{Francesco Cianfrani$^{12}$, Orchidea Maria Lecian$^1$, Giovanni Montani$^{134}$.}

%\address{$^1$ICRA-International Center for Relativistic Astrophysics, Physics Department (G9), University of Roma ``Sapienza", Piazzale Aldo Moro 5, 00185 Rome, Italy.}

%\address{$^2$ School of Mathematical Sciences, Queen Mary, University of London, London E1 4NS, UK.}

%\address{$^3$ENEA C.R. Frascati (Dipartimento F.P.N.), via Enrico Fermi 45, 00044 Frascati, Rome, Italy.}

%\address{$^4$ICRANET C. C. Pescara, Piazzale della Repubblica, 10, 65100 Pescara, Italy.}

%\eads{\mailto{francesco.cianfrani@icra.it}, \mailto{lecian@icra.it}, \mailto{montani@icra.it}}

%\maketitle

%%%%%%%%%%%%%%%%%%%%%%%%%%%%%%%%%%%%%%%%%%%%%%%%%%%%%%%%%%%%%%%%%%%%%%%%%%%%%%%%%%%%%%%%%%%
\pagestyle{empty}

\begin{center}

%\null\hfill\\

\vspace{3cm}

\Huge{
\textbf{Fundamentals and recent developments in non-perturbative canonical Quantum Gravity}
}

\vspace{3cm}

{\Large Francesco Cianfrani$^{12}$, Orchidea Maria Lecian$^1$, Giovanni Montani$^{134}$.
}

\vspace{2cm}

{\normalsize$^1$ICRA-International Center for Relativistic Astrophysics, Physics Department (G9), University of Roma ``Sapienza", Piazzale Aldo Moro 5, 00185 Rome, Italy.

\vspace{0.5cm}
$^2$ School of Mathematical Sciences, Queen Mary, University of London, London E1 4NS, UK.

\vspace{0.5cm}

$^3$ENEA C.R. Frascati (Dipartimento F.P.N.), via Enrico Fermi 45, 00044 Frascati, Rome, Italy.

\vspace{0.5cm}

$^4$ICRANET C. C. Pescara, Piazzale della Repubblica, 10, 65100 Pescara, Italy.

\vspace{0.5cm}

email: francesco.cianfrani@icra.it, lecian@icra.it, montani@icra.it}

\end{center}

\cleardoublepage

\pagestyle{plain}

\pagenumbering{roman}

%\include {headings}
%\input{riassunto}
%\pagestyle{empty}
%\include{tesine}
%Fa l'indice%%%%%%%%%%%%%%%%%%%%%%%%%%%%%%

\tableofcontents

\cleardoublepage

\chapter*{Preface}
The request of a coherent quantization for the
gravitational field dynamics emerges as a natural
consequence of the Einstein Equations:
the energy-momentum of a field is source of the space-time
curvature and therefore its microscopic quantum
features must be reflected onto microgravity effects.
The use of expectation values as sources is well-grounded as a first approximation only, and does not fulfill the requirements for a fundamental theory. 

However, as well-known, the achievement of a consistent
Quantum Gravity theory remains a complete open task,
due to a variety of different subtleties, which are here
summarized in the two main categories:
-General Relativity is a background-independent theory
and therefore any analogy with the non-Abelian gauge
formalisms must deal with the concept of a dynamical
metric field;
-the implementation in the gravitational sector 
of standard prescriptions, associated with the quantum
mechanics paradigms, appears as a formal procedure,
lacking of real physical content.

There exists a qualitative consensus on the idea that a convincing
solution to the Quantum Gravity problem will not arise
before General Relativity and Quantum Mechanics
are both deeply revised in view of a converging picture.
Indeed, over the last ten years, the three most promising
approaches to Quantum Gravity (i.e. String Theories,
Loop Quantum Gravity and Non-commutative Geometries)
revealed common features in defining a ''lattice''
nature for the microphysics of space-time.
This consideration makes clear that, up to the best
of our present understanding, the main effort to improve
fundamental formalisms must be the introduction of
a ''cut-off'' physics, able to replace the notion of
a space-time continuum by a consistent discrete scenario.
The correctness of such a statement will be probably regarded as
the main success of the end of the last century reached in
Theoretical Physics. A proper task for the present century
is now to constrain, as properly as possible, the
morphology of such a discrete microstructure of space-time,
to get, at least, a phenomenological description
for Quantum Gravity phenomena.
From a theoretical point of view, an important aim
would consist of a unified picture containing
common features of the present approaches, but synthesized
into a more powerful mathematical language.
On the one hand, it would be important to recognize
non-commutative properties in the loop representation of
space-time. On the other hand, 
 transporting the background independence of the
''spin networks'' into the interaction framework
characterizing String Theories would take a crucial role. The viability of these
two goals is an intuitive perspective, but it contrasts with the
rigidity of the corresponding formalisms, which confirms
the request for more general investigation tools.

The simplest approach to the quantization of General
Relativity relies on the implementation of the canonical
method on the phase-space structure associated with
the gravitational degrees of freedom. This attempt is
the one originally pursued by B. DeWitt in 1967 and
it immediately revealed all the pathologies contained in the
canonical quantum geometrodynamics. However, it was just
this clear inconsistence of the canonical quantum procedure
referred to the second order formalism that attracted
the attention of a large number of researchers, active on
the last four decades.
This strong effort of re-analysis, which could give
the feeling of an overestimation of the real chances
allowed by the canonical method, found its
''merry ending'' in the recent developments in Loop
Quantum Gravity and its applications. In fact, the
formulation of the Hamiltonian problem for General Relativity
provided by A. Ashtekar in 1986, allowed for a new
paradigm for the canonical method, in close analogy with
an $SU(2)$ gauge group. The main success of Loop
Quantum Gravity is recognizing a discrete
structure of space-time, by starting from continuous
variables in the phase space of the theory.
The origin of such an excellent result consists in the
discrete nature acquired by the spectra of areas and
volumes, reflecting to some extent the compactness of the
$SU(2)$ group emerging in this formulation.\\
The merit of the Loop Quantization of gravity can be
identified in the possibility to deal with non-local
geometrical variables, like holonomies and fluxes,
instead of the simple metric analysis faced by DeWitt.
In fact, using these physical properties associated with
the connection and electric fields, the background independence
of the theory naturally emerges, and the quantum structure
of the space-time comes out, as far as the Hilbert space
is characterized via the spin-network basis.

The aim of this review is to provide a detailed account
of the physical content emerging from this story of the
canonical approach to Quantum Gravity. All the crucial
steps in our presentation have a pedagogical character,
providing the reader with the necessary tools to
become involved in the field. Such a pedagogical aspect is
then balanced  and completed by subtle discussions on
specific topics which we regard as relevant for the physical
insight they outline on the treated questions.
Our analysis is not aimed at convincing the reader about
a pre-constituted point of view, bu instead our principal
goal is to review the picture of Canonical Quantum Gravity
on the basis of the concrete facts at the ground of its
clear successes, but also of its striking shortcomings.\\
In order to focus our attention  on the physical questions affecting the
the consistence of a canonical quantum model, when extended to
the gravitational sector, we provide a critical discussion
of all the fundamental concepts of quantum physics,
by stressing the peculiarity of the gravitational case in
their respect. All the key features of Loop Quantum
Gravity are described is some details, presenting the
specific technicalities, but privileging their
physical interpretation.
Finally, all the open questions concerning the faced
topics are clearly outlined by a precise criticism on
the weak points present in the new quantum construction.

The detailed structure of the review is

\begin{itemize}

{\item in the first chapter, we  recall the basic features of quantum mechanics (sections 1.1 and 1.2), and focus our attention on the problem of time (sections 1.3 and 1.4). Furthermore, we also introduce the Weyl-quantization method (section 1.5), based on a different representation of operators in canonical commutation relations, and outline the main steps of the GNS construction (section 1.6).  }

{\item in the second chapter, the standard Hamiltonian formulation of General Relativity is reviewed: a discussion on the fundamental role of Einstein-Hilbert action and of its possible modifications is presented in section 2.1, while, in 2.2, the Gauss-Codazzi equations governing the space-time slicing are inferred. The full Hamiltonian description is given in 2.3, and, in 2.4, the corresponding Hamilton-Jacobi equation is discussed. Finally, the reduction to the canonical form is presented in section 2.5.}

{\item the third chapter outlines similarities and differences between General Relativity and gauge theories. In 3.1, the framework of gauge theories is presented, while, in 3.2, the main properties of a Palatini-like formulation for gravity are pointed out. Then, starting from the behavior of spinors in a curved space, a comparison between these two frameworks is made in 3.3. An example of gravity as a gauge theory is emphasized in section 3.4, where Poincar\'e gauge theory is presented. Section 3.5 deals with the Holst formulation of General Relativity, for which a phase space similar to the one of gauge theories comes out as soon as Ashtekar-Barbero-Immirzi connections are taken as configuration variables. In this respect, the full set of constraints is explicitly evaluated. Finally, in section 3.6, the intriguing role of the Kodama state, which is the only up-to-now known solution of such constraints, is outlined.}

{\item in fourth chapter, the quantization of the gravitation field is analyzed. After revising the WDW equation (section 4.1), we reappraise the problem of time in some quantization schemes (section 4.1). The interpretation of the wave function of the Universe is discussed in section 4.3, and the idea of third quantization is then proposed (section 4.4).}

{\item the fifth chapter is devoted to Loop Quantum Gravity. Properties of holonomy and flux variables are analyzed in section 5.1: after some speculation on the physical meaning of holonomies in quantum theories in subsection 5.1.1, their role in lattice gauge theories is outlined in 5.1.2, while their application to the Loop Quantum Gravity framework is presented in 5.1.3. The spectra of the area and volume operators are evaluated in section 5.2, and their discreteness is pointed out. Section 5.3 describes attempts towards the resolution of the quantum dynamics. This issue leads, in 5.4, to a discussion on the main open points of Loop Quantum Gravity and the possibility to solve them in the context of the Master Constraint Analysis and of Algebraic Quantum Gravity is sketched in subsection 5.4.1. As an application to a minisuperspace model, where the dynamics can be solved, isotropic loop quantum cosmology is described in section 5.5. Other issues are presented in sections 5.6, where the role of Immirzi parameter is outlined, and in 5.7, where the boost invariance in Loop Quantum Gravity is discussed. Finally, in section 5.8, a comparison between the picture of the quantum space-time coming out from the Loop Quantum Gravity and the Wheeler DeWitt frameworks is performed.}
\end{itemize}

\vspace{3cm}

We would like to thank Abhay Ashtekar for interesting discussions on Loop Quantum Gravity we had with him during the First Stueckelberg Workshop (Pescara, June 25-July 1 2006), and which inspired this review work.

\vspace{3cm}

We would also like to thank Richard Arnowitt, Stanley Deser and Charles W. Misner for their comments and discussion on the Hamiltonian formulation of Generel  Relativity and for having called out attention on the specific relevance of reference \cite{ADM62}.

\newpage

\pagenumbering{arabic}

\chapter{Quantization methods}

\section{Classical and quantum dynamics}\label{classquan}

In this section we will revise some fundamental features of Quantum Mechanics, in order to fix the notation and to give a clear insight of the basic formalism underlying the quantization procedure \cite{Coh,Mes,Sak}.

\paragraph{Dirac observables}
Observables are described by Hermitian operators that act on vectors in Hilbert spaces; these vectors describe the set of all the possible states the system can be found in. A ket $|\alpha >$ in a Hilbert space $H$ can be decomposed along a basis of the Hilbert space, $\left\{ |a>\right\}$, satisfying the identity $\sum_{a} |a><a|=I$, so that 
\begin{equation}\label{decomp}
| \alpha >=\sum_{a}|a><a|\alpha >.
\end{equation}
The result of a measure does not depend on the characterization of the Hilbert space or on the choice of the basis, but only on the definition of observables. On the one hand, from a physical point of view, the equivalence of a description of an observable $A_{1}$ in a Hilbert space $H_{1}$ with that of $A_{2}$ in $H_{2}$ is given by the unitary transformation $U$, $U:H_{1}\rightarrow H_{2}$, such that $A_{1}=U^{-1}A_{2}U$. On the other hand, the crucial operation is the definition of self adjoint operators from classical quantities.\\
When a measure of an observable $A$ is performed, the system, which was previously described as a linear combination of the eigenstates of $A$, (\ref{decomp}), falls into one of these states $|a'>\in \left\{ |a>\right\}$, which corresponds to the spectral value $a'$ of $A$, {\it i.e.} the measure operation changes the state of the system unless the system was already in an eigenstate of the observable. Since (\ref{decomp}) describes the system before the measure, it is not possible \slshape a priori \upshape to know which state $|a'>\in \left\{ |a>\right\}$ will be the result of the measure, but the probability $P_{a'}$ of a particular result $|a'>$ is given by
\begin{equation}\label{prob}
P_{a'}=|<a'|\alpha >|^{2}.
\end{equation}
The expectation value of an observable $A$ with respect to $|\alpha >$, $<A>_{\alpha}$ is defined as
\begin{equation}\label{expe}
<A>_{\alpha}\equiv <\alpha |A|\alpha >=\sum_{|a'>\in \left\{ |a>\right\} }a'P_{a'}.
\end{equation}
From a phenomenological point of view, the probability (\ref{prob}) has to be determined after a lot of measures on systems of identical preparation, all characterized by the same ket $|\alpha >$, {\it i.e.} e set of pure states. A mixture is a set of states $|\alpha_{i}>$, each one with a percentage weight $w_{i}$, such that $\sum_{i}w_{i}=1$. The ensemble average of an operator $A$ on a mixture is defined as
\begin{equation}
[A]=\sum_{i}w_{i}<\alpha_{i}|A|\alpha_{i}>=\sum_{i}w_{i}<A>_{\alpha_{i}},
\end{equation}
where where the expectation value (\ref{expe}) is weighted by the $w_{i}$'s. If the density operator $\rho$ is introduced,
\begin{equation}
\rho\equiv\sum_{i}w_{i}|\alpha_{i}><\alpha_{i}|,
\end{equation}  
the ensemble average rewrites
\begin{equation}
[A]=tr(\rho A).
\end{equation}
The density operator is self adjoint, and $tr(\rho)=1$. For a pure state, $|\alpha_{j}>$, $w_{j}=1$, the density operator $\rho=|\alpha_{n}><\alpha_{n}|$ is idempotent, $\rho^{2}=\rho$, so that $tr(\rho^{2})=1$.\\
It's worth remarking that probabilities (\ref{prob}) are given by squared amplitudes, and the vector $|\alpha>$ in $H$ cannot be measured: the vector $|\alpha>$ in $H$ and the ray $\lambda|\alpha>$, $\lambda\in Z, |\lambda|^{2}=1$ are a representation of the same physical state, as they give the same contribution to a measure. It is rays, rather then vectors, that represents physical states in $H$.

\paragraph{Poisson brackets and commutators}\label{poisson}
Quantum states are characterized by the canonical commutation relations, which hold between those operators that define the quantum system.\\
The canonical commutation relations
\begin{equation}\label{cancommrel}
[x_{i},x_{j}]=0,\ \ [p_{i},p_{j}]=0,\ \ [x_{i},p_{j}]=i\hbar \delta_{ij}
\end{equation}
can be inferred from the quantum properties of translations alone: such a derivation can be performed also for quantum objects that have no classical analogue. However, quantum-mechanical commutators $[\ \,\ \ ]$ can be obtained from the corresponding classical Poisson brackets $[\ \,\ \ ]_{cl}$
\begin{equation}
[A(p,q),B(p,q)]_{cl}\equiv\sum_{s}\left(\frac{\partial A}{\partial q_{s}}\frac{\partial B}{\partial p_{s}}-\frac{\partial A}{\partial p_{s}}\frac{\partial B}{\partial q_{s}}\right)
\end{equation}
by the replacement 
\begin{equation}\label{pass}
[\ \,\ \ ]_{cl}\rightarrow\frac{[\ \,\ \ ]}{i\hbar}.
\end{equation}
This identification is possible because both Poisson brackets and commutators obey the algebraic properties
\begin{equation}
[A,A]=0,\ \ [A,B]=-[B,A],\ \ [A,c]=0, c\in Z,
\end{equation}
\begin{equation}
[A+B,C]=[A,C]+[B,C],\ \ [A,BC]=[A,B]C+B[A,C],
\end{equation}
\begin{equation}
[A,[B,C]]+[B,[A,C]]+[C,[A,B]]=0,
\end{equation}
the factor $1/i\hbar$ being necessary in (\ref{pass}), because, while the Poisson brackets of two real functions are real, the commutator of two Hermitian operator is anti-Hermitian. This way, the third of (\ref{cancommrel}) can be derived from the classical $[x_{i},p_{j}]_{cl}=\delta_{ij}$ through (\ref{pass}).\\
The motion equation for an operator $A^{H}$ in the Heisenberg picture, whose Schroedinger representation does not depend on time explicitly,
\begin{equation}\label{heis}
\frac{dA^{H}}{dt}=\frac{1}{i\hbar}[A^{H},H]
\end{equation}
can be derived, from a quantum-mechanical point of view, from the properties of the time-evolution operator and the definition of $A^{H}$ alone. On the other hand, such a motion equation could be obtained, from a classical point of view, for a function $A(p,q)$ that does not depend on time explicitly,
\begin{equation}
\frac{dA(p,q)}{dt}=[A,H]
\end{equation}
through the replacement (\ref{pass}). Again, the replacement (\ref{pass}) allows one to obtain quantum relations from classical ones, but (\ref{heis}) can be worked out from quantum properties even for operators that have no classical analogue, thus pointing out once more that classical mechanics can be derived as the limit of quantum mechanics, but the converse is not true.

\section{Quantum operators and wave functions}
After implementing the canonical commutation relations from a quantum point of view, it is possible to find out the representations of such operators.\\
In the coordinate representation, the wave function $\psi_\alpha(x)$ for the state $|\alpha>$ is given by
\begin{equation}
\psi_\alpha(x)=<x|\alpha>
\end{equation}
while, in the momentum representation, we get
\begin{equation}
\psi_\alpha(p)=<p|\alpha>.
\end{equation}
These two representations are linked by Fourier duality, i.e., it is possible to pass from a representation to the other one by Fourier transform.
The probability of finding the system in a given state, say $|\alpha'>$, is given by \ref{expe}
\begin{equation}
P_{\alpha'}=\int_D dx |\psi_{\alpha'}(x)|^2<\infty,
\end{equation}
and the equivalent version in the momentum representation, so that wave functions belong to the Hilbert space $\mathcal{H}=L^2(D,dx)$, where the integration domain $D$ (and therefore, the specification of $L^2$) depend on the physics to be described.\\
Since the momentum $p$ is the generator of translations, its action in the coordinate and in the momentum representations is given by $P=-i\hbar \partial/\partial x$ and $i\hbar\partial/\partial p$. respectively, while $x$ can be shown to behave as a multiplicative operator. 

\section{Difference operators versus differential operators}
It is also possible to implement a quantum-mechanical system by introducing difference operators instead of differential operators \cite{dob02}, i.e.,
\begin{equation}
D_{x_1,x_2}f(x)=\frac{f(x_2)-f(x_1)}{x_2-x_1},\ \ x_2\neq x_1,\ \  x_1, x_2\in \mathds{R},
\end{equation}
and the differential case should be recovered in the limiting process $x_2\rightarrow x_1$.\\
 In particular, the derivative can be replaced by two kinds of operators
\begin{itemize}
	\item additive operators $D^a$, i.e.
	\begin{equation}
	D^a f(x)=\frac{f(x+a)-f(x-a)}{(x+a)-(x-a)}=\frac{f(x+a)-f(x-a)}{2a},\ \ a\in \mathds{R}
\end{equation}
  \item multiplicative operators $D^q$, i.e.
	\begin{equation}
	D^q f(x)=\frac{f(qx)-f(q^{-1}x)}{qx-q^{-1}x}=\frac{1}{x}\frac{f(qx)-f(q^{-1}x)}{q-q^{-1}},\ \ q\in \mathds{R}.
\end{equation}
\end{itemize}
Considering difference operators instead of differential operators can be consistent for those cases, where differential operators do not exist, or where a discretized underlying structure is hypothesized, i.e., if a lattice is considered. According to the previous definitions, two kinds of lattices can be recognized, respectively, i.e.
\begin{itemize}
	\item uniform $a$-lattices, i.e. $\mathds{L}_a=\left\{x_0+ja|j\in\mathds{Z}, x_0\in\mathds{R}\right\}$
	\item uniform $q$-lattices, i.e. $\mathds{L}_q=\left\{x_0 q^j|j\in\mathds{Z}, x_0\in\mathds{R}, x_0\neq0\right\}$.
\end{itemize}
The introduction of a scale is closely related to the definition of a scale, and the consequence continuum limit. The relevance in introducing a scale is the possibility to focus the attention from the points of the lattice to the intervals defined by the lattice, with the aim of approximating continuous functions on $\mathds{R}$ with functions that are constants on such intervals. For any given scale, one can approximate functions on the lattice, and one can pass from one scale to the next one by a coarse-graining map \cite{CVZ07a}.

\paragraph{The Eherenfest theorem}
If motion equations are looked for, the proper Hamiltonian operator has to be found. In Quantum Mechanics, for a physical system, which has a classical analogue, the variables $x$ and $p$ have to be replaced by the correspondent operators.\\
The Heisenberg equations of motion can be evaluated for a free particle, whose Hamiltonian is taken of the form
\begin{equation}\label{ham}
H=\frac{p^{2}}{2m}=\sum_{i}\frac{p_{i}^{2}}{2m},
\end{equation}
where $i=1,2,3$, and $m$ is the mass of the particle. In the Heisenberg picture, one has
\begin{equation}\label{commp}
\frac{dp_{i}}{dt}=\frac{1}{i\hbar}[p_{i},H]=0,
\end{equation}
\begin{equation}\label{commx}
\frac{dx_{i}}{dt}=\frac{1}{i\hbar}[x_{i},H]=\frac{p_{i}}{m}=\frac{p_{i}(0)}{m}:
\end{equation}
the momentum operator is a constant of the motion, $p_{i}(0)=p_{i}(t)$, while for the position operator, the solution
\begin{equation}
x_{i}(t)=x_{i}(0)+\left(\frac{p_{i}(0)}{m}\right)t
\end{equation}
is found.\\
If a potential $V(x)$ is added to the Hamiltonian (\ref{ham}), {\it i.e.}
\begin{equation}
H=\frac{p^{2}}{2m}+V(x),
\end{equation}
eqs. (\ref{commp}) and (\ref{commx}) rewrite
\begin{equation}\label{commp2}
\frac{dp_{i}}{dt}=\frac{1}{i\hbar}[p_{i},V(x)]=-\frac{\partial}{\partial x_{i}}V(x),
\end{equation}
\begin{equation}
\frac{dx_{i}}{dt}=\frac{1}{i\hbar}[x_{i},H]=\frac{p_{i}}{m},
\end{equation}
so that
\begin{equation}
\frac{dx^{2}_{i}}{dt^{2}}=\frac{1}{m}\frac{dp_{i}}{dt},
\end{equation}
which, together with (\ref{commp2}), gives the quantum analogue of the Newton law,
\begin{equation}
m\frac{d^{2}x}{dt^{2}}=-\nabla V.
\end{equation}
Its expectation value reads
\begin{equation}
m\frac{d^{2}}{dt^{2}}<x>=-<\nabla V>,
\end{equation}
and is known as the Eherenfest theorem: expectation values are independent of the picture, and there is no relic of $\hbar$. While, in the case of the free Hamiltonian, the time evolution of a wave-packet describes a delocalization as time goes by, in the case of the potential $V(x)$, the motion of center of the wave-packet is that of a classical particle with a potential $V(x)$.\\
For an application of the Ehrenfest theorem in the framework of canonical quantum gravity, see \cite{gre91}.

\paragraph{Hamilton-Jacobi formalism}
In the case that the Hamiltonian $H$ depends explicitly on the time $t$, a canonical transformation \cite{Arn} can be looked for, such that the coordinates and the momenta $(q,p)$ evaluated at the time $t$ can be written as constants quantities $(q_{0},p_{0})$ at the time $t$,  {\it i.e.}
\begin{equation}
q=q(q_{0}, p_{0}, t),\ \ p=p(q_{0}, p_{0}, t):
\end{equation}
if the new variables are constant in time, the transformed Hamiltonian $K=H+\partial F/\partial t$, where $F$ is the generating function of the canonical transformation, vanishes identically, and the equations of motion read
\begin{equation}
\frac{\partial K}{\partial P_{i}}=\dot{Q_{i}}=0,\ \ -\frac{\partial K}{\partial Q_{i}}=\dot{P_{i}}=0.
\end{equation}
If $F\equiv F(q,P,t)$, then $p_{i}=\partial F/\partial q_{i}$, and $K$ rewrites
\begin{equation}
H\left(q_{1}, ...,q_{n},\partial F/\partial q_{1},...,\partial F/\partial q_{n}\right)=\partial F\partial t=0,
\end{equation}
which is known as the Hamilton-Jacobi equation, and its solution, $S$, is the principal Hamilton function. $S$ is the generating function of a canonical transformation that leads to constant coordinates and momenta, and defines an equivalence between the $2n$ first-order differential equations of motion and the first-order partial-derivative Hamilton-Jacobi equation. The Hamilton principal function is related to the Lagrangian $L$ by
\begin{equation}
S=\int dt L + const,
\end{equation}
since
\begin{equation}
\frac{dS}{dt}=\frac{\partial S}{\partial q_{i}}\dot{q}_{i}+\frac{\partial S}{\partial t}=p_{i}\dot{q}_{i}-H.
\end{equation}
It is always possible to split the function $S$ into two parts, one that depends on the $\left\{ q_{i}\right\}$ only, and one that depends on time only, if the Hamiltonian $H$ does not depend on time explicitly: in this case, the Hamilton-Jacobi equations reads
\begin{equation}\label{918}
\frac{\partial S}{\partial t}+H\left( q_{i},\frac{\partial S}{\partial q_{i}}\right)=0,
\end{equation}
so that it can be guessed that the expression of $S$ as a function of the new momenta $\left\{ \alpha_{i}\right\}$  should write
\begin{equation}\label{919}
S(q_{i}, \alpha_{i},t)=W(q_{i}, \alpha_{i})-\alpha_{1}t.
\end{equation}
After direct substitution of (\ref{919}) in (\ref{918}), one finds
\begin{equation}
H\left(q_{i},\frac{\partial W}{\partial q_{i}}\right)=0,
\end{equation}
which is independent of time, and the integration constant $\alpha_{1}$ corresponds to the constant value of the Hamiltonian $H$. The function $W$, the characteristic Hamilton function, is defined by the condition
$H(q_{i}, p_{i})=\alpha_{i}$ , and generates a canonical transformation where all the coordinates are cyclic.

\paragraph{The WKB method}\label{wkb}
From the Schroedinger equation in 1 dimension 
\begin{equation}\label{schro}
\left(\frac{p^{2}}{2m}+V(x)\right)\psi=H\psi=E\psi,
\end{equation}
the probability density
\begin{equation}
\rho(x,t)=|\psi(x,t)|^{2}
\end{equation}
and the probability flux
\begin{equation}
j(x,t)=\left(\frac{\hbar}{m}\right)Im(\psi^{+}\frac{d\psi}{dx})=\frac{\rho\nabla S}{m},
\end{equation}
which obey the continuity equation
\begin{equation}\label{conti}
\frac{\partial\rho}{\partial t}+\frac{dj}{dx}=0,
\end{equation}
lead to the expression of the wave function
\begin{equation}\label{strange}
\psi(x,t)=\sqrt{\rho(x,t)}e^{\frac{iS(x,t)}{\hbar}}.
\end{equation}
After direct substitution of (\ref{strange}) in the time dependent Schroedinger equation (\ref{schro}), it is easy to recognize that the solution of the classical Hamilton-Jacobi equation is
\begin{equation}
S(x,t)=W(x)-Et=\pm\int_{x}dx'\sqrt{2m[e-V(x')]}-Et:
\end{equation}
we stress that the phase of the wave function can be interpreted as the solution of the Hamilton-Jacobi equation from a classical point of view.\\
For a stationary state $\partial\rho/\partial t=0$: by means of the continuity equation (\ref{conti}), an expression for $\rho$ is found, which, substituted in the wave function (\ref{strange}), leads to the approximated solution
\begin{equation}
\psi(x,t)\simeq\left(\frac{const}{[E-V(x)]^{1/4}}\right)e^{\pm\left(\frac{i}{\hbar}\right)\int_{x}dx'\sqrt{2m[e-V(x')]}-\frac{iEt}{\hbar}},
\end{equation}
which is known as the WKB (Wemtzel, Kramers, Brillouin) solution.

\section{Time}\label{1.4}
The definition of time is not unique, despite its feature of commonly-experienced quantity. The measure of time does not shed light on its physical nature, both from a classical and from a quantum-mechanical perspective. In GR, time is generalized as a coordinate, which is not, however, the physical time. The puzzle is far from being solved \cite{rov91,ber92}, and the inadequacy of every a priori characterization leads to different possibilities of treating time in canonical quantum gravity \cite{Is92}, as well as the definition of partial observables \cite{rovlib,Ro03,dit04,Th06}.  
\paragraph{The definition of time}
Even from a classical point of view, in Newtonian physics, time  $t$ is a paramater, and can be measured by means of clocks. Clocks, however, do not measure time directly, but they display its representation $T(t)$, which might have a linear functional dependence on $t$, {\it i.e.}, $T(t)=\alpha t$. The correspondence is nevertheless not perfect, as $-\infty<t<+\infty$, while $T(t)$ is defined in within a finite interval, and is always affected by an experimental uncertainty. Furthermore, the measure of a time-dependent physical quantity $Q$ depends on $T$ rather than on $t$, {\it i.e.}, $Q=Q(T)$.\\
The same idea of external parameter that describes the dynamics of a system is present in QM too. In fact, the Schroedinger equation treats the time variable $t$ as disconnected from the other physical coordinates.\\
\paragraph{Time in QG}
The problem of the definition of time in Quantum Gravity arises from the conflict between the classical definition of time as a fixed parameter, and the invariance of General Relativity under 4-dimensional diffeomorphisms, which might lead to contradictory quantized models. In fact, on one hand, it is difficult to introduce the quantum analogue of time, because time is not a physical observable, and, on the other hand, time lies at the basis of equal-time canonical commutation relations, as well as of the notion of a Hilbert space, where the scalar product is conserved. Since GR is invariant under 4-dimensional diffeomorphism, time is treated as a mere coordinate, but, in order to obtain the Newtonian limit, it would be necessary to choose a particular foliation as fundamental, to label events on a manifold according to some physical clock.\\
The emergence of the classical role of time from a quantum theory can be achieved either before or after the quantization, or by some phenomenological considerations, in a model where time plays no precise role.\\
If time is regarded to as fundamental, the canonical constraints are to be solved before the quantization of the system, where the internal time is expressed as a functional of the canonical variables, but picked up from the set of all the other dynamical variables by suitable canonical transformations. Alternatively, the interaction of matter fields with the space-time can be used as physical clocks, or a notion of cosmological time can be found in the variable conjugate to a dynamical cosmological constant.\\
If one wants to discover the role of time after the quantization, it is physical states, written as functional of a fixed background geometries, that carry such an information, in the WDW approach: these states can be treated as operator, in the third-quantization scheme, or interpreted in their semi-classical limit.\\
In the third approach, observables can be chosen as physical clocks, if the solutions of the WDW equations are interpreted as probability densities, or may be associated to operators which commute with all the constraints. 
Also, many efforts have been made in order to define the history of a quantum system, without involving the notion of time.\\ 
For a further discussion of these tantalizing topics see section \ref{probltime}.

\paragraph{Partial Observables}
The failure to apply to General Relativity the Newtonian description of a fixed background with an external time as a scenario where the particle dynamics takes place has lead to a new concept of the universe, whose state is not represented by a given field configuration, but by an equivalence class of field configurations under active diffeomorphisms (which do not involve coordinates ), which imply invariant field equations under coordinate changes: the new role of coordinates allows for a description of physical phenomena as the reciprocity between physical objects and the gravitational field, and the time coordinate cannot be assigned any peculiar role in the notion of states and observables. Because of that, no privileged-reference body or clock can help interacting with the gravitational field, thus enforcing the need to explain mechanics as the connection between partial observables.\\
Given a Gelfand triple $S\in K\in S'$, partial observables are associated to commuting operators in the kinematical space $K$, and their kinematical eigenstates are in $S$ \cite{rovlib}. The equivalence of the scalar product between kinematical states and dynamical states is established by the ''projection operator'', whose matrix elements are the transition amplitudes. This way, the result of the measurement operation is the description of the spectral properties of an operator in the kinematical space, while the dynamical space accounts for the correlation between different sets or measurements.\\
The idea of partial observables can be even further generalized in order to define Dirac observables in gauge systems, and techniques have been developed to perform a reduced-phase-space quantization on diffeomorphism-invariant states. 
   
\section{Quantization of Hamiltonian constraints}
First-class constraints and second-class constraints can be quantized, according to their nature \cite{HT}.\\
Second-class constraints $\chi_{\alpha}$ can be expressed in the quantum theory by solving them classically, as functions of some classical variables, ($x^{\alpha}$), in terms of some other independent variables, ($y^{i}$), {\it i.e.} $\chi_{\alpha}=0\Leftrightarrow x^{\alpha}=f^{\alpha}(y^{i})$, and then defining the quantum representation of the Dirac brackets, $[\hat{y}^{i},\hat{y}^{j}]=i\hbar\sigma^{ij}(\hat{y}^{k})$. Since it is often difficult to find irreducible representations of the Dirac brackets, it might be convenient to turn second-class constraints into first-class constraints by adding extra-variables, and then using Poisson brackets.\\
First-class constraints, on the other hand, can be quantized by means of the reduced-phase-space method, the Dirac Method, and the Dirac-Fock method.\\
The reduced-phase-space method consists in eliminating the gauge degrees of freedom by quantizing only a complete set of gauge invariant functions, to which a complete set of gauge-invariant observables is associated, so that the quantization of second-class constraints becomes equivalent to the reduced phase-space, under canonical gauge conditions. The difficulties arising in this quantization paradigm are due to the fact that the elimination of the gauge degrees of freedom might destroy the invariance under some symmetry, as well as locality in space.\\
In the Dirac method, gauge degrees of freedom are eliminated, but kept as operators in a larger space, where unphysical features are eliminated by defining gauge-invariant physical states. The definition of a scalar product in this space contains a gauge condition, able to eliminate the integration over the gauge degrees of freedom, so that the physical scalar product coincides with the scalar product in the reduced phase space.\\
The Dirac-Fock quantization of first-class constraints consists in assigning to each couple of constraints and to each couple of their conjugated variables a couple of destruction and creation operators, thus adopting a Fock representation with an indefinite metric. As a result, the whole system, composed of the direct product of the Fock space times the standard representation of the physical degrees of freedom, is invariant under gauge transformations because destruction operators only annihilate physical states, but invariance for creation operators is assured by the presence of null states, whose scalar product with any state vanishes.\\
In the next paragraphs, we will apply these quantization techniques to relevant examples, where we will appreciate the powerful tools of Dirac quantization. 
   
\paragraph{Non-relativistic particle}

The action for a single non-relativistic particle in a potential $V$ can be written as a function of the canonical variables $\left\{x^{i},p_{i};\right\}$ , $i=1,2,3$, and the absolute time $t$:
\begin{equation}\label{one}
S[x^{i},p_{i}]=\int dt \left(p_{i}\frac{dx^{i}}{dt}-H\left(x^{i},p_{i},t\right)\right).
\end{equation}
An equivalent expression for (\ref{one}) can be obtained by considering an arbitrary label time $\tau$ and the set of variables $\left\{x^{\mu}=\left(t,x^{i}\right),p_{\mu}=\left(p_{0},p_{i}\right)\right\}$, $\mu=0,1,2,3$, through the introduction of the constraint $\mathcal{H}\equiv p_{0}+H\left(x^{\mu},p_{i}\right)=0$ and a Lagrange multiplier $N$, which are needed in order to restore the right number of degrees of freedom,
\begin{equation}\label{two}
s[x^{\mu},p_{\mu};N]=\int d\tau\left(p_{\mu}\dot{x}^{\mu}-N\mathcal{H}\right):
\end{equation}
the variation with respect to $p_{0}$ leads to the physical connotation of $N$, $N=\dot{t}$, while the variations with respect to $x^{i}$, $p_{i}$ and $t$ lead to the Hamilton equations, expressed for the parameter $\tau$, and to the energy-balance equation $\dot{H}=(\partial H/\partial t)\dot{t}$, respectively.\\
The quantization of the constraints consists in the quantization of the variables involved in the constraints, and in the definition of the spaces where these operators live in.\\
After defining the operator analogues, obeying standard commutation relations, of the canonical variables, the super-Hamiltonian, which becomes an operator itself as an effect the substitution, $\mathcal{H}=0\rightarrow\hat{\mathcal{H}}=0$, picks up the physical states, $\hat{\mathcal{H}}\psi=0$, for which a suitable Hilbert space and appropriate self-adjoint operators have to be defined. If the coordinate representation is chosen for the operators, {\it i.e.}, $x^{\mu}\rightarrow \hat{x^{\mu}}\equiv x^{\mu}$, and $p_{\nu}\rightarrow \hat{p}_{\nu}\equiv -i\partial_{\nu}$, the wavefunctions are $\psi(x^{\mu})$. Since the operators $\hat{x}^{\mu}$ and $\hat{p}_{\mu}$ satisfy the commutation relation $[\hat{x}^{\mu},\hat{p}_{\nu}]=i\delta^{\mu}_{\nu}$, the factor ordering of these operators becomes crucial in the requirement of preserving the covariance and the algebra of constraints, as well as in the definition of the inner product.\\
As an example for the definition of the inner product, the dynamics of a non relativistic particle can be investigated, and the quantum constraint for eq. (\ref{two}) can be expressed as invariant under diffeomorphisms in the $x^{i}$-space: s factor ordering for $\hat{\mathcal{H}}$ can be chosen, such that a Laplacian on the field $\psi(x^{i})$ appears,
\begin{equation}
\hat{\mathcal{H}}=\hat{p}_{0}+\frac{1}{2}{^{3}g}^{-1/4}\hat{p}_{i}\sqrt{^{3}g}g^{ij}\hat{p}_{j}{^{3}g}^{-1/4}+V.
\end{equation}
Accordingly, the inner product 
\begin{equation}
<\psi_{1},\psi_{2}>\equiv\int_{t=const}d^{3}x\psi^{*}_{1}(x,t)\psi_{2}(x,t)
\end{equation} 
is formally consistent with the space of the constraints $\hat{\mathcal{H}}\psi=0$ and defines the norm of the Hilbert space of the solutions; in fact, the two fields obey the constraint $\mathcal{H}\psi_{1}=\mathcal{H}\psi_{2}=0$  and the continuity equation $\rho_{12}+j^{i}_{12}=0$, where $\rho_{12}\equiv\psi_{1}^{*}\psi_{2}$, and $j^{i}_{12}\equiv\frac{1}{2}ig^{ij}\left(\psi_{1}^{*}\orch{\partial}_{j}\psi_{2}\right)$. It is worth noting that the constraint $\hat{\mathcal{H}}\psi=0$ is the Schroedinger equation, and the inner product between two wave-function does not depend, by construction, on any particular time slice.\\

\paragraph{Relativistic particle}
The definition of an inner product is not always straightforward, and one-particle states have not always a precise physical meaning. This is the case of a relativistic particle $\phi(x^{\mu})$, with a parametrized action of the form (\ref{two}), in the $x^{\mu}$ representation in curved space-time, whose constraint
\begin{equation}
\mathcal{H}\equiv \frac{1}{2m}\left(g^{\mu\nu}p_{\mu}p_{\nu}+m^{2}\right)=0
\end{equation}
has to be quantized covariantly under space-time diffeomorphisms.\\
A factor ordering of the super-Hamiltonian can be found, such that the D'Alembert operator operates on scalars, {\it i.e.}
\begin{equation} \hat{\mathcal{H}}\phi(x^{\mu})=\left[(^{4}g)^{-1/2}\hat{p}_{\mu}\sqrt{^{4}g}g^{\mu\nu}\hat{p}_{\nu}+m^{2}\right]\phi(x^{\mu})=0,
\end{equation}
which is the Klein-Gordon equation. As in the previous case, a continuity equation can be worked out,
\begin{equation} 
\nabla_{\mu}j^{\mu}_{12}=0,\ \   j^{\mu}_{12}\equiv\frac{1}{2}g^{\mu\nu}\phi_{1}\orch{\partial}_{\nu}\phi_{2}
\end{equation}
for two fields $\phi_{1}$ and $\phi_{2}$. Nonetheless, the functional
\begin{equation}
\Omega\left[\phi_{1},\phi_{2}\right]=\frac{1}{2}\int_{\sigma}d\sigma_{\mu}g^{\mu\nu}\phi_{1}\orch{\partial}_{\nu}\phi_{2},
\end{equation}
though independent of the space-time hypersurface taken into account, does not define the Hilbert space of the solutions, because it is antisymmetric: it is, indeed, the symplectic form of such a space.\\
It is possible, however, to build a complex Hilbert space from the solutions of the real Klein-Gordon equation in stationary space-time, endowed with a time-like hypersurface orthogonal Killing vector field $t^{\mu}$
\begin{equation}
t^{\mu}=N^{2}g^{\mu\nu}t,_{\nu},
\end{equation}
where $x^{\mu}=(t,x^{i})$, $N$ and $N^{i}$ being the lapse function and the shift vector, respectively. In such coordinates, the Klein-Gordon equation reads
\begin{equation}
N^{2}\hat{\mathcal{H}}_{\phi}\equiv \ddot{\phi}+\hat{H}_{N}\phi=0,
\end{equation}
\begin{equation}
\hat{H}_{N}\phi\equiv N\left[-g^{-1/2}\left(Ng^{1/2}g^{ij}\phi,_{j}\right),_{i}+Nm^{2}\phi\right].
\end{equation}
In the Hilbert space whose inner product reads
\begin{equation}
(\phi_{1},\phi_{2})\equiv\int d^{3}xg^{1/2}N^{-1}\phi_{1}\phi_{2},
\end{equation}
the operator $\hat{H}_{N}$ is symmetric and positive definite, and has a complete set of eigenfunctions $u_{E}(x^{a})$ obeying the eigenvalue equation
\begin{equation}
\hat{H}_{N}u_{E}(x^{i})=E^{2}u_{E}(x^{i}),\ \ E\geq E_{0}>0:
\end{equation}
the projection of the solutions $\phi(t,x^{i})$ of the Klein-Gordon equation along this basis allows one to separate positive and negative frequencies,
\begin{equation}
\phi(t,x^{i})=\phi^{+}(t,x^{i})+\phi^{-}(t,x^{i}),\ \ \phi^{\pm}(t,x^{i})=\int_{E_{0}}^{\infty}dEu_{E}(x^{i})e^{\mp iEt}.
\end{equation}  
The map $J$, {\it i.e.} the complex structure $J^{2}=-1$ of the Hilbert space, sends any real solution of the Klein-Gordon equation in another real solution,
\begin{equation}
\phi\rightarrow J\phi\equiv i\phi^{+}-i\phi^{-},
\end{equation}
does not depend on time, is compatible with $\Omega$, {\it i.e.}
\begin{equation}
\Omega[\phi_{1}, J\phi_{2}]=\Omega[\phi_{2}, J\phi_{1}],\ \ \Omega[\phi, J\phi]\geq0,
\end{equation}
thus expressing the space of the solution of the Klein-Gordon equation as a complex vector space, where the antilinear scalar product reads
\begin{eqnarray}
<\phi_{1},\phi_{2}>\equiv \Omega[\phi_{1}, J\phi_{2}]+i\Omega[\phi_{1}, J\phi_{2}],\\ <\phi_{1},J\phi_{2}>=i<\phi_{1},\phi_{2}>,\ \ <J\phi_{1},\phi_{2}>=-i<\phi_{1},\phi_{2}>,
\end{eqnarray}
so that norms are positive definite, $<\phi,\phi>\equiv \Omega[\phi, J\phi]$.

\paragraph{Scalar field}
Let $\mathcal{M}^{4}$ be a pseudo-Riemannian manifold, endowed with the metric tensor $g_{\mu\nu}(y^{\rho})$; a $(3+1)$-slicing can be performed, such that a one-parameter family of space-like hypersufaces $\Sigma^{3}_{t}:y^{\rho}=y^{\rho}(t,x^{i})$ is uniquely defined by the value of $t$. After the coordinate transformation $y^{\mu}\rightarrow y^{\mu}(t,x^{i})$, the line element reads
\begin{equation}
ds^2 = g_{\mu \nu }dy^{\mu }dy^{\nu }
= -N^2dt^2 + h_{ij}
(dx^i + N^i dt)(dx^j + N^jdt),  
\label{b} 
\end{equation}
where $N$ and $N^{i}$ are the lapse function and the shift vector, respectively, that decompose the the deformation vector $N^{\mu}=\partial_{t}p_{\mu}$ along the basis $\{ {\bf n }, {\bf e}_i \}$, composed of the normal vector $n^{\mu}$ to the family of hypersurfaces $\Sigma^{3}_{t}$ and the three tangent vectors $e^{\mu}_{i}$, and $h_{ij}\equiv g_{\mu\nu}e_{i}^{\mu}e^{\nu}_{j}$ is the metric tensor induced on $\Sigma^{3}_{t}$ by the coordinate transformation. From the line element (\ref{b}), it is possible to obtain the expression of the contravariant and of the covariant components of the normal vector ${\bf n}$ in the coordinates $(t,x^{i})$: $n^{\mu}=(1/N; -N^{i}/N)$, $N_{\mu}=(-N, \vec{0})$. The so-called kinematical action reads
\begin{equation}    
S^k (p_{\mu }, y^{\mu }) = \int_{{\cal M}^4}d^3xdt\left\{   
p_{\mu }\partial _t y^{\mu } - N^{\mu }p_{\mu } \right\}.   
\label{e} 
\end{equation}
The action of a self-interacting scalar field on a fixed background,
\begin{equation}  
S^{\phi }(\pi_{\phi }, \phi ) = 
\int_{{\cal M}^4} \left\{ \pi _{\phi }\partial _t\phi  - NH^{\phi } - 
N^iH^{\phi }_i \right\}d^3xdt, 
\label{c} 
\end{equation}
where $\pi_{\phi}$ is the momentum conjugate to $\phi$, and the Hamiltonian terms read

\begin{equation}  
H^{\phi } \equiv 
\frac{1}{2\sqrt{h}}{\pi _{\phi }}^2 + 
\frac{1}{2}\sqrt{h}h^{ij}\partial _i \phi \partial _j\phi + \sqrt{h}V(\phi )  
\,\quad H^{\phi }_i \equiv \partial _i \phi {\pi }_{\phi },  
\label{d} 
\end{equation}
can be quantized by adding the kinematical action (\ref{e}). In fact, without the kinematical action, the variation has to be performed with respect to $\phi$ and $\pi_{\phi}$, but no precise role is assigned to $N$, $N^{i}$, and $h_{ij}$, since the background is fixed. If the kinematical action $S^{k}$ is taken into account, so that
\begin{equation}  
S^{\phi k}\equiv S^{\phi } + S^k = 
\int_{{\cal M}^4} \left\{ \pi _{\phi }\partial _t\phi 
+ p_{\mu }\partial _ty^{\mu } - 
N(H^{\phi } + H^k) - N^i(H^{\phi }_i + H^k_i) \right\} d^3xdt, 
\label{f} 
\end{equation}
where $H^k \equiv p_{\mu }n^{\mu } \, \quad H^k_i \equiv p_{\mu }e^{\mu }_i$, the field equations for $\phi$ remain unchanged, but the Hamiltonian constraints $H^{\phi } = - p_{\mu }n^{\mu } $ and $ 
H^{\phi }_i = - p_{\mu }e^{\mu }_i $ are obtained. \\
The canonical quantization is accomplished by the assumption that the states of the system are functionals of the variables, $y^{\mu}$ and $\phi$, $\Psi [y^{\mu }(x^i), \phi (x^i)]$, and by the implementation of the canonical variables to operators, $\{ y^{\mu }, p_{\mu }, \phi , \pi _{\phi } \}\rightarrow \{ \hat{y}^{\mu } ,  \; 
\hat{p}_{\mu } = -i\hbar \delta (\; ) /\delta y^{\mu } , 
\; \hat{\phi } , \; 
\hat{\pi }_{\phi } 
= -i\hbar \delta (\; )/\delta \phi \}$, the quantum dynamics being described by the equations
\begin{equation}   
i\hbar n^{\mu }\frac{\delta \Psi }{\delta y^{\mu }} 
= \hat{H}^{\phi } \Psi = \left[  
-\frac{\hbar ^2}{2\sqrt{h}} 
\frac{\delta  }{\delta \phi } \frac{\delta  }{\delta \phi } +  
\frac{1}{2}\sqrt{h}h^{ij}\partial _i \phi \partial _j\phi 
+ \sqrt{h}V(\phi ) \right] \Psi  
\end{equation}
\begin{equation}
i\hbar e^{\mu }_i\frac{\delta \Psi }{\delta y^{\mu }} 
= \hat{H}^{\phi }_i \Psi = -i\hbar \partial _i \phi \frac{\delta \Psi }
{\delta \phi }.
\label{mm} 
\end{equation}  
The space of the solutions of (\ref{mm}) can be cast into a Hilbert space by defining the inner product
\begin{equation}   
\langle \Psi _1 \mid \Psi _2 \rangle \equiv 
\int_{y^{\mu } = y^{\mu }(x^i)} \Psi_ 1^*\Psi _2 D\phi \, \quad 
\frac{\delta \langle \Psi _1 \mid \Psi _2 \rangle }
{\delta y^{\mu }} = 0,
\label{n} 
\end{equation}
which implies the conserved functional probability distribution $\varrho \equiv \langle \Psi \mid \Psi \rangle$.\\
The semiclassical limit, $\psi=\exp{iS^{\phi k}}$ is recovered by substituting the wave functional $   
\Psi = \exp\left\{ \frac{1}{\hbar }\Sigma (y^{\mu }, \phi )\right\}$ in (\ref{mm}) and taking the zero-th order of the series-expansion, with $\hbar\rightarrow 0$ (see section \ref{classquan}).\\
It is work remarking that (\ref{mm}) has $5\infty^{3}$ degrees of freedom, given by the scalar field $\phi$ and the four components of $y^{\mu}$.

\section{Weyl Quantization}
Weyl quantization \cite{Weyl} consists in assuming canonical commutation relation for two operators $\hat{p}$, $\hat{q}$, as \ref{cancommrel}, and in establishing a different (Weyl) representation of the operators. One can thereafter implement a quantization programme, and then recover information about the standard quantization method via the so-called GNS construction.

\paragraph{Weyl Systems}
Given a symplectic vector space $(E,\omega)$, i.e., a vector space $E$ endowed with a symplectic (non-degenerate, skew-symmetric, bilinear ) form $\omega$, a Weyl system is the strongly-continuous map $W$ from $E$ to unitary transformations on some Hilbert space $\mathcal{H}$
\begin{equation}
W:E\rightarrow \mathcal{U}(\mathcal{H})
\end{equation}
and the Weyl form of the commutation relations reads
\begin{equation}\label{weyl}
W(e_1)W(e_2)=e^{\frac{i}{\hbar}\omega(e_1,e_2)}W(e_2)W(e_1),
\end{equation}
where the cocycle of the representation is determined by the the symplectic structure $w$.\\
Complex coordinates, and the construction of a Fock space, with creation and annihilation operators, can be defined by the introduction of a complex form $J:E\rightarrow E$, $J^2=-1$. An inner product on $E$ can be defined by using $J$ and $\omega$.\\ 
It is possible to decompose the vector space $E$ as $\mathcal{L}\oplus \mathcal{L}^*$, where $\mathcal{L}\subset E$ is a Lagrangian (both isotropic and coisotropic) subspace of $E$. According to the von neumann theorem, the Hilbert space $\mathcal{H}$ is the space of square-integrable functions $\phi$ on $\mathcal{L}$ endowed with the translation-invariant Lebesgue measure $d\mu$, i.e., $\mathcal{H}=L^2(d\mu, \mathcal{L})$. In this decomposition, vectors on $E$ can be defined as $e=(\alpha, \beta)$, $\beta\in \mathcal{L}$, $\alpha\in \mathcal{L}^*$, and the action of $W$ on the functions $\phi$ reads
\begin{eqnarray}
U(\alpha)\phi(q)\equiv W((\alpha,0))\phi(q)=e^{\frac{i}{\hbar}\alpha q}\phi(q)\\
V(\beta)\phi(q)\equiv W((0, \beta))\phi(q)=\phi(q-\beta).
\end{eqnarray} 
The vacuum expectation values of the operators $U$ and $V$ depends on the metric $g$ constructed out of $J$, i.e., $g(e_1, e_2)=w(e_1, Je_2)$.

\paragraph{The Stone-von Neumann Uniqueness theorem}
The Stone-von Neumann Uniqueness theorem \cite{VN31} states that any unitaty irreducible representation of the Weyl commutation relation on $\mathds{C}^n$ is isomorphic to the Schroedinger represntation. Furthermore, as corollary, it is also possible to show that any representation of the Weyl commutation relation on $\mathds{C}^n$ is the direct sum of copies of the Schroedinger representation.

\section{GNS Construction}\label{GNS}
The GNS construction allows one to gain insight onto different representations of a given algebra \cite{haag}.
From a mathematical point of view, a \itshape{state} \normalfont is a normalized positive linear form. 

Given a function $\phi$ and an algebra $\mathcal{A}$, $\phi$ is called a linear form over $\mathcal{A}$ if
\begin{equation}
\phi(\alpha A +\beta B)=\alpha\phi(A)+\beta\phi(B),\ \ \forall A,B\in\mathcal{A}, \forall \alpha, \beta \in \mathds{C}
\end{equation}
If
\begin{equation}
\left\| xy \right\|\leq\left\| x\right\| \left\| y\right\|,\ \ \forall x,y \in \mathcal{A},
\end{equation}
then $\mathcal{A}$ is a Banach algebra. If $\mathcal{A}$ is a Banach algebra, $\phi$ is bounded if
\begin{equation}
\mid\phi(A)\mid\leq c \left\| A\right\|
\end{equation}
and the lowest bound for $c$ is the norm of $phi$.\\
If
\begin{equation}
(Ax,y)=\overline{(y,Ax)}=(x,A^* y),\ \ \forall A\in\mathcal{A},
\end{equation}
then $\mathcal{A}$ is a ${}^{*}$ algebra. If $\mathcal{A}$ is a ${}^{*}$ algebra, with unit, the linear form $\phi$ is a state if it is 
\begin{eqnarray}
real\ \  \phi(A^*)=\phi(A)\\
positive\ \  \phi(A^* A)\geq0\\
normalized\ \  \left\| \phi\right\|=1.
\end{eqnarray}

As a result, a positive linear form over a Banach ${}^{*}$ algebra with unit is bounded, and $\left\| \phi\right\|=\phi(I)$. Furthermore, it satisfies the Schwarz inequality
\begin{equation}
\mid\phi(AB)\mid ^2\leq \phi(AA^*)\phi(BB^*).
\end{equation} 
In fact, if we assume that the self-adjoint elements of $\mathcal{A}$ correspond to physical observables, and that the unit element $I$ correspond to the trivial observable, whose value is 1 for any physical state, then such a linear form can be interpreted as an expectation functional over physical observables.\\

Each positive linear form $\omega$ over a ${}{*}$ algebra $\mathcal{A}$ defines a Hilbert space $\mathcal{H}_\omega$ and a representation $\pi_\omega$ of $\mathcal{A}$ by linear operators acting on $\mathcal{A}$.\\
Since $\mathcal{A}$ is a linear space over the field $\mathds{C}$, $\omega$ defines an Hermitian semi-definite product on $\mathcal{A}$, i.e. 
\begin{equation}\label{scalprod}
<A|B>=\omega (A,B),\ \ <A|A>\geq0,\ \ \mid<A|B>\mid ^2\leq <B|B><A|A>,\ \ \forall A,B\in\mathcal{A}.
\end{equation}
The set $\mathcal{J}\subset\mathcal{A}$, $\mathcal{J}=\left\{ X\in\mathcal{A}:\omega(X^*X)=0\right\}$, is a left ideal, and is called the Gelfand ideal of the state. Eliminating this set from $\mathcal{A}$, i.e., considering $\mathcal{A}/\mathcal{J}$ allows us to obtain linear space equipped with Hermitian positive-definite scalar product, and a vector $\psi$ in this space corresponds to the equivalence class $[A]$ of the elements of $\mathcal{A}$ modulo $\mathcal{J}$, $\psi=\left\{\mathcal{A}+\mathcal{A}\right\}$, with  
\begin{equation}
<\psi|\psi>\equiv \left\| \psi\right\|^2 >0.
\end{equation}
It is worth remarking that the scalar product defined in (\ref{scalprod}) does not depend on  $[A]$. Because of that, the action of the representation $\pi_\omega(A)$, defined on $\mathcal{A}/\mathcal{J}\subset\mathcal{H}_\omega$, is
\begin{equation}
\pi_\omega(A)\psi=[AB]\ \  if\ \  \psi=[B].
\end{equation} 
A representation $\pi$ is called cyclic if a cyclic vector $\Omega\in\mathcal{H}$ exists. A vector $\Omega$ is called cyclic if $\pi(\mathcal{A})\Omega$ is dense in $\mathcal{H}$. If $\mathcal{A}$ has a unit, $\Omega=[I]$. Furthermore
\begin{equation}
\omega(A)=<\Omega|\pi_\omega(A)|\Omega>,
\end{equation}
and it is sometimes referered to as the vacuum state. Similarly, any vector $\psi\in\mathcal{H}_\omega$ defines a state
\begin{equation}
\omega_\psi(A)=<\psi|\pi_\omega(A)|\psi>.
\end{equation}

The GNS theorem (named after Gel'fand, Najmark and Segal) states that, given a C* algebra $\mathcal{A}$ endowed with unit, and a linear positive functional $\omega$ in a compact subset of $\mathcal{A}$, the triple $(\mathcal{H}(\omega), \pi_{\omega}, \Omega)$ exists and is unique.  
\newpage

\chapter{Hamiltonian formulation of the geometrodynamics}

\section{The action for the gravitational field}\label{actiongrav}
Guiding principles in the development of the Lagrangian density for the gravitational field are the Equivalence Principle and the General Covariance. The latter imposes the action be invariant under diffeomorphism, while the former states that, by a coordinate transformation, the metric tensor can always be reduced to a Minkowskian one locally, thus first derivatives of the metric can be made to vanish in any local region. Therefore, if combined together, they forbid the existence of a sensible action for the gravitational field with only first-order derivatives. Hence, second-order derivatives have to be contained in the Lagrangian, but only trough a surface term, to avoid the appearance of third derivatives in the equations of motion. This request rules out some possible actions.\\
Let us consider a 4-dimensional space-time manifold endowed with a metric $g_{\mu\nu}$, the simplest Lagrangian satisfying the above mentioned properties is the Einstein-Hilbert \cite{H15,E16} one, {\it i.e.}
\begin{equation}           
\Lambda=\frac{c^4}{16\pi G}\sqrt{g}R
\end{equation}
being $G$ Newton coupling constant, $R$ the scalar curvature and $g$ the determinant of the metric tensor. As far as $R$ is concerned, in the Einstein formulation its expression in terms of the metric can be calculated from Christoffel symbols 
\begin{equation}
\Gamma^\rho_{\mu\nu}=\{^\rho_{\mu\nu}\}=\frac{1}{2}g^{\rho\sigma}(\partial_\nu g_{\mu\sigma}+\partial_\mu g_{\nu\sigma}-\partial_\sigma g_{\mu\nu}).
\end{equation}
We just have to introduce the Riemann tensor
\begin{equation}
R_{\mu\nu\rho\sigma}=g_{\mu\tau}(-\partial_\sigma\Gamma^\tau_{\nu\rho}+\partial_\rho\Gamma^\tau_{\nu\sigma}-\Gamma^\tau_{\theta\sigma}\Gamma^\theta_{\nu\rho}+\Gamma^\tau_{\theta\rho}\Gamma^\theta_{\nu\sigma})
\end{equation}
and the Ricci tensor $R_{\mu\nu}=g^{\rho\sigma}R_{\mu\rho\nu\sigma}$, such that $R=g^{\mu\nu}R_{\mu\nu}$.\\ 
From these definitions one recognizes that Christoffel symbols are symmetric with respect to the exchange of lower indexes, while the Riemann tensor satisfies 
\begin{equation}
R_{\mu\nu\rho\sigma}=-R_{\mu\nu\sigma\rho}\qquad R_{\mu\nu\rho\sigma}=R_{\rho\sigma\mu\nu}.
\end{equation}
Other properties of $R_{\mu\nu\rho\sigma}$ are the cyclic identity 
\begin{equation}
R_{\mu\nu\rho\sigma}+R_{\mu\rho\sigma\nu}+R_{\mu\sigma\nu\rho}=0\label{cyclic identity}
\end{equation}
and the Bianchi identity
\begin{equation}
\nabla_{\tau}R_{\mu\nu\rho\sigma}+\nabla_{\rho}R_{\mu\nu\sigma\tau}+\nabla_{\sigma}R_{\mu\nu\tau\rho}=0\label{bianchi1}.
\end{equation}
By varying the action with respect to the metric tensor, Einstein equations come out
\begin{equation}
\delta S=-\frac{c^3}{16\pi G}\bigg[\int_M d^4x \bigg(R_{\mu\nu}-\frac{1}{2}g_{\mu\nu}R\bigg)\delta g^{\mu\nu}-2\int_{\partial M} d^3x \delta K\bigg]\label{varEH}
\end{equation}
once the second term in the last relation disappears; in order to impose this condition, we should require the variation of the metric and of its first derivatives vanish on the boundary \cite{P06,SL06a}. Hence, in general, a term is added to the Lagrangian density, in order to cancel the surface piece.\\ 
This second order formulation is equivalent to the Palatini one, where the metric tensor and connections are treated like independent fields (see section \ref{1order}). In fact, the additional equations we obtain from the variation of the Einstein-Hilbert action with respect to $\Gamma^\mu_{\nu\rho}$ imply them be equal to Christoffel connections.\\
A very useful reformulation is one in which we introduce vier-bein vectors $e^\alpha_\mu$, a set of  four orthonormal vectors for each point of the space-time; in fact, we can rewrite the action as 
\begin{equation}
S_{G}=-\frac{c^{3}}{16\pi G}\int d^{4}x \det(e^\alpha_\mu) e_\alpha^\mu e^\nu_\beta R^{\alpha\beta}_{\mu\nu}\quad R^{\alpha\beta}_{\mu\nu}=\partial_{[\mu}\omega^{\alpha\beta}_{\nu]}+\omega^\alpha_{[\mu|\gamma|}\omega^{\gamma\beta}_{\nu]},
\end{equation}
being $\omega^{\alpha\beta}_{\mu}$ Lorentz connections. Now, we can perform variations with respect to $\omega^{ab}_{\mu}$ to obtain the I structure equation
\begin{equation}
\partial_{[\mu} e^\alpha_{\nu]}-\omega^{\alpha\gamma}_{[\mu}e_{\nu]\gamma}=0\label{1streq}
\end{equation}
while from $\delta e^\alpha_\mu$ we get again Einstein equations. By solving the equation (\ref{1streq}), the following expression for connections is obtained $\omega^{\alpha\beta}_\mu=e^{\beta\nu}\nabla_\mu e^\alpha_\nu$.\\
Even though GR is in agreement with experiments, there are hints (appearance of singularities and of closed time-like loops, difficulties in the quantization, dark matter, dark energy) that it has to be changed in the strong field limit.\\  Modified Lagrangian densities consist of a power series in $R$, with both negative, for large scale corrections, and positive powers, which become relevant in a quantum setting. The most general case is that of a generic function $f(R)$ (see \cite{FS08} for a review), {\it i.e.}
\begin{equation}
S=-\frac{c^{3}}{16\pi G}\int d^{4}x \sqrt{-g} f(R),
\end{equation}
from which generalized Einstein equations are found:
\begin{equation}
-\frac{1}{2}g_{\mu\nu}f(R)+f'(R)R_{\mu\nu}-\nabla_{\nu}\nabla_{\nu}f'(R)+g_{\mu\nu}\nabla_{\rho}\nabla^{\rho}f'(R)=\frac{8\pi G}{c^{4}}T_{\mu\nu}.
\end{equation}
These models are equivalent to General Relativity plus additional fields \cite{M95}; in fact by the following conformal rescaling of the metric tensor
\begin{equation}
g_{\mu\nu}\rightarrow f'(R)g_{\mu\nu}=e^{\sqrt{2/3}\varphi}g_{\mu\nu},
\end{equation}
the Lagrangian density becomes
\begin{equation}
L=\frac{c^3}{16\pi G}(R-g^{\mu\nu}\partial_\mu\varphi\partial_\nu\varphi-V(\varphi))
\end{equation}
being the potential V
\begin{equation}
V=e^{-\sqrt{2/3}\varphi}R(e^{-\sqrt{2/3}\varphi}g_{\mu\nu})-e^{-2\sqrt{2/3}\varphi}f(R(e^{-\sqrt{2/3}\varphi}g_{\mu\nu}))
\end{equation}
the only relic of the function $f$.\\
In this sense, several proposals have been made, from terms of the form $\frac{1}{R}$, to explain dark energy \cite{V03}(but these models suffer of instabilities, while the Newtonian limit \cite{DK03} and the evolution of scalar cosmological perturbations \cite{BBPST06} are not reproduced), to Lagrangian $R+\alpha R^2$ with $\alpha$ relevant in the early universe dynamics. However, no $f(R)$ theory exists, up to now, which is able to pass all experimental tests \cite{APT06}.\\
As expected, the new field equations contain higher-order derivatives, and, in particular, forth-order derivatives of the metric tensor appear.
To avoid the appearance of derivatives up to the second order, metric-affine theories \cite{SL06a} can be considered, {\it i.e.} theories in which metrics and connections are independent fields. In particular, this first-order formulation is equivalent to GR plus a cosmological constant term for the free gravitational field, while, in presence of matter, connections are different from Christoffel symbols \cite{M95} and they lead, in general, to different conclusions about instability \cite{S06}.\\
However there are also more radical modifications, based on the introduction of scalars built from the Ricci or the Riemann tensor. An example is given by the Gauss-Bonnet invariant, 
\begin{equation}
R^2-4R_{\mu\nu}R^{\mu\nu}+R_{\mu\nu\rho\sigma}R^{\mu\nu\rho\sigma}
\end{equation}
which arises in many brane-world scenarios \cite{KKL00} and accounts for possible topological changing in the the space-time manifold. The addition of a topological invariant ensures that no modification occurs in the equations of motion.\\
Other models, instead of adding new terms, are based on taking something `less' than General Relativity: to solve the issue related to the second term in relation (\ref{varEH}) and in connection with the holographic principle, there are several attempts to develop a theory for gravity with only the surface term of the Einstein-Hilbert action \cite{SL06b}, \cite{P06}; even if they have problems with General Covariance, they get a theory in which the metric is insensitive to a cosmological constant term, thus explaining why vacuum energy does not contribute to gravity.\\
However, up to now there is no convincing substitute for the Einstein-Hilbert action, even though there are some formulations equivalent to General Relativity from a classical perspective, but leading to inequivalent results after the quantization procedure.

\section{The space-time slicing}
The description of the gravitational field dynamics requires the identification of a time parameter, with respect to which the evolution occurs. This point seems to conflict with the request of General Covariance, but we can introduce a formal splitting of the space-time in order to avoid any breaking of the diffeomorphism invariance (ADM splitting, named after Arnowitt, Deser and Misner \cite{ADM59,ADM60a,ADM60b}, see \cite{ADM62} for a review).\\
One refers to a global hyperbolic space-time, which is a manifold endowed with a Cauchy surface, {\it i.e.} a surface such that the evolution backwards and forward of initial conditions gives the full manifold. This request ensures the possibility of having a well-posed initial-value formulation for the gravitational field. Geroch demonstrated that a global hyperbolic space-time is diffeomorphic to a manifold $\Sigma\otimes \textbf{R}$, where $\Sigma$ is the hypersurface of equal time \cite{G70}.\\ 
The identification of spatial hypersurfaces $\Sigma_{x^0}$, giving a slicing of the full space-time, requires simply the introduction of a time-like vector field $\eta$; in fact, the Frobenius theorem ensures that, under very general assumptions, vectors ortogonal to $\eta$ define globally a sub-manifold.\\ 
Let us denote with $u^\mu=u^\mu(x^i;x^0),\quad i=1,2,3$,  the equation for spatial hypersurfaces, being $x^0$ a parameter, which characterizes each $\Sigma$, and $x^i$ coordinates on it. We can use $x$ as coordinates on the manifold and study the relation with basis vectors $\vec{f}_\mu$. The spatial character of $\Sigma$ implies following conditions on the normal and the tangential vector ($\vec{\eta}$ and $\vec{e}_i=\frac{\partial u^\mu}{\partial x^i}\vec{f}_\mu$, respectively)
\begin{eqnarray}
\left\{\begin{array}{c} \vec{\eta}\cdot\vec{\eta}=-1\\ \vec{\eta}\cdot\vec{e}_i=0 \\ \vec{e}_i\cdot\vec{e}_j=h_{ij} 
\end{array}\right.\Rightarrow \left\{\begin{array}{c} \eta^\mu\eta^\nu g_{\mu\nu}=-1\\ \eta^\mu e^\nu_i g_{\mu\nu}=0 \\ e^\mu_i e^\nu_j g_{\mu\nu}=h_{ij} 
\end{array}\right.
\end{eqnarray}   
being $h$ a positive-definite symmetric matrix. Hence, we can express the deformation vector, {\it i.e.} the time-like basis vector adapted to $x$ coordinates, in terms of $\vec{\eta}$ and $\vec{e}_i$
\begin{eqnarray}
\vec{e}_0=\frac{\partial u^\mu}{\partial x^0}\vec{f}_\mu=N\vec{\eta}+N^i\vec{e}_i,\label{e0}
\end{eqnarray}
while, obviously, space-like components of the basis are $\vec{e}_i$. We will refer to $N$ and $N^i$ as Lapse function and Shift vector, respectively, and we can give them geometrical interpretation, by observing that $\vec{e}_0$ relates points on the different $\Sigma$ with the same $x^i$. Moreover, $N$ must be non-vanishing, since $\vec{e}_0$ is a time-like vector, and we will take it as positive.\\
Hence, from the basis vectors, we obtain the following relations
\begin{eqnarray}
\left\{\begin{array}{c} g_{00}=\vec{e}_0\cdot\vec{e}_0=-N^2+h_{ij}N^iN^j\\ g_{0i}=\vec{e}_0\cdot\vec{e}_i=h_{ij}N^j \\ g_{ij}=\vec{e}_i\cdot\vec{e}_j=h_{ij} 
\end{array}\right.\label{metrsplit}
\end{eqnarray}
giving the 3+1 splitting of the metric tensor \cite{K81}, \cite{KT90}. The action of the diffeomorphism group implies the choice of different hypersurfaces $\Sigma$, so different values for the Lapse function and the Shift vector. Therefore General Covariance is not violated in the ADM splitting as far as $N$ and $N^i$ are not specified.\\
For the sake of calculations, it is very useful to define a tensor $q_{\mu\nu}$, which projects tensors on each $\Sigma$: because  the set $(\vec{\eta};\vec{e}_i)$ forms a complete basis, we can write completeness relations, which read as follows
\begin{equation}
-\eta^\mu\eta^\nu+h^{ij}e_i^\mu e_j^\nu=g^{\mu\nu}.\label{comp}
\end{equation}
From the last relation the tensor $q^{\mu\nu}=h^{ij}e_i^\mu e_j^\nu$ can be recognized as the projector one is looking for, since it annihilates $\eta_\mu$, so any component orthogonal to $\Sigma$, and it behaves as a metric tensor on the hypersurface itself.\\
Now, let us consider the splitting of the covariant derivative on $\Sigma$: given a spatial vector $\vec{A}=A^i\vec{e}_i$, we have
\begin{equation}
\partial_j\vec{A}=(\partial_jA^i)\vec{e}_i+A^i\partial_j\vec{e}_i=(\partial_jA^i)\vec{e}_i+A^i(\Gamma^k_{ij}\vec{e}_k+\Pi_{ij}\vec{\eta}),
\end{equation}
where $\Gamma^k_{ij}$ are 3-dim affine connections, which define covariant derivatives on $\Sigma$
\begin{equation}
D_jA^k=\partial_jA^k+\Gamma^k_{ij}A^i.
\end{equation}
The last derivative coincide with covariant one on $\Sigma$, in fact we have
\begin{equation}
\vec{e}_i\cdot(\partial_j\vec{A})=D_jA_i\label{De}
\end{equation}
so it gives variations of $\vec{A}$ along directions tangential to $\Sigma$. The relation with the covariant derivative on the full space-time manifold can be obtained by rewriting the expression above in terms of space-time indexes,
\begin{equation}
D_\mu A_\nu=\frac{\partial x^i}{\partial u^\mu}\frac{\partial x^j}{\partial u^\nu}D_jA_i=e^i_\mu e^j_\nu D_jA_i,
\end{equation}
and, with the help of condition (\ref{De}), one can show the following condition
\begin{equation}
D_\mu A_\nu=q^\rho_\mu q^\sigma_\nu \nabla_\rho A_\sigma,\label{3D}
\end{equation}
so that the projection of the 4-dim covariant derivative is the 3-dim one, defined by relation (\ref{De}).\\ 
Another interesting quantity is the extrinsic curvature
\begin{equation}
K_{\mu\nu}=q^\rho_\mu q^\sigma_\nu \nabla_\rho \eta_\sigma\label{excurv}
\end{equation}
whose 3-d projections can be written as
\begin{equation}
K_{ij}=(\partial_i\vec{\eta})\cdot\vec{e}_j,
\end{equation}
from which its symmetry can be demonstrated, and, by virtue of definition (\ref{e0}), the following relation holds
\begin{equation} 
K_{ij}=\frac{1}{2N}[-\partial_0h_{ij}-D_iN_j-D_jN_i]\label{excurv1}.
\end{equation}
For the geometrical interpretation of $K_{ij}$ we stress that 
\begin{equation}
\partial_i \vec{A}=(D_{i}A^k)\vec{e}_k-K_{ki}A^k\vec{\eta},
\end{equation}
so that it gives the curvature of $\Sigma$ as it is seen from a 4-dimensional perspective.\\
Let us turn to the splitting of the Riemann tensor in terms of the 3-dimensional one, whose definition is the following 
\begin{equation}
{}^{(3)}\!R^\sigma_{\phantom1\rho\mu\nu}A_{\sigma}=[D_\mu;D_\nu]A_\rho;
\end{equation}
from relations (\ref{comp}) and (\ref{3D}), we have
\begin{eqnarray}
D_\mu D_\nu A_\rho=q^\alpha_\mu q^\beta_\nu q^\gamma_\rho\nabla_\alpha(q^{\beta'}_\beta q^{\gamma'}_\gamma\nabla_{\beta'}A_{\gamma'})=q^\alpha_\mu q^\beta_\nu q^\gamma_\rho(q^{\beta'}_\beta q^{\gamma'}_\gamma\nabla_\alpha\nabla_{\beta'}A_{\gamma'}-\nonumber\\-q^{\beta'}_\beta\nabla_\alpha(\eta_\gamma\eta^{\gamma'})\nabla_{\beta'}A_{\gamma'}-q^{\gamma'}_\gamma\nabla_\alpha(\eta_\beta\eta^{\beta'})\nabla_{\beta'}A_{\gamma'})
\end{eqnarray}
and from the definition of the extrinsic curvature (\ref{excurv}), we obtain
\begin{equation}
{}^{(3)}\!R^\sigma_{\phantom1\rho\mu\nu}A_{\sigma}=q^\alpha_\mu q^\beta_\nu q^\gamma_\rho q_\lambda^\delta R_{\delta\gamma\alpha\beta}A^{\lambda}+(K_{\mu\lambda}K_{\nu\rho}-K_{\nu\lambda}K_{\mu\rho})A^\lambda.
\end{equation}
Therefore, for the 3-dimensional curvature scalar, the following relation stands
\begin{equation}
{}^{(3)}\!R=q^{\mu\nu}q^{\lambda\rho}R_{\nu\rho\mu\lambda}-(K^2-K_{\mu\nu}K^{\mu\nu}),\label{3R}
\end{equation}
being K the trace of $K_{\mu\nu}$, while for the 4-dimensional one we get from the completeness relation (\ref{comp})
\begin{equation}
R=q^{\mu\nu}q^{\lambda\rho}R_{\nu\rho\mu\lambda}+2q^{\mu\nu}\eta^\lambda\eta^\rho R_{\nu\rho\mu\lambda}\label{R}
\end{equation}
where the second term in the right hand side can be rewritten as
\begin{equation}
2\eta^\mu[\nabla_\nu;\nabla_\mu]\eta^\nu=2(K^2-K_{\mu\nu}K^{\mu\nu})+2\nabla_\nu(\eta^\mu\partial_\mu\eta^\nu-\eta^\nu K)\label{pregc}.
\end{equation}
Combining the expressions (\ref{3R}), (\ref{R}) and (\ref{pregc}), the Gauss-Codazzi equation comes out, {\it i.e.}
\begin{equation} 
R={}^{(3)}\!R+(K^2-K_{\mu\nu}K^{\mu\nu})+2\nabla_\nu(\eta^\mu\partial_\mu\eta^\nu-\eta^\nu K),
\end{equation}
which relates the curvature of the full space-time manifold with that of spatial hypersurfaces. We stress that the last term in the relation above is a divergence, so it gives the surface term we discussed on in the previous section. Calculations involving the extrinsic curvature can be performed by summing on 4-dimensional or on 3-dimensional indices, since relations $K_{\mu\nu}K^{\mu\nu}=K_{ij}K^{ij}$ and $K=K^i_i=K^\mu_\mu$ stand.\\

\section{The Hamiltonian structure}\label{hamstruct}

Hence, we can develop the Hamiltonian formulation of GR; once recognized from the form (\ref{metrsplit}) for the metric tensor that $\sqrt{-g}=-N\sqrt{h}$, $h$ being the determinant of $h_{ij}$, the full action reads as follows
\begin{equation}
S=-\frac{c^4}{16\pi G}\int dtd^3x N\sqrt{h}(K^2-K_{ij}K^{ij}+{}^{(3)}\!R)\label{azsplit}
\end{equation}
where we neglect surface terms (however, as mentioned in section \ref{actiongrav}, their treatment is non-trivial). The lapse function, the shift vector and the 3-dimensional metric can be identified as the variables of this formulation, so that the conjugate momenta $\pi$, $\pi^i$ and $\pi^{ij}$, respectively, are determined. First of all, the absence of $N$ and $N^i$ time derivatives in the Lagrangian density implies primary constraints,
\begin{equation}
\pi=0\qquad\pi^i=0,\label{prcon}
\end{equation}
while for $\pi^{ij}$ we obtain
\begin{equation}
\pi^{ij}=\frac{c^4}{16\pi G}\sqrt{h}(Kh^{ij}-K^{ij}). 
\end{equation} 
The Hamiltonian can now be calculated; according with the Dirac prescription for a Hamiltonian formulation of constrained system, we introduce Lagrange multipliers $\lambda$ and $\lambda_i$ and for the action (\ref{azsplit}) we have 
\begin{equation}
S=\int dtd^3x \bigg\{\pi^{ij}\partial_0h_{ij}+\pi^N\partial_0N +\pi^i\partial_0N_i+\lambda\pi+\lambda_i\pi^i+\frac{c^4}{16\pi G}N\sqrt{h}(K^2-K_{ij}K^{ij}+{}^{(3)}\!R)\bigg\};
\end{equation}
from variations of $S$ with respect to $\lambda$ and $\lambda_i$ primary constraints come out.\\ 
Hence, using relation (\ref{excurv1}) and after an integration by part (we neglect surface terms), we can rewrite the expression above as follows
\begin{equation}
S=\int dtd^3x \{\pi^{ij}\partial_0h_{ij}+\pi^N\partial_0N +\pi^i\partial_0N_i-[\lambda\pi+\lambda_i\pi^i+N\mathcal{H}+N_i\mathcal{H}^i]\},
\end{equation}
being $\mathcal{H}^i$ the super-momentum
\begin{equation}
\mathcal{H}^i=-2D_j\pi^{ij}\label{supmom}
\end{equation}
and $\mathcal{H}$ the super-hamiltonian
\begin{equation}
\mathcal{H}=\frac{16\pi G}{2c^4\sqrt{h}}G_{ijkl}\pi^{ij}\pi^{kl}+\sqrt{h}{}^{(3)}\!R,\label{supham}
\end{equation}
with $G_{ijkl}=h_{ik}h_{jl}+h_{il}h_{jk}-h_{ij}h_{lk}$ the Supermetric.\\ 
We can now refer to the symplectic structure, obtained by imposing standard Poisson brackets between variables and their conjugated momenta, {\it i.e.}
\begin{eqnarray}
\{\pi^{ij}(x^0;x);h_{lm}(x^0;y)\}=\frac{16\pi G}{c^4}\delta^i_{[l}\delta^j_{m]}\delta^3(x-y)\\
\{\pi(x^0;x);N(x^0;y)\}=\frac{16\pi G}{c^4}\delta^3(x-y)\\\{\pi^{i}(x^0;x);N_j(x^0;y)\}=\frac{16\pi G}{c^4}\delta^i_{j}\delta^3(x-y),
\end{eqnarray}
and to the Hamiltonian
\begin{equation}
H=\int dtd^3x[\lambda\pi+\lambda_i\pi^i+N\mathcal{H}+N_i\mathcal{H}^i],
\end{equation}
in order to infer the dynamics.\\
If we introduce smeared functions $f$ and $f^i$, we can work with smeared quantities $\Pi(f)=\int_\Sigma d^3x f\pi$ and $\vec{\Pi}(\vec{f})=\int_\Sigma d^3x f_i\pi^i$, and, by equations of motion, we obtain
\begin{eqnarray} 
\partial_0\Pi(f)=\{H;\Pi(f)\}=\frac{16\pi G}{c^4}\int d^3x f\mathcal{H}=\frac{16\pi G}{c^4}\mathcal{H}(f)\\
\partial_0\vec{\Pi}(\vec{f})=\{H;\vec{\Pi}(\vec{f})\}=\frac{16\pi G}{c^4}\int d^3x f_i\mathcal{H}^i=\frac{16\pi G}{c^4}\vec{\mathcal{H}}(\vec{f}),
\end{eqnarray}
therefore the consistency of constraints with the dynamics imposes the vanishing of the super-momentum and super-Hamiltonian as secondary constraints, {\it i.e.}
\begin{equation}
\mathcal{H}=0\qquad \mathcal{H}^i=0.\label{secconstr}
\end{equation}
The following relations 
\begin{equation}
G_{\mu\nu}\eta^\mu\eta^\nu=-\frac{\mathcal{H}}{2\sqrt{h}}\qquad G_{\mu\nu}e^\mu_ie^\nu_i=\frac{\mathcal{H}_i}{2\sqrt{h}}
\end{equation}
emphasize that conditions (\ref{secconstr}) are equivalent to Einstein equations $G_{0\mu}=0$, which in fact constitute a set of non-evolutionary constraints ({\it i.e.} if initial conditions are such that these constraints are satisfied, then they stand at any time).\\
Let us turn to the algebra generated by these constraints, {\it i.e.}
\begin{eqnarray} 
\{\vec{\mathcal{H}}(\vec{f});\vec{\mathcal{H}}(\vec{f'})\}=\frac{16\pi G}{c^4}\vec{\mathcal{H}}({\cal L}_{\vec{f}} \vec{f'})\label{alcons1}\\
\{\vec{\mathcal{H}}(\vec{f});\mathcal{H}(f')\}=\frac{16\pi G}{c^4}\vec{\mathcal{H}}({\cal L}_{\vec{f}}f')\label{alcons2}\\
\{\mathcal{H}(f);\mathcal{H}(f')\}=\frac{16\pi G}{c^4}\mathcal{H}(\vec{N}(f;f';h)),\label{alcons3} 
\end{eqnarray} 
being ${\cal L}$ the Lie derivative, while the expression of the function $\vec{N}(f;f';h)$ is the following
\begin{eqnarray}
N^i(f;f';h)=h^{ij}(f\partial_jf'-f'\partial_jf).
\end{eqnarray}
Relations $(\ref{alcons1})-(\ref{alcons3})$ show that these constraints are first class (their Poisson brackets are linear combinations of them), thus they do not modify the symplectic structure; or in other words, the sub-manifold of the full phase space, where constraints hold, is preserved during the evolution. However the algebra of constraints is not a Lie one, because coefficients of those linear combinations are not constant.\\
The study of the dynamics of $N$ and $N^i$ gives 
\begin{eqnarray}
\partial_0N=\lambda\qquad\partial_0N_i=\lambda_i
\end{eqnarray}
so classical dynamics never fixes their values and does not depend on them, since $\lambda$ and $\lambda_i$ are Lagrange multipliers, so totally arbitrary quantities. Because $N$ and $N^i$ determine unambiguously the spatial hypersurface, this feature implies the motion in the phase space to be independent of how the space-time splitting is performed. Therefore, this result is not surprising, but a consequence of the General Covariance.\\
The super-momentum constraints show the arbitrariness in the choice of the coordinate system on each $\Sigma$, in fact, under an infinitesimal diffeomorphism $x'^i=x^i-\xi^i$, the transformation induced on the 3-metric is given by
\begin{equation}
h'_{ij}=h_{ij}+D_{i}\xi_{j}+D_{j}\xi_{i},
\end{equation}
while, for the variation of the action, we have 
\begin{equation}
\delta S=\int\pi^{ij}\partial_t\delta h_{ij}d^3x=\int\pi^{ij}(D_{i}\xi_{j}+D_{j}\xi_{i})d^3x=-2\int\nabla_i\pi^{ij}\xi_jd^3x=0
\end{equation}
A solution of these constraints can thus be obtained simply by requiring $S$ be a function of 3-geometry equivalence classes, which we indicate with $\{h_{ij}\}$, {\it i.e.} $\{h_{ij}\}$ and $\{h'_{ij}\}$ coincide if they are related by a spatial diffeomorphism. This configuration space is known as Superspace.\\  
In particular, being $h_{ij}$ the only dynamical degrees of freedom, the full dynamics is encoded in the constraint on the super-hamiltonian, {\it i.e.}
\begin{equation}
\frac{16\pi G}{2c^4\sqrt{h}}G_{ijkl}\pi^{ij}\pi^{kl}+\sqrt{h}{}^{(3)}\!R=0.\label{dyn3metr}
\end{equation} 
In an analogous way, one can show that on-shell this constraint implies the invariance under diffeomorphisms orthogonal to $\Sigma$ \cite{Th01}.\\

\section{The Hamilton-Jacobi equation}
A Hamilton-Jacobi (HJ) formulation for the gravitational field is possible and it can be used as a first step towards quantization (see paragraph \ref{wkb}). Let us introduce the action functional $S=S[N;N_i;h_{ij}]$, such that momenta can be rewritten as functional derivatives of $S$ with respect to the corresponding variables, {\it i.e.},
\begin{equation}
\pi^{ij}=\frac{\delta S}{\delta h_{ij}}\qquad\pi=\frac{\delta S}{\delta N}\qquad\pi^i=\frac{\delta S}{\delta N_i}. 
\end{equation} 
Primary constraints ensure the validity of secondary ones; in fact, from conditions (\ref{prcon}) the vanishing of the super-Hamiltonian and of the super-momentum follows, since $\frac{\delta S}{\delta N}=\mathcal{H}$ and $\frac{\delta S}{\delta N_i}=\mathcal{H}^i$.\\
In Super-space, the full dynamics is described by the super-Hamiltonian constraint, which can be rewritten as the Einstein-Hamilton-Jacobi equation \cite{P62}
\begin{equation}
\frac{16\pi G}{2c^4\sqrt{h}}G_{ijkl}\frac{\delta S}{\delta h_{ij}}\frac{\delta S}{\delta h_{kl}}+\sqrt{h}{}^{(3)}\!R=0,
\end{equation}
thus reducing the problem to that of $\infty^3$ particles moving in a space-time manifold, with the metric tensor given by the Supermetric itself, subjected to a potential $\sqrt{h}{}^{(3)}\!R$.\\
In particular, the identification of a direction in which the Supermetric is negative definite provides a way to introduce a time-like variable. An example can be given by rewriting the 3-metric as
\begin{equation} 
h_{ij}=\eta^{4/3}u_{ij},\qquad det(u_{ij})=1
\end{equation}
and taking $\eta$ and $u_{ij}$ as configuration variables; in this case, the Hamiltonian reads as follows
\begin{equation}
H=-\frac{3}{2c^4}\pi G p_\eta^2+\frac{16\pi G}{c^4\eta^2}u_{ik}u_{jl}p^{ij}p^{kl}-\frac{c^4}{16\pi G}
{}^{(3)}\!R(u_{ij};\nabla\eta;\nabla u_{ij}),
\end{equation}
where it is clear that the variable $\eta$, giving the determinant of the metric, is time-like.\\   
This result is well-known in cosmological settings; for example in Freedmann-Robertson-Walker space-times the scale factor is as an appropriate time variable \cite{DeW67}.

\section{Reduction to the canonical form}\label{2.5}
The super-Hamiltonian constraint (\ref{dyn3metr}) encodes information on the dynamics of the metric tensor $h_{ij}$. However, the presence of additional constraints signals that $(h_{ij},\pi^{ij})$ are a set of redundant variables. Hence, in view of giving a better physical characterization of the gravitational field dynamics, one must identify into the metric tensor physical degrees of freedom $\phi_A$. In this respect, the Lagrangian density has to be written in the canonical form, {\it i.e.} 
\begin{equation}
L=\int d^3x [\pi^{A}\partial_t\phi_A-H_{true}(\phi_A,\pi^B)].
\end{equation} 

This reduction has been performed in \cite{ADM62} into the framework of a Palatini-like formulation. The metric tensor has been split into the transverse traceless component $h_{ij}^{TT}$, the trace of the transverse part $h^T$ and the longitudinal part $h_i$. The main steps of the procedure adopted are
\begin{itemize}
{\item the imposition of constraints, by which $h^T$ and $\pi^i$ can be evaluated.}
{\item the choice of a system of coordinates, which fixes $g_i$ and $\pi^T$.}
\end{itemize} 

The investigation on the form of the generating functional underlying the Hamiltonian framework allow Arnowitt, Deser and Misner to identify $h^{TT}_{ij}$ and conjugate momenta as variables describing physical degrees of freedom. Finally, the full Lagrangian density reads
\begin{equation}
\mathcal{L}=\pi^{ijTT}\partial_th_{ij}^{TT}+\textsl{T}^0_0,\qquad \textsl{T}^0_0=\nabla^2h^T(\pi^{TTij},h_{ij}^{TT},g_i(\pi^{TTkl},h_{kl}^{TT}),\pi^{T}(\pi^{TTkl},h_{kl}^{TT})),
\end{equation}      

By choosing a different system of coordinates, one finds different relations fixing $g_i$ and $\pi^T$, such that the Hamiltonian density $\textsl{T}^0_0$ takes a new expression as a function of $(\pi^{TTij}, h_{ij}^{TT})$. This feature outlines that the Hamiltonian density depends on the frame. 

The canonical form makes GR similar to a field theory formulation, such that it is possible to define the energy-momentum of the gravitation field in terms of generators of translations $\textsl{T}^\mu_0$, as follows
\begin{equation} 
P^\mu=-\int_\tau d^3x\textsl{T}^\mu_0.
\end{equation}

In an asymptotically-flat space-time, by restricting to those coordinate transformations which do not modify the flatness at infinity ({\it i.e.} $g'_{\mu\nu}-\eta_{\mu\nu}$ goes like $1/r$) and averaging over oscillatory terms, it can be shown that $P^\mu$ is invariant. Nevertheless, the form of $\textsl{T}^\mu_0$ is affected by choosing a different reference. Such a dependence disappears as far as only transformations between Heisenberg frames are considered, which means that the full metric $g_{\mu\nu}$ can be expressed in terms of canonical variables only, without any explicit coordinate dependence. 

Therefore, within this scheme $P^\mu$ fulfills the requirement for a well-defined energy-momentum for the gravitational field in an asymptotically flat space-time.

Furthermore, proper conditions can be fixed such that a wave-like behavior comes out for canonical variables in a certain space-time region, where non-linearities can be neglected. 

Although such a scenario looks tantalizing in view of a quantum description, nevertheless the quantization of the Hamiltonian in the canonical form has not been achieved. Main difficulties are due to ambiguities coming out from non linear terms and to the issue of reproducing classical canonical transformations by unitary transformations. Therefore, the attempts towards quantization are based on promoting to operators $\mathcal{H}$ and $\mathcal{H}_i$ too.    
\newpage

\chapter{Gravity as a gauge theory}

All interactions but gravity are described by gauge theories; for such kinds of models, it is possible to probe renormalizability, so that requirements for a predictive Quantum Field Theory can be accomplished. Therefore, giving a gauge formulation for gravity would allow one to perform a quantization procedure for such a field.\\ 
In this section, we will develop a formulation of GR similar to a gauge one, but we will also stress differences, that lead to the conclusion that gravity in its present formulation cannot be interpret as a gauge interaction.

\section{Gauge theories} 
Gauge theories are a mathematical tool that describe interactions through the invariance of the action $S(\phi, \partial_{\mu}\phi)$ for the field $\phi(x) \equiv \left\{ \phi_{r}(x)\right\}$ under  Lie groups of transformations ($U(\epsilon)$) \cite{GelNai,NaiSter}. The invariance of the action is expressed by
\begin{equation}
0=\delta S = \delta  \int d^{4}x L(\phi(x),\partial_{\mu}\phi (x))= \int d^{4}x \delta L,
\end{equation}
which implies,
\begin{equation}\label{deltal}
\delta L = 0 \Rightarrow \frac{\partial L}{\partial \phi_{r} }\delta \phi_{r}+\frac{\partial L}{\partial (\partial_{\mu} \phi_{r} )}\delta \partial_{\mu}\phi_{r}=0.
\end{equation}
It is therefore crucial to know the expressions for $\delta \phi_{r}$ and $\delta \partial_{\mu}\phi_{r}$.\\
Let's consider the action of the operator $U(\epsilon)$ on $\phi$
\begin{equation}\label{u}
\phi \rightarrow \phi' =U(\epsilon)\phi,\ \  U(\epsilon)=e^{ig\epsilon^{a}\tau_{a}}=I+ig\epsilon^{a}\tau_{a},\ \ \epsilon^{a}<<1 
\end{equation}
which induces on each component the transformation
\begin{equation}
\phi_{r}(x) \rightarrow \phi'_{r}(x) =\phi_{r}(x) + \delta \phi_{r}(x),\ \ \delta \phi_{r}(x)=\phi'_{r}(x)-\phi_{r}(x)=ig \epsilon^{a}\tau_{a}^{rs}\phi_{s},
\end{equation}
where $g$ is the coupling constant, $\epsilon^{a}$ a set of parameters, and $\tau_{a}$ the generators, that obey the commutation rule
\begin{equation}
[\tau_{a},\tau_{b}]=iC^{c}_{ab}\tau_{c}:
\end{equation} 
vanishing structure constants $C^{c}_{ab}$ define Abelian groups, while non-Abelian groups have non-vanishing structure constants.\\
If the parameters $\epsilon^{a}$ are constant, the transformation is called global, and $\partial_{\mu}\delta\phi_{r}=\delta\partial_{\mu}\phi_{r}$. This way, after substitution of the Euler-Lagrange equation in (\ref{deltal}), the conserved current  $j^{\mu}$, $\partial_{\mu}j^{\mu}=0$, is found,
 \begin{equation}\label{quantitàconservata}
j^{\mu}_{a} \equiv \frac{\partial L}{\partial (\partial_{\mu} \phi_{r}) }\delta \phi_{r}=\frac{\partial L}{\partial (\partial_{\mu} \phi_{r}) }\tau_{a}^{rs}\phi_{s},
\end{equation}
which allows one to define the conserved charges $Q_{a}$
\begin{equation}
 Q_{a} = \int d^{3}x j^{0}_{a}= \int d^{3}x \frac{\partial L}{\partial (\partial_{\mu} \phi_{r}) }\tau_{a}^{rs}\phi_{s},
\end{equation}
according to the Noether theorem.\\
If the parameters $\epsilon^{a}$ are not constant, $\epsilon^{a}=\epsilon^{a}(x)$, the transformation is called local, {\it i.e.} a gauge transformation, and $\partial_{\mu}\delta\phi_{r}\neq\delta\partial_{\mu}\phi_{r}$. To restore the invariance of the Lagrangian density, it is therefore necessary to define a new derivative, the covariant derivative $D_{\mu}$, that commutes with the variation operation. In fact, the ordinary derivative 
\begin{equation}
d \phi = \phi (x+dx)-\phi (x)=dx^{\mu} \partial_{\mu}\phi
\end{equation}
is ill-defined under a local (\ref{u}) because the fields in two different points $x$ and $x+dx$ transform according two different laws.  The transport operator $T(x,y)$, when applied to a field,
\begin{equation}\label{trasporto}
T(x,y)\phi(y) \rightarrow T'(x,y)\phi'(y)=U(x)T(x,y)\phi(y),
\end{equation}
generates an object with the same transformation properties of the field itself, so that
\begin{equation}\label{trnsf}
T(x,y) \rightarrow T'(x,y)=U(x)T(x,y)U^{+}(y).
\end{equation}
Since the transport operator is an element of the transformation group, it can be expressed as a function of the generators, and, for infinitesimal transformations one has
\begin{equation}\label{carta}
T(x,x+dx)=I+igdx^{\mu}A^{i}_{\mu}(x)\tau_{i},
\end{equation}\\
where the vector fields $A^{i}_{\mu}$, which are the gauge fields, in the combination $dx^{\mu}A^{i}_{\mu}$, play the role of $\epsilon^{a}(x)$. The covariant derivative $D_{\mu}$,
\begin{equation}\label{derivatacovariante}
D_{\mu}\phi(x)=(\partial_{\mu} +igA^{i}_{\mu}\tau_{i})\phi(x),
\end{equation}
as $T(x,y)$, transforms as (\ref{trnsf}) under a local (\ref{u}). The transformation law for the gauge fields can be obtained from the definition (\ref{derivatacovariante}), (\ref{trasporto}) and (\ref{carta}), and read
\begin{equation}\label{trasfgauge}
A^{i}_{\mu}(x) \rightarrow A'^{i}_{\mu}(x)=U(x)A^{i}_{\mu}(x)U(x)^{+}+\frac{1}{ig}U(x)\partial_{\mu}U(x)^{+}
\end{equation}
and, for infinitesimal $\epsilon^{i}$
\begin{equation}\label{infinitesime}
A^{i}_{\mu}(x) \rightarrow A'^{i}_{\mu}(x)=A^{i}_{\mu}+\delta A^{i}_{\mu}=      A^{i}_{\mu}+C^{i}_{jk}A^{j}_{\mu}\epsilon^{k}-\partial_{\mu}\epsilon^{i}.
\end{equation} 
The properties of the transport operator along a closed loop allow one to verify that the Lagrangian density for the gauge fields,
\begin{equation}\label{gimunugimunu}
L=-\frac{1}{4}G^{i}_{\mu\nu}G_{i}^{\mu\nu},
\end{equation}
where
\begin{equation}
G^{i}_{\mu\nu}(x)=\partial_{\nu}A^{i}_{\mu}-\partial_{\mu}A^{i}_{\nu}+gA^{j}_{\mu}A^{k}_{\nu}C^{i}_{jk},
\end{equation}
is invariant under the transformation (\ref{trasfgauge}).\\
The same result can be achieved by defining a covariant derivative that commutes with the variation operation via the introduction of compensating fields, the gauge fields, {\it i.e.}
\begin{equation}
\delta D_{\mu}\phi =ig\epsilon^{a}(x)\tau_{a}D_{\mu}\phi.
\end{equation}
The transformation law (\ref{infinitesime}) follows from direct calculation, while the expression for $G^{i}_{\mu\nu}$ is given by the commutator of the covariant derivatives,
\begin{equation}\label{Jacobi}
[D_{\nu},D_{\mu}]\phi =igG_{\mu\nu}^{i}\tau_{i}\phi.
\end{equation}
Applying the Jacobi identity to (\ref{Jacobi}), it is easy to verify that $G^{i}_{\mu\nu}$ obeys the Bianchi identity,
\begin{equation}
D_{\lambda}G^{i}_{\mu\nu}+D_{\nu}G^{i}_{\lambda\mu}+D_{\mu}G^{i}_{\nu\lambda}=0.
\end{equation}

\section{First-order formulation for the gravitational field}\label{1order}

\paragraph{The Palatini method} The first difficulty in recognizing gravity as a gauge theory is the presence of second derivatives of the configuration variables in the Einstein-Hilbert action, which do not arise in the Lagrangian (\ref{gimunugimunu}). We have seen in  section \ref{actiongrav} that this feature is deeply connected with the Equivalence Principle, one of the cornerstone of the geometrical interpretation. \\
However, as we have already seen in a vier-bein-connections framework, a first-order formulation is formally possible: one treats the metric and its first derivatives as independent variables. The first example of this kind is the Palatini formulation of General Relativity, which is still based on Einstein-Hilbert action, but in a configuration space built up by the metric components and by connections $\Gamma^\rho_{\mu\nu}$. So when we perform the variation of the action, two contributions must be considered
\begin{equation}
\delta S=-\frac{c^3}{16\pi G}\int \bigg[\frac{\delta(\sqrt{g}R(g;\Gamma))}{\delta g_{\mu\nu}}\delta g_{\mu\nu}+\frac{\delta(\sqrt{g}R(g;\Gamma))}{\delta \Gamma^\rho_{\mu\nu}}\delta \Gamma^\rho_{\mu\nu}\bigg]d^4x.
\end{equation}
In absence of matter this formulation is completely equivalent to ``second order'' General Relativity. In fact, equations 
\begin{equation}
\frac{\delta(\sqrt{g}R(g;\Gamma))}{\delta \Gamma^\rho_{\mu\nu}}=0
\end{equation}
imply that connections are equal to Christoffel symbols, which, if substituted in 
\begin{equation}
\frac{\delta(\sqrt{g}R(g;\Gamma))}{\delta g_{\mu\nu}}=\sqrt{g}\bigg(R_{\mu\nu}(\Gamma)-\frac{1}{2}g_{\mu\nu}R(\Gamma)\bigg), 
\end{equation}
reproduce Einstein equations for the metric tensor $g_{\mu\nu}$. In presence of matter, the first-order formulation gives different results only if the matter Lagrangian contains connections.\\ 
However a first objection against this approach is the use of non-tensorial variables, thus of quantities with an unclear geometrical meaning, since they can be made to vanish by a diffeomorphism.\\ 
Therefore, it looks more appropriate to work with other variables containing first-order derivatives of the metric and transforming as tensors. In this respect, a formulation based on Lorentz connections $\omega^{\alpha\beta}_\mu$ and on vier-bein vectors $e^\alpha_\mu$ is the right one (see section \ref{actiongrav} for notations).\\ 
Moreover, also from a physical point of view there are some problems. In fact, connections are not fundamental fields in General Relativity, but they are related to vier-bein by the first structure equation. This feature enforces a second order formulation, but we can enrich the geometrical structure of space-time, such that connections acquire an independent character. This additional structure is given by torsion.\\  

\paragraph{Torsion}
Torsion is defined as the antisymmetric part of connections
\begin{equation}
T^\rho_{\mu\nu}=\frac{1}{2}(\Gamma^\rho_{\mu\nu}-\Gamma^\rho_{\nu\mu})
\end{equation}
and its behavior under space-time transformations is that of a tensor.\\ 
In GR standard assumptions are the vanishing of torsion and that connections are metric-compatible, {\it i.e.} 
\begin{equation}
\nabla_\rho g_{\mu\nu}=\partial_\rho g_{\mu\nu}-\Gamma^\sigma_{\rho\mu}g_{\sigma\nu}-\Gamma^\sigma_{\rho\nu}g_{\mu\sigma}=0.
\end{equation}
By virtue of these hypothesis, it can be easily demonstrated that connections result to be equal to Christoffel symbols
\begin{equation}
\Gamma^\rho_{\mu\nu}=\{^\rho_{\mu\nu}\}=\frac{1}{2}g^{\rho\sigma}(\partial_\nu g_{\mu\sigma}+\partial_\mu g_{\nu\sigma}-\partial_\sigma g_{\mu\nu}).
\end{equation}
If torsion is present, connections differ from the expression above and we have the following modification 
\begin{equation}
\Gamma^\rho_{\mu\nu}=\{^\rho_{\mu\nu}\}-K^\rho_{\mu\nu}\qquad K^\rho_{\mu\nu}=-\frac{1}{2}(T^\rho_{\mu\nu}-T_{\mu\nu}^{\phantom1\phantom2\rho}-T_{\nu\mu}^{\phantom1\phantom2\rho})\label{connmetrcom},
\end{equation}
being $K^\rho_{\mu\nu}$ the contortion tensor. The previous relation clarifies that the introduction of torsion provides new degrees of freedom, such that connections are no longer determined only by the metric tensor. Moreover, we want to stress that, in general, a modification is produced in the symmetric part, too.\\
From a geometrical point of view, the presence of torsion implies that infinitesimal parallelograms do not close.\\
Once a first order formulation is performed in terms of vier-bein and Lorentz connections, the torsion-less condition arises from the first structure equation (\ref{1streq}). In fact, $\omega^{\alpha\beta}_\mu$ must be vectors containing first derivatives of vier-bein, while, from properties of the Riemann tensor, it follows they have to be antisymmetric in indices $\alpha$ and $\beta$. The only expression satisfying such requests is given by $\omega^{\alpha\beta}_\mu=-e^{\beta\nu}\nabla_\mu e^\alpha_\nu$. Hence, by substituting the expression for $\omega^{\alpha\beta}_\mu$ in the equation (\ref{1streq}), the torsion vanishes, {\it i.e.} 
\begin{equation}
\partial_{[\mu} e^\alpha_{\nu]}-e^{\gamma\rho}(\nabla_{[\mu}e^\alpha_\rho) e_{\nu]\gamma}=\Gamma^\rho_{[\mu\nu]}e^\alpha_\rho=0. 
\end{equation}
From calculations above, it is clear that if a term in the right side is present, torsion cannot be avoided.\\   
Therefore, we have seen that as soon as torsion is present, we have additional degrees of freedom, in such a way that connections are independent from vier-bein vectors. This way, a non-vanishing torsion forces us to give a first order formulation. However, not all kind of matter fields produces torsion, but only those ones whose Lagrangian density contains Lorentz connections $\omega^{\alpha\beta}_\mu$. For instance, let us consider the case of spinors.

\paragraph{Spinors in curved space-time}
The introduction of spinors in a curved background suggests the possibility to interpret $\omega^{\alpha\beta}_\mu$ as Lorentz connections. The formalism suitable for the description of spinors is based on the Dirac algebra, whose main properties can be summarized as follows:
\begin{itemize}
\item the existence of a Clifford algebra, {\it i.e.} an algebra generated by four matrices $\gamma^\mu$ (Dirac matrices) satisfying $\{\gamma^\mu;\gamma^\nu\}=2\eta^{\mu\nu}$;
\item the independence of Dirac matrices by coordinates, {\it i.e.} $\partial_\mu \gamma^\nu=0$;
\item the conjugation relations ${\gamma^\mu}^\dag=\gamma^0\gamma^\mu\gamma^0$.
\end{itemize}
The implementation of this formulation on a curved background is non-trivial \cite{BW57}, since the extension of the first condition gives 
\begin{equation}
\{\gamma^\mu;\gamma^\nu\}=2g^{\mu\nu},\label{commgamma}
\end{equation}
so that, being the right-hand side coordinate dependent, in general Dirac matrices are not constant. However, in order to define the analogue of the second condition, a new covariant derivative $D_\mu$ has to be defined for spinors, such that
\begin{equation}
D_\mu\gamma^\nu=\nabla_\mu\gamma^\nu-[\Gamma_\mu;\gamma^\nu]=0.
\end{equation}
From this request, the expression for $\Gamma_\mu$ results to be as follows
\begin{equation} 
\Gamma_\mu=-\frac{1}{4}\gamma_\nu\nabla_\mu\gamma^\nu=-\frac{i}{2}\omega^{\alpha\beta}_\mu\Sigma_{\alpha\beta},
\end{equation}
with $\Sigma_{\alpha\beta}=\frac{i}{4}[\gamma_\alpha;\gamma_\beta]$ and $\gamma_\alpha=e_\alpha^\mu\gamma_\mu$. Since the $\gamma_\alpha$'s are the projection of the $\gamma_\mu$'s on vier-bein indices, thus, on the tangent space, they result to coincide with Dirac matrices of the flat case. It can be demonstrate simply by multiplying relation (\ref{commgamma}) times $e_{\mu\alpha}e_{\nu\beta}$.\\
Therefore, $\Sigma_{\alpha\beta}$ are the generators of the Lorentz group transformations on the tangent space, from which one argues $\omega^{\alpha\beta}_\mu$ as associated connections. In fact, we can write Dirac Lagrangian density for spinors on a curved space-time as
\begin{equation}
\Lambda_\psi=\frac{i\hbar c}{2}\bigg[\bigg(\partial_\mu\bar{\psi}+\frac{i}{2}\omega^{\alpha\beta}_\mu\bar{\psi}\Sigma_{\alpha\beta}\bigg)\gamma^\mu\psi-\bar{\psi}\gamma^\mu\bigg(\partial_\mu\psi-\frac{i}{2}\omega^{\alpha\beta}_\mu\Sigma_{\alpha\beta}\psi\bigg)\bigg].\label{dirgrav}
\end{equation}

\section{Gravity as a gauge theory of the Lorentz group?}
Starting from the expression of the Lagrangian density for spinors, one recognizes $\omega^{\alpha\beta}_\mu$ having the same coupling with spinors as gauge bosons for the Lorentz group. Hence, the formal development of the Dirac theory in curved spaces seems to indicate that $\omega^{\alpha\beta}_\mu$ implements the local Lorentz invariance. This invariance would be manifest in a non flat space-time, since in this case one can choose different basis vectors of the tangent space in each point. Therefore, new connections have to be introduced in order to ensure invariance under this local symmetry. In a Minkoswkian space-time this invariance would not arise, because one usually identifies the tangent space with the manifold itself (in other words one identifies all tangent spaces), thus the only Lorentz symmetry is a global one.\\
Torsion arises as a consequence of the back-reaction of spinors on the space-time. In fact, being the variation with respect to $\omega^{\alpha\beta}_\mu$ into the Lagrangian density (\ref{dirgrav}) non vanishing, the following modification of the structure equation (\ref{1streq}) is provided
\begin{equation}
\partial_{[\mu}e^\alpha_{\nu]}-\omega^\alpha_{[\mu |\beta|}e^\beta_{\nu]}=\frac{1}{4}\epsilon^{\alpha}_{\phantom1 \beta\gamma\delta}e^\beta_\nu e^\gamma_\mu J_A^\delta\qquad J_A^\alpha=\bar{\psi}\gamma^\alpha\gamma_5\psi,
\end{equation}  
so the axial current provides us with a torsion term $T^\rho_{\mu\nu}=\frac{1}{4}\epsilon^\rho_{\phantom1\nu\mu\sigma}e^\sigma_\delta J_A^\delta$. Hence, if we substitute the solution of the structure equation in the Einstein-Dirac Lagrangian density, we obtain the so-called Einstein-Cartan theory, with the well-known (and non renormalizable) four fermions interaction term
\begin{equation}
S=\int\bigg(-\frac{c^3}{16\pi G}R+\frac{i\hbar c}{2}\bar{\psi}\gamma^\mu{}^{(0)}D_\mu\psi+c.c.-\frac{3\pi G\hbar^2}{c^3}\eta_{\alpha\beta}J^\alpha_AJ^\beta_A\bigg)ed^4x.
\end{equation}
Although the coupling between Dirac spinors and the gravitational field is the same one as a gauge theory of the Lorentz group, nevertheless we cannot conclude that General Relativity is a gauge theory of the Lorentz group. First of all, the free Lagrangian density (the Einstein-Hilbert one) differs significantly from the one of gauge bosons, because instead of a term $R_{\mu\nu}^{\alpha\beta}R^{\mu\nu}_{\alpha\beta}$ we have $e_\alpha^\mu e_\beta^\nu R_{\mu\nu}^{\alpha\beta}$. Then, connections are not the only variables, but we have additional fields, vier-bein vectors. There are several attempts (see section \ref{PGT}) to interpret them as connections associated with translations, so to develop a gauge theory of the Poincar\'e group, but conceptual problems are still present. In particular, for infinitesimal transformations, Lorentz rotations, {\it i.e.}
\begin{equation}
\delta x^\mu=\epsilon^{\mu\nu}(x)x_\nu=\theta^\mu(x)\qquad \epsilon^{\mu\nu}=-\epsilon^{\nu\mu},
\end{equation}
cannot be distinguished by translations
\begin{equation}
\delta x^\mu=\chi^\mu(x)
\end{equation}
being in previous relations $\epsilon^{\mu\nu}$, thus $\theta^\mu$, and $\chi$ arbitrary functions of space-time coordinates.
This reduction of local Lorentz transformations to translations explains while, in the standard treatment, connections associated with the former,  $\omega^{\alpha\beta}_\mu$, can be obtained from those associated with the latter, $e_\mu^\alpha$.\\ 
Finally, we want to stress that in order to give a real physical character to connections, they cannot be obtained by other fields, either a gravitational one, nor a matter one. The reason for this request is clear, for example, in the Einstein-Cartan theory: in this framework, torsion is non-vanishing only in points where a spin density is present. This way, it does not propagate, so it cannot be detected, because it is ``overwhelmed'' by the spin density itself in any place and any time it is. Such kind of a field has no physical meaning.\\ 
The previous speculations lead to conclude that a first order formulation requires the introduction of a fundamental torsion field, which must be present also in vacuum \cite{LM}. For these reasons, the development of a gauge theory for gravity requires a modification of GR. 

\section{Poincar\'{e} gauge theory}\label{PGT}

Poincar\'e Gauge Theory (PGT) \cite{bla02}\cite{bla03} is aimed at describing local Poincar\'e transformations within the framework of the gauge formalism, {\it i.e.}, by the introduction of covariant derivatives and conserved currents.\\
Let us consider an infinitesimal global Poincar\'{e} transformation in Minkowski space
\begin{equation}\label{poincare global}
x^{\mu}\rightarrow x^{\prime \mu}=x^{\mu}+\tilde{\epsilon}^{\mu}_{\phantom1\nu}x^{\nu}+\tilde{\epsilon}^{\mu},
\end{equation}
and the consequent transformation law for spinor fields
\begin{equation}\label{new poincare global}
\psi\left(x\right)\rightarrow\psi^{\prime}\left(x\right)=\left(1+\frac{1}{2}\tilde{\epsilon}^{\mu\nu}M_{\mu\nu}+\tilde{\epsilon}^{\mu}P_{\mu}\right)\psi\left(x\right),
\end{equation}
where the generators $M_{\mu\nu}=L_{\mu\nu}+\Sigma_{\mu\nu}$ and $P_{\mu}$ obey Lie-algebra commutation relations. If the matter Lagrangian density is assumed to depend on the spinor field and on its derivatives only, $L=L(\psi, \partial_{a}\psi)$, and if the equations of motion are assumed to hold, the conservation law $\partial_{\mu}J^{\mu}=0$ is found, where
\begin{equation}\label{j}
J^{\mu}=\frac{1}{2}\hat{\epsilon}^{\nu\lambda}M^{\mu}_{\ \nu\lambda}-\hat{\epsilon}^{\nu}T^{\mu}_{\ \nu},
\end{equation}
where the canonical energy-momentum and angular-momentum tensors are defined, respectively, as
\begin{equation}
T^{\mu}_{\ \nu}=\frac{\partial L}{\partial \psi,_{\mu}}\partial_{\nu}\psi-\delta^{\mu}_{\ \nu}L,
\end{equation}
\begin{equation}
M^{\mu}_{\ \nu\lambda}=\left(x_{\nu}T^{\mu}_{\ \lambda}-x^{\lambda}T^{\mu}_{\ \nu}\right)-S^{\mu}_{\ \nu\lambda}\equiv \left(x_{\nu}T^{\mu}_{\ \lambda}-x^{\lambda}T^{\mu}_{\ \nu}\right)+\frac{\partial L}{\partial \psi,_{\mu}}\Sigma_{\nu\lambda}\psi.
\end{equation}
Because the parameters in (\ref{j}) are constant, according to Noether's theorem, the conservation laws for the energy-momentum current and for the angular-momentum currents, together with the related charges, are established:
\begin{equation}
\partial_{\mu}T^{\mu}_{\ \nu}=0\rightarrow P^{\nu}=\int d^{3}x T^{0\nu}
\end{equation}
\begin{equation}\label{m}
\partial_{\mu}M^{\mu}_{\ \nu\lambda}=0\rightarrow M^{\nu\lambda}=\int d^{3}xM^{0\nu\lambda}.
\end{equation}
When the theory is locally implemented, eq.s (\ref{j})-(\ref{m}) do not hold any more, and compensating gauge fields have to be introduced in order to restore local invariance. As a first step, a covariant derivative $D_{k}\psi$ is defined as 
\begin{equation}\label{derivatives}
D_{k}\psi=e_{k}^{\ \mu}D_{\mu}\psi=e^{k}_{\ \mu}\left(\partial_{\mu}+A_{\mu}\right)\psi=e^{k}_{\ \mu}\left(\partial_{\mu}+\frac{1}{2}A_{\mu}^{ij}\Sigma_{ij}\right)\psi,
\end{equation}
where the compensating fields $e^{k}_{\ \mu}$ and $A_{\mu}^{ij}$, and the generator $\Sigma_{ij}$ have been taken into account. This way, the Lagrangian density depends on the covariant derivative of the fields, instead of the ordinary one, $L=L(\psi,D_{k}\psi)$; covariant derivatives (\ref{derivatives}) don't commute, but satisfy the commutation relation
\begin{equation}
[D_{\mu},D_{\nu}]\psi=\frac{1}{2}F^{ij}_{\ \ \mu\nu}\Sigma_{ij}\psi,\ \ [D_{k},D_{l}]\psi=\frac{1}{2}F^{ij}_{\ \ kl}\Sigma_{ij}\psi-F^{s}_{\ kl}D_{s}\psi,
\end{equation}
where $F^{ij}_{\ \ \mu\nu}$ and $F^{s}_{\ kl}$ are the Lorentz field strength and the translation field strength, respectively.\\
Covariant energy-momentum and spin currents, $T'^{\mu}_{\ \nu}$ and $S'^{\mu}_{\ ij}$, can be found, in analogy with the global case, after the substitution $\partial_{\mu}\rightarrow D_{\mu}$, and are found to be equivalent to the dynamical currents $\tau^{\mu}_{\ \nu}$ and $\sigma^{\mu}_{\ ij}$,
\begin{equation}
T'^{\mu}_{\ \nu}=\tau^{\mu}_{\ \nu}=e_{k}^{\ \mu}\frac{\partial L}{\partial e_{k}^{\ \nu}},
\end{equation}
\begin{equation}\label{spin}
S^{\mu}_{\ ij}=\sigma^{\mu}_{\ ij}=-\frac{\partial L}{\partial A^{ij}_{\ \ \mu}},
\end{equation}
whose meaning will be outlined throughout the rest of this section.\\
A simple and illuminating example by Hehl \emph{et al.} \cite{hehvdhker76} illustrates the inadequacy of special relativity to describe the behavior of matter fields under global Poincar\'{e} transformations. Global Poincar\'{e} transformations preserve distances between events and the metric properties of neighboring matter fields: comparing field amplitudes in nearby points before performing the transformation, and then transforming the result, or comparing the transformed amplitudes of the fields is equivalent. This property is known as rigidity condition, as matter fields behave as rigid bodies under this kind of transformations. On the contrary, it can be shown that the action of local Poincar\'{e} transformations can be interpreted as an irregular deformation of matter fields, thus predicting different phenomenological evidences for the field and for the transformed field. The compensating gauge fields $e_{\mu}^{k}$ and $A^{ij}_{\ \ mu}$, introduced to restore local invariance, describe geometrical properties of the space-time: it can be demonstrated that PGT has the geometrical structure of a Riemann-Cartan space-time.\\
The geometrical approach to PGT can be carried out by considering the most general metric-compatible linear connections, with 24 independent components, which can be written as a function of the torsion field $T^{\mu}_{\ \nu\rho}$ (\ref{connmetrcom}).\\
Geometric covariant derivatives are defined as
\begin{equation}
D_{\mu}\psi=\left(\partial_{\mu}+\omega_{\mu}\right)\psi=\left(\partial_{\mu}+\frac{1}{2}\omega^{ij}_{\ \ \mu}\Sigma_{ij}\right)\psi,
\end{equation}
where spin connections $\omega^{ij}_{\ \ \mu}$ consists of the bein projection of the Ricci rotation coefficients and the contortion field, respectively: $\omega_{ij\mu}=R_{ij\mu}+K_{ij\mu}$.\\
The gauge potentials $e^{\phantom1\alpha}_i$ are generally interpreted as the connection between the orthonormal frames (denoted by Greek indices) and the coordinate frames (denoted by Latin indices), while the introduction of the gauge potentials $\omega_i^{\alpha\beta}$ is connected with the relative rotations of the orthonormal basis at neighboring points: this induces a change in the derivative operator, {\it i.e.}
\begin{equation}
\partial_i\rightarrow D_i\equiv\partial_i+\frac{1}{2}\omega_i^{jk}\Sigma_{jk}.
\end{equation}
The comparison between the gauge approach and the geometrical approach leads to the identification of the gauge field $A^{ij}_{\ \mu}$ with spin connections $\omega^{ij}_{\ \ \mu}$, and the field $e_{k}^{\ \mu}$ with the components of the tetrad field. This way,  the identification of the Lorentz field strength with the curvature, and that of the translation field strength with torsion are straightforward.\\
Torsion contributes to the gravitational dynamics, according to its gravitational action: it has been illustrated that \cite{hashi} the most general form for a Lagrangian $L_{T}$, which allows for equations of motion that are at most of second order in the field derivatives, is
\begin{equation}
L_{T}=AT_{ijk}T^{ijk}+BT_{IJK}T^{JIK}+CT_{i}T^{i},
\end{equation}
where $T_{i}=T^{j}_{\ ji}$. The values of the parameters $A,B,C$ are to be determined according to the Physics that has to be described, and some relevant examples are discussed in \cite{rel1,rel2,rel3,re4}.\\

For later purposes, it will be convenient to restate the description of PGT in a slightly different formalism, which allows for a better explication of the role of spin.\\
Eq. (\ref{poincare global}) can be written as 
\begin{equation}\label{new poincare global 2}
\psi\left(x\right)\rightarrow\psi^{\prime}\left(x\right)=\left(1+\frac{1}{2}\epsilon^{\mu\nu}\Sigma_{\mu\nu}+\epsilon^{\mu}P_{\mu}\right)\psi\left(x\right),\ \ \epsilon^{\gamma}\equiv\widetilde{\epsilon}^{\gamma}+\widetilde{\epsilon}_{\alpha}^{\phantom1\beta}\delta^{\alpha}_ix^i,\qquad\epsilon^{\alpha\beta}=\widetilde{\epsilon}^{\alpha\beta},
\end{equation}
the generators of translations and spin rotations satisfying the relations:
\begin{equation}
\left[\Sigma_{\alpha\beta},\Sigma_{\gamma\delta}\right]=\eta_{\gamma[\alpha}\Sigma_{\beta]\delta}-\eta_{\delta[\alpha}\Sigma_{\beta]\gamma},\ \ \left[\Sigma_{\alpha\beta},P_{\gamma}\right]=-\eta_{\gamma[\alpha}\partial_{\beta]},\ \ \left[P_{\alpha},P_{\beta}\right]=0.
\end{equation}
The advantage of eq. (\ref{new poincare global 2}) consists in keeping pure rotations distinguished from translations. The orbital angular momentum is this way kept independent of the spin angular momentum: the former is strictly related with the energy-momentum, thus with the rotation-dependent part of $\epsilon^{\mu}$, while the latter is connected with the pure-rotation parameter $\epsilon^{\mu\nu}$. In fact, if the analogy is drawn between a generic diffeomorphism and a global Poincar\'{e} transformation, it is impossible to perform translations and rotations independently, but, when a localized symmetry is considered, this becomes possible, because the parameters defining the transformation are allowed to vary freely.\\
It is worth noting that, after the geometrical identification of the covariant gauge derivative, eq. (\ref{spin}) becomes an algebraic relation between spin and torsion: since the relation is not differential, torsion is not predicted to propagate, but its existence is bound to the presence of spin-$\frac{1}{2}$ matter fields.
Finally, the field equations read:
\begin{subequations}
\begin{align}
&\frac{1}{e}D_j\left(e e^i_{\alpha}e^j_{\beta}\right)=\mathcal{S}_{\phantom1\alpha\beta}^{i},
\\
&R_{\alpha}^{\phantom1i}-\frac{1}{2}\,e^i_{\phantom1\alpha}R=\frac{8\pi G}{c^4}T_{\alpha}^i.
\end{align}
\end{subequations}
The first equation is the first Cartan structure equation, which provides one with the expression of connections of the group of rotations as a function of connections of the group of translation and matter fields, while the second equation is the Einstein dynamical equation: tensor fields involved in the equations above must satisfy the identities (\ref{bianchi1}) and (\ref{cyclic identity}), thus predicting a non-propagative behavior for torsion.

\section{The Holst formulation}\label{holst}
If one wants to extend the formalism of GR, one must not contradict its well-tested predictions. An easy way to accomplish this task is to give simply a reformulation of gravity, which does not modify the equations of motion, so that classical dynamics is unchanged. However some new features can arise as far as the quantization procedure is concerned.\\ 
In this respect, one can add to the action a topological term, {\it i.e.} a piece that vanishes as soon as equations of motion stand. This is the case of the Holst reformulation \cite{Ho96} of GR.\\
Holst took the following action for the gravitational field
\begin{equation}
S_{G}=-\frac{c^{3}}{16\pi G}\int d^{4}x e e_\alpha^\mu e^\nu_\beta\bigg(R^{\alpha\beta}_{\mu\nu}-\frac{1}{2\gamma}\epsilon^{\alpha\beta}_{\phantom1\phantom2\gamma\delta} R^{\gamma\delta}_{\mu\nu}\bigg)    
\end{equation}    
being $\gamma$ a free parameter (the Immirzi parameter). If we perform variations with respect to connections $\omega^{\alpha\beta}_\mu$, we get
\begin{equation}
\delta S=-\frac{c^3}{16\pi G}\int d^4xee^\mu_\alpha e^\nu_\beta\bigg(\delta R^{\alpha\beta}_{\mu\nu}-\frac{1}{2\gamma}\epsilon^{\alpha\beta}_{\phantom1\phantom2\gamma\delta} \delta R^{\gamma\delta}_{\mu\nu}\bigg)=\frac{c^3}{16\pi G}\int d^4x\delta {}^{(\gamma)}\!A_\nu^{\alpha\beta}\textit{D}_\mu(ee^\mu_\alpha e^\nu_\beta)
\end{equation}  
where derivatives $\textit{D}_\mu$ act on space-time and Lorentz indexes, {\it i.e.}
\begin{equation}
\textit{D}_\mu e^\nu_\alpha=\nabla_\mu e^\nu_\alpha-\omega_{\mu\alpha}^{\phantom1\phantom2\beta}e_\beta^\nu,
\end{equation}
and new connections ${}^{(\gamma)}\!A_\mu^{\alpha\beta}$ (Barbero-Immirzi connections \cite{B95}) are as follows
\begin{equation}
{}^{(\gamma)}\!A_\mu^{\alpha\beta}=\omega^{\alpha\beta}_{\mu}-\frac{1}{2\gamma}\epsilon^{\alpha\beta}_{\phantom1\phantom2\gamma\delta}\omega^{\gamma\delta}_{\mu}.
\end{equation}
The last relation can be inverted for $\gamma\neq\pm i$, giving $\omega_\mu^{\alpha\beta}=\frac{\gamma^2-1}{\gamma^2}\bigg({}^{(\gamma)}\!A^{\alpha\beta}_{\mu}+\frac{1}{2\gamma}\epsilon^{\alpha\beta}_{\phantom1\phantom2\gamma\delta}{}^{(\gamma)}\!A^{\gamma\delta}_{\mu}\bigg)$, therefore arbitrary variations of $\omega_\mu^{\alpha\beta}$ provide arbitrary variation of $A_\mu^{\alpha\beta}$, so that a stationary action must provide for 
\begin{equation}
\textit{D}_\mu(ee^\mu_\alpha e^\nu_\beta)=0.\label{hstreq}
\end{equation}
By substituting the relation above into the action, which is equivalent to the first Cartan structure equation (\ref{1streq}), one finds the second-order formulation. In this framework, the cyclic identity for the Riemann tensor (\ref{cyclic identity}) provides us with the vanishing of the Holst modification, in fact we have
\begin{equation}
\epsilon^{\alpha\beta}_{\phantom1\phantom2\gamma\delta} e e_\alpha^\mu e^\nu_\beta R^{\gamma\delta}_{\mu\nu}=\epsilon^{\mu\nu\rho\sigma}R_{\mu\nu\rho\sigma}=0.
\end{equation} 
The last relation demonstrates that the Holst action differs from the Einstein-Hilbert one by a term vanishing ``on-shell''. Being equations of motion the same as GR, the two formulations are equivalent in a classical framework.\\
The cases $\gamma=i$ and $\gamma=-i$ (Ashtekar connections \cite{As86,As87,ART89}) are very peculiar, since ${}^{(\pm i)}\!A^{\alpha\beta}_\mu$ turns out to be the self-dual and the anti-self-dual part of $\omega^{\alpha\beta}_\mu$, respectively, {\it i.e.} they satisfy 
\begin{equation}
\pm\frac{i}{2}\epsilon^{\alpha\beta}_{\phantom1\phantom2\gamma\delta}{}^{(\pm i)}\!A^{\gamma\delta}_{\mu}={}^{(\pm i)}\!A^{\alpha\beta}.
\end{equation}
This way, by assigning components ${}^{(\pm i)}\!A^{0a}_\mu$, the full connections $\omega^{\alpha\beta}_\mu$ are determined (${}^{(\pm i)}\!A^{\alpha\beta}_\mu$ are complex quantities, so ${}^{(\pm i)}\!A^{0a}_\mu$ and $\omega^{\alpha\beta}_\mu$ have the same number of degrees of freedom).\\
Moreover, the term $R^{\alpha\beta}_{\mu\nu}\pm\frac{i}{2}\epsilon^{\alpha\beta}_{\phantom1\phantom2\gamma\delta} R^{\gamma\delta}_{\mu\nu}$ contains only ${}^{(\pm i)}\!A^{\alpha\beta}_\mu$, not ${}^{(\mp i)}\!A^{\alpha\beta}_\mu$. Thus, being ${}^{(\pm i)}\!A^{\alpha\beta}_\mu$ the only variables appearing in the Lagrangian density, one performs variations with respect to them directly. This way the I Cartan structure equations (\ref{hstreq}) still come out as equations of motion.\\
Furthermore, we can rewrite the full action in terms of ${}^{(\pm i)}\!A^{0a}_\mu$ only, as follows
\begin{equation}
S=-\frac{c^3}{16\pi G}\int d^4x e (e_0^\mu e_a^\nu\pm\frac{i}{2}\epsilon_a^{\phantom1 bc}e_b^\mu e_c^\nu){}^{(\pm i)}\!F^a_{\mu\nu}
\label{azash}\end{equation}
being ${}^{(\pm i)}\!F^a_{\mu\nu}=\partial_{[\mu}{}^{(\pm i)}\!A^{0a}_{\nu]}\pm \frac{i}{2}\epsilon^a_{\phantom1bc}{}^{(\pm i)}\!A^{0b}_{[\mu}{}^{(\pm i)}\!A^{0c}_{\nu]}$ the curvature associated to SU(2) connections. This result is not surprising, since it is well-know that the Lorentz group is isomorphic to the direct product of two SU(2) groups. The self-dual and the anti-self-dual parts of SO(1;3) correspond precisely to the projection on these two SU(2) groups, which are related by a complex conjugation. Therefore, we can replace the Lorentz group by an $SU(2)\otimes SU(2)$ one. We also want to stress that the Holst action for $\gamma=\pm i$ is simply the self-dual and antiself-dual projection of the Einstein-Hilbert one, respectively. It is the above mentioned possibility of the splitting of the Lorentz symmetry into $SU(2)\otimes SU(2)$, which provides an explanation for the fact that only ${}^{(i)}\!A^{\alpha\beta}_\mu$ or ${}^{(-i)}\!A^{\alpha\beta}_\mu$ are present in the action. Conversely, for $\gamma\neq\pm i$, the full action cannot be rewritten in terms of ${}^{(\gamma)}\!A^{\alpha\beta}_\mu$ only, but also ${}^{(-\gamma)}\!A^{\alpha\beta}_\mu$ is required, even though it has no evolutionary character.\\ 
The considerations above outline the special role played by Ashtekar connections. They appear as the basic configuration variables in which we can split $\omega^{\alpha\beta}_\mu$ and for this reason, historically, they arose first.\\ 
Although, we stressed how the Holst formulation is equivalent to GR, nevertheless, after the ADM splitting, new canonical variables arise, together with a redefinition of constraints. This last feature will be very useful in view of a canonical quantization.

\paragraph{3+1 splitting} The splitting procedure we presented in section \ref{hamstruct} has been performed in metric variables. Hence, we have to rewrite this formulation in terms of vier-bein vectors. Since at the end, the only dynamical coordinates are 3-geometries, one usually simply introduces a set of 3-bein vectors $e^a_i$ on spatial hypersurfaces and takes the configuration space as given by $N$, $N_i$ and $e^a_i$. However this choice is equivalent to the following identification 
\begin{equation}
e_\mu^{\phantom1\alpha}=\left(\begin{array}{cc} N & N^ie_i^a \\ 0 & e^a_i \end{array}\right)\label{3+1bein}
\end{equation}
this way, one is fixing the $e^0$ vector as normal to $\Sigma$ (time gauge). Therefore, in this framework, boost transformations are frozen out. As usual, one expects this procedure not to lead to different results with respect to a fully covariant formulation (see section \ref{timegauge}), because a gauge symmetry has been fixed.\\

Hence, one performs the splitting by substituting the form of 4-bein (\ref{3+1bein}) in the action, thus obtaining
\begin{eqnarray*}
S=-\frac{c^4}{16\pi G}\int dt d^3x\bigg[2E_a^i\bigg(\frac{1}{2}\partial_t{}^{(\gamma)}\!A^a_i-\frac{1}{2}(\partial_i\omega_t^{0a}-\frac{1}{2\gamma}\epsilon^{a}_{bc}\partial_i\omega^{bc}_t)+\omega^{ab}_{[t}\omega_{i]b}^a-\frac{1}{2\gamma}\epsilon^{a}_{bc}\omega^{b0}_{[t}\omega_{i]0}^c-\\-\frac{1}{2\gamma}\epsilon^a_{bc}\omega^{bd}_{[t}\omega_{i]d}^c\bigg)-2N^iE^j_a\bigg(R^{0a}_{ij}-\frac{1}{2\gamma}\epsilon^{a}_{bc}R^{bc}_{ij}\bigg)+ee^i_ae^j_b\bigg(R^{ab}_{ij}-\frac{1}{\gamma}\epsilon^{ab}_{\phantom1c}R^{c0}_{ij}\bigg)\bigg]
\end{eqnarray*}
with ${}^{(\gamma)}\!A^a_i={}^{(\gamma)}\!A^{0a}_i$ and $E^i_a$ densitized 3-bein $ee^i_a$.\\
From the last expression, it can be easily recognized that ${}^{(\gamma)}\!A^a_i$ are the only variables with an evolutionary character, while time derivatives of $\omega^{ab}_t$, $\omega_t^{0a}$ ${}^{(-\gamma)}\!A^a_i$ do not appear, therefore they behave as Lagrangian multipliers. By solving the equations of motion that come from variations with respect to them and after a long manipulations, one ends up with the following expression for the action  
\begin{equation}
S=-\frac{c^4}{16\pi G}\int dt d^3x (-\partial_t{}^{(\gamma)}\!A^a_i E^i_a+\frac{1}{\gamma}\Lambda^aG_a+N^i\mathcal{H}_i+N\mathcal{H})
\end{equation}
being $\Lambda^a=\frac{1}{2}\epsilon^a_{\phantom1bc}{}^{(\gamma)}\!A^{bc}_0-\frac{1}{\gamma}{}^{(\gamma)}\!A_0^{0k}$, while the new constraints take the following form
\begin{eqnarray}
G_a=\partial_iE^i_a+\gamma\epsilon^c_{ba}{}^{(\gamma)}\!A^b_iE^i_c\label{Gcon}\\
\mathcal{H}_i=E^j_a{}^{(\gamma)}\!F^a_{ij}\label{supmome}\\
\mathcal{H}=\frac{1}{\gamma}\epsilon_i^{\phantom1jk}e^i_a({}^{(\gamma)}\!F^a_{jk}-\frac{\gamma^2+1}{2\gamma}R_{jk}^{\phantom1\phantom2a})\label{supHam}
\end{eqnarray}
with ${}^{(\gamma)}\!F^{a}_{\mu\nu}=\partial_{[i}{}^{(\gamma)}\!A^a_{j]}+\frac{\gamma}{2}\epsilon^a_{\phantom1bc}{}^{(\gamma)}\!A^b_{[i}{}^{(\gamma)}\!A^c_{j]}$, while $R_{ij}^a=\partial_{[i}\Gamma^a_{j]}+\epsilon^a_{\phantom1bc}\Gamma^b_i\Gamma^c_j$, being $\Gamma^a_i=-\frac{1}{2}\epsilon^a_{\phantom1bc}e^b_j\nabla_ie^{jc}$.\\
Therefore, in this framework connections ${}^{(\gamma)}\!A^a_i$ turn out to be configuration variables, while densitized triads are conjugate momenta. The Hamiltonian is still a combination of constraints, with a super-momentum $\mathcal{H}_i$ and a super-Hamiltonian $\mathcal{H}$, but, since $\Lambda_a$ behave as Lagrange multipliers, new constraints arise, whose form is that of Gauss constraints in a $SU(2)$ gauge theory. This is one of the key point of this reformulation, since it will allow for the use of quantization techniques proper of gauge models to gravity.\\ Indeed, since the super-Hamiltonian is a constraint not linear in momenta, in order to have a gauge invariant action a boundary term must be added \cite{HTV92}.\\ 
For further applications, it will be useful to rewrite the super-Hamiltonian constraint as follows
\begin{equation}
\mathcal{H}=\epsilon_{abc}\frac{E^{aj}E^{bk}{}^{(\gamma)}\!F^c_{jk}}{\gamma e}-\frac{\gamma^2+1}{4\gamma^2e}E^{[i}_aE^{j]}_bK^a_iK^b_j\label{sHam}
\end{equation}
being $K^b_i=K_{ij}e^{jb}$. The last expression differs from the relation (\ref{supHam}) by a term which is proportional to the Gauss constraint.\\ 

\paragraph{} 
In the case $\gamma=\pm i$, an important simplification occurs for the super-Hamiltonian, because the second term in relation (\ref{supHam}) disappears.\\ 
This way, the full set of constraints, except for a square root of the 3-metric in the super-Hamiltonian, is polynomial in configuration variables and it results to be an important feature in view of the quantization. The reason for this simplification, as pointed out by Samuel \cite{Sa01}, deals with the geometrical meaning of Ashtekar connections. In fact, ${}^{(\pm i)}\!A^a_{i}$ are the pull-back on $\Sigma$ of space-time connections, {\it i.e.} ${}^{(\pm i)}\!A^{ab}_\nu$, the self-dual and the anti-self-dual connections of the Lorentz group. In general, ${}^{(\gamma)}\!A^{ab}_{i}$ do not behave as connections, because transformations they are related to do not form a subgroup inside the Lorentz group itself (recently, there are attempts \cite{FFR07} to reproduce Barbero-Immirzi connections from that of a space-time related to a reduced $SU(2)$ group, but the physical interpretation of such a group is not clear). This way, they have a much more complicated behavior under time diffeomorphisms, which is induced by the super-Hamiltonian itself. Other consequences are that holonomies depends on the slicing \cite{Sa00} and the time gauge condition turns out to be crucial for the splitting above. As far as the last statement is concerned, Alexandrov \cite{Al00} stressed how, in a general Lorentz frame, second class constraints arise.\\ 
Despite the above-mentioned features of the Ashtekar case, nevertheless the development of a theory with real connections is free of reality conditions we have to impose on the phase space, in order to insure that 3-beins are real \cite{Ro91,Me95}. These conditions are preserved by the Hamiltonian flow, but they induce deep complications in a quantum regime. For this reason, the case $\gamma$ real is preferred.
People working on LQG often regards issues about transformation properties of configuration variables as aesthetic problems \cite{Sa01} \cite{Th06a}, since the mathematical formulation should allow for results invariant under some kind of symmetries, even though configuration variables are not ``suitable'' to those symmetries ({\it i.e.} non gauge-invariant or with complex behavior under transformations associated). However we want to emphasize that while, from a classical point of view, the use of variables is just a matter of simplification, being in principle configurations variables arbitrary, when quantization procedures are applied, different choices of variables are inequivalent (for a complete discussion on the quantization of constrained systems see \cite{HT}).

Furthermore, the Ashtekar formulation of GR looks very promising in view of performing a unification with the electro-weak model. For instance, in the work of Nesti and Percacci \cite{NP07} they recover the Lagrangian density of the electro-weak $SU(2)$ sector and the one proper of Ashtekar reformulation (\ref{azash}) starting from a gauge theory of the complexified Lorentz group.

Finally, if $\Sigma$ has no boundary, no term has to be added to have a gauge-invaraint action for $\gamma=\pm i$ \cite{MV01}.

\section{The Kodama state}
One of the main issue in the determination of the gravitational field dynamics is the the search for observables, {\it i.e.} of phase space functionals which Poisson commute with the full set of constraints (\ref{Gcon}), (\ref{supmome}) and (\ref{supHam}). However, in presence of a cosmological constant, a solution is known, the so-called Kodama state \cite{Ko88,Ko90}. 
Let us consider the Ashtekar case $\gamma=i$: the introduction of a non-vanishing cosmological constant $\Lambda$ provides us with the following super-Hamiltonian constraint 
\begin{equation}
\mathcal{H}=-i\epsilon_{abc}\frac{E^{ai}E^{bj}}{e}(F_{ij}^c+\frac{\Lambda}{3}\epsilon_{ijk}E^{ck})
\end{equation}
while other constraints are not modified. Hence one can easily demonstrate \cite{Sm02} that a solution is given by the state functional
\begin{equation}
\psi[A]=e^{\frac{2}{3\Lambda}\int_\Sigma Y_{CS}[A]\sqrt{h}d^3x}
\end{equation}
$Y_{CS}$ being the Chern-Simons class of the connections, {\it i.e.}
\begin{equation}
Y_{CS}[A]=\frac{1}{2}\epsilon^{ijk}\delta_{ab}A^a_i\nabla_jA^b_k+\frac{2}{3}\epsilon_{abc}\epsilon^{ijk}A^a_iA^b_jA^c_k.
\end{equation}
By studying the full system of equations of motion \cite{Sm02}, it can be demonstrated that these solutions describe a De-Sitter space-time, {\it i.e.} an homogeneous and isotropic manifold with constant curvature, given by the coefficient $\Lambda$.\\ 
The Kodama state arises clearly after the quantization as a solution of the full set of constraints, but, in this framework, problems with its normalization arise\cite{MM95}. Nevertheless, as soon as a WKB viewpoint is taken (see section \ref{wkb}), one can still see how this state describes a De-Sitter space-time in the semi-classical approximation.\\ 
The main applications of the Kodama functional are devoted to cosmological models and to the study of the quantum behavior for homogeneous space-times \cite{Ez96,PG00}. This class of models suffers same problems (due to the complex nature of variables involved) as LQG. A convincing link with the LQG framework is still missing, even though recently the extension to arbitrary values of the Immirzi parameter has been given \cite{Ra06a,Ra06b}.\\ 
Finally, although the Kodama state is the only exact solution to all quantum constraints, nevertheless its role in a canonical quantization of gravity has not been completely understood \cite{GGP97}.\\  
\newpage

\chapter{Quantization of the gravitational field}

\section{The WDW equation}
If one wants to quantize the gravitational action, the procedure to be followed is one analogous to that for the relativistic particle, because of the super-Hamiltonian has hyperbolic features.\\
If the metric representation is chosen, the canonical variables $h_{ij}$ and $\pi^{ij}$ have to be implemented as operators, {\it i.e.} $h_{ij}(x)\rightarrow \hat{h}_{ij}(x)\equiv h_{ij}(x)$ and $\pi^{ij}(x)\rightarrow \hat{\pi}^{ij}(x)\equiv -i\frac{\delta}{\delta h_{ij}(x)}$, and canonical commutation relations must be established, such as
\begin{equation}
[\hat{h}_{ij}(x),\hat{\pi}^{kl}(x')]=i\delta^{kl}_{ij}\delta(x-x'),
\end{equation}
while other commutators vanish. The super-Hamiltonian (\ref{supham}) and the super-momentum (\ref{supmom}) must be turned into operators as well, $\mathcal{H}\rightarrow \hat{\mathcal{H}}$ and $\mathcal{H}_{i}\rightarrow \hat{\mathcal{H}}_{i}$, and their action on the wave function has to be fixed, {\it i.e.}
\begin{equation}\label{superham}
\hat{\mathcal{H}}\psi\equiv ''G_{ijkl}(x)[h_{mn}]\frac{\delta^{2}\psi}{\delta h_{ij}(x)\delta h_{kl}(x)}''-\sqrt{h}{}^{(3)}\!R(x)\psi=0,
\end{equation}
\begin{equation}\label{supermom}
\hat{\mathcal{H}}_{i}\psi\equiv\left[ \frac{\delta \psi}{\delta h_{ij}(x)}\right]_{;j}=0,
\end{equation}
where $\psi\equiv \psi[h_{ij}(x)]$. In eq (\ref{superham}), which is usually called the Wheeler-DeWitt (WDW) equation, the factor ordering for the Kinetic term is not specified, while, in eq. (\ref{supermom}), a factor ordering can be chosen, such that the wave functional should depend on the 3-geometry $\{h_{ij}\}$ only, rather than on any specific representation. Nevertheless, a factor ordering must be adopted, such that the Dirac algebra ({\it i.e.} relations (\ref{alcons1}), (\ref{alcons2}) and (\ref{alcons3})) is preserved.

\section{The problem of time}\label{probltime}
As discussed in section 1.2, one of the main issues in the quantization of the gravitational field is the problem of time. In this respect, here we point out the main features of the two standard procedures, by which a time parameter arises. 
%This issue reflects the different concepts of time in quantum mechanics and in any diffeomorphism invariant theory like GR. In fact, while in Quantum Mechanics time appears as an external parameter which labels different states a physical system is, in GR it is one of the space-time manifold coordinates and no unique determination of it can be given without violating diffeomorphism invariance. This property is expressed by the vanishing of the Hamiltonian for the gravitational field, since the Hamiltonian flow gives the transformation phase space functions are subjected to under time reparametrizations.\\ 
%Behind these differences, there is the role the measurement process plays in the two frameworks. In QM, we need an external (classical) observable with an external clock, whose relation with the evolution of the system under investigation is provided by the Schrodinger equation. In GR, one treats as dynamical variables the full geometry of the space-time. Therefore, in trying to combine together the two cornerstone of Theoretical Physics, one must to account for the different features time has in the two frameworks.\\
%The standard approach consists in a clear distinction between time as a space-time coordinate and time as the external parameter, which provides the quantum description and which has to be identified among time-like diffeomorphism invariant quantities. In this sense, we identify two main approaches, if time is identified before or after the quantization procedure is performed.\\ 
\paragraph{Time before the quantization}
The first approach we discussed is based on the following procedure:
\begin{itemize} 
\item the constraints are solved classically, 
\item a functional $T$ of the configuration variables is identified with the time parameter,
\item the Hamiltonian associated with $T$ is quantized. 
\end{itemize}
A proper feature of these models is the choice of the time-functional. Such functional must be a monotonically increasing function, at least locally, and, after the quantization, has to provide a well-defined and conserved probability density. Moreover, the separation between the time and other variables can be performed in different ways, which provide us with different scenarios.\\ 
Among these approaches, we point our attention on the Brown-Kucha$\breve{r}$ one. Their standard work  \cite{BK95} (see also \cite{SS04a,SS04b}) is based on the introduction of a dust, whose world-line identifies a preferred time-like direction without violating General Covariance. This direction plays the role of time. In terms of constraints, the super-Hamiltonian and the super-momentum are modified by terms due to the matter field, {\it i.e.}    
\begin{equation}
\mathcal{H}'=\mathcal{H}+\mathcal{H}^{D}\qquad \mathcal{H}'_i=\mathcal{H}_i+\mathcal{H}^D_i.
\end{equation}
Hence, Brown and Kucha$\breve{r}$ demonstrated that, by using the new super-momentum constraint, the super-Hamiltonian can be rewritten as follows
\begin{equation}
P+h(h_{ij};\pi^{ij})=0,
\end{equation}
$T$ and $P$ being the proper time of dust flow lines and its conjugate momentum, respectively, while for $h$ they got the expression
\begin{equation}
h=-\sqrt{G}\qquad G=\mathcal{H}^2-h^{ij}\mathcal{H}_i\mathcal{H}_j.
\end{equation}
Therefore, by taking $P$ as a time parameter, an evolution described by the hamiltonian $h$ follows. This hamiltonian turns out to be positive-definite and, starting from the corresponding Schrodinger equation, a quantum description for the system can be given, together with a definition for an inner product (but it could be only formal).\\ 
The Brown-Kucha$\breve{r}$ approach relays on the dualism existing between time and matter \cite{Mo02}, \cite{MM03} in GR. In this respect, the properties a matter field should have to be a good clock are still under investigation.\\ A different proposal is that of Thiemann \cite{Th06b}, who claimed to be able to introduce an internal time. In fact his work consists in the application of the Brown-Kucha$\breve{r}$ formulations to a K-essence, which comes out as a relic of the quantum description of space-time geometry.\\ 
There is however a more general, even though rather formal, approach in order to derive an internal time parameter for the gravitational field, the multi-time formalism \cite{Is92}. This procedure is based on the reduction of the Lagrangian to a canonical form (see section \ref{2.5}). For gravity, one identifies among the components of the metric tensor $h_{ij}$, some variables $H_r\hspace{0.2cm}(r=1,2)$ describing the two degrees of freedom. Hence, by performing a canonical transformation from $\{h_{ij};\pi^{ij}\}$ to $\{H_r;P^r\}$ plus a set $\{\xi^\mu;\pi_\mu\}\hspace{0.2cm}(\mu=0,\ldots,3)$ of embedding variables, {\it i.e.} with no physical meaning, the action can be rewritten as
\begin{equation} 
S=-\frac{c^3}{16\pi G}\int d^4x(\pi_\mu\partial_t\xi^\mu+P^r\partial_tH_r-N\mathcal{H}-N^i\mathcal{H}_i)\label{acmt}
\end{equation}
being $\mathcal{H}=\mathcal{H}(\xi^\mu;\pi_\mu;H_r;P^r)$ and $\mathcal{H}_i=\mathcal{H}_i(\xi^\mu;\pi_\mu;H_r;P^r)$. We can use the constraints $\mathcal{H}=0$ and $\mathcal{H}_i=0$ to get an expression for the momenta $\pi_\mu$ in terms of other phase space coordinates, and then substitute into the action (\ref{acmt}), so having
\begin{equation}
S=-\frac{c^3}{16\pi G}\int d^4x(P^r\partial_tH_r-\pi_\mu(\xi^\mu;H_r;P^r)\partial_t\xi^\mu).
\end{equation}   
Finally, the multi-time idea consists in a canonical quantization of $\xi^\mu$ and $H_r$ variables, taking wave functionals $\psi=\psi(\xi^\mu;H_r)$ whose evolution is provided by the following set of Schrodinger-like equations
\begin{equation}
i\hbar\frac{\delta\psi}{\delta\xi^\mu}=\pi_\mu\bigg(\xi^\mu;H_r;\frac{\delta}{\delta H_r}\bigg)\psi.
\end{equation}   
This formalism finds application especially in cosmological settings \cite{BM07}.

\paragraph{Time after the quantization}
The second approach is based on the Dirac prescription for the quantization of constrained systems (see \cite{HT} for a review on this topic). This method consists in imposing constraints after the quantization procedure: one promotes all variables in a suitable Hilbert space (kinematical Hilbert space) as operators, then, being constraints translated into operators acting on wave functional, physical states are imposed to be those states which are annihilated by constraints. We want to stress that, even though one would prefer to quantize variables with a gauge-invariant meaning, nevertheless symmetries play an important role in Quantum Physics. For instance, particles are classified as irreducible representation of the Lorentz groupor of gauge groups. This example explains how the preservation of symmetries after the quantization turns out to be very useful. However, many complications could arise, like anomalies, requests of regularization or problems with the self-adjoint character of constraints operators, which are usually solved by a suitable choice of the kinematical Hilbert space.\\
In dealing with the super-Hamiltonian constraints, this formulation is the so-called ``frozen'' one, since the total Hamiltonian is constrained to vanish, therefore no evolution at all is provided. This result reflects the absence of an external time parameter, proper of a diffeomorphism-invariant theory. Therefore, time has to arise from a quantum degree of freedom, in a relational way.\\
An example of this approach is provided by the work of Rovelli and Smolin \cite{RS94}, where, in the framework of the Ashtekar formulation (see section 3.5), they perform the quantization of gravity in presence of a scalar field $T(x)$. By a partial fixing of the time coordinate, they demonstrate that the Hamiltonian constraint reduces to the following condition
\begin{equation}
\frac{1}{\sqrt{2\mu}}\int_\Sigma\pi d^3x+\int_\Sigma \sqrt{-\mathcal{H}}d^3x=0
\end{equation} 
being $\mu$ a constant introduced for dimensional reasons, $\pi$ the momentum conjugated to $T$, while $\mathcal{H}$ is the super-Hamiltonian constraint (\ref{sHam}) for $\gamma=i$. This constraint is translated by the quantization procedure in the Schroedinger equation
\begin{equation}
i\hbar\frac{\delta\Psi}{\delta T}=\sqrt{2\mu}\hat{\mathcal{H}}\Psi
\end{equation}
which thus provide an evolution in terms of the scalar field $T$. Moreover, a regularization procedure is found such that $\hat{H}$ turns out to be finite and diffeomorphism invariant.\\ 
In more recent works, a standard procedure to provide a relational time has been developed in terms of evolutions of partial observables among each others \cite{Mo01,Ro03} (see section \ref{1.4}). A proper evolution has been inferred in simple cases \cite{MRT99}, while in \cite{MR00} properties of a general covariant statistical mechanics has been outlined. The application to gravity is the next task \cite{Di06}.

\paragraph{Other approaches}
Despite approaches discussed above, there are many heuristic arguments for a solution of the problem of time based on a reformulation of Quantum Mechanics. Such speculations involve the physical characterization of the measurement process in a quantum setting and the definition of observables in a diffeomorphism invariant framework. A promising tool in this direction is the consistent-history approach \cite{Ha91}, which leads to spin-foam models (see section \ref{latticegauge}) or to topos theory \cite{IB00}, which, however, turn out to provide deep conceptual and technical complications.\\

\section{Interpretation of the wave function}
Even though a time parameter could be introduced, nevertheless the implementation of the quantum description is highly non-trivial.\\
For instance, the wave function of the Universe \cite{vil1989,par97}, as a solution of the WDW equation, on one hand, should maintain its role of defining the probability density of events happening in the Universe, and, on the other hand, should be able to reproduce the physical features of the Universe itself, both quantum and classical.\\
On of the most striking differences between the quantum and the cosmological wave function if that, while the former depends on time explicitly and allows, under very reasonable hypotheses, for the definition of a positive-definite probability distribution $dp$, {\it i.e.}
\begin{equation}
\psi=\psi(q_{i},t)\rightarrow dp=|\psi(q_{i},t)|^{2},
\end{equation}
the latter does not. In fact, it depends only on 3D metrics and on matter fields, {\it i.e.} $\psi=\psi\left(h_{ij}(\vec{x}), \phi_{A}(\vec{x}\right)$, while the dependence on ''time'' would make little sense, because of the invariance under arbitrary reparametrizations of the label time.\\
As far as the probabilistic interpretation is concerned, two main cases can be distinguished, when the superspace variables are all semiclassical, or when a quantum subsystem is taken into account. \\
If we restrict the investigation to homogeneous superspace models, whose action reads
\begin{equation}
S+\int dt \left(p_{\alpha}\dot{h}^{\alpha}-N\left[g^{\alpha\beta}p_{\alpha}p_{\beta}+U(h)\right]\right),
\end{equation}
where $h^{\alpha}$ labels generalized superspace variables, and $P_{\alpha}$ their conjugate momenta, the superpotential $U=h^{1/2}[V(\phi)-{^{3}R}]$ defines the WDW equation
\begin{equation}\label{ewdwsc}
\left(\nabla^{2}-U\right)\psi=0.
\end{equation}
In the case of semiclassical variables, the wave function can be written as the superposition of functions of the action  $S(h)$, as
\begin{equation}\label{cwf}
\psi=A(h)e^{iS(h)},
\end{equation}
which admits a WKB expansion and leads to the conserved current
\begin{equation}\label{conscurr}
j^{\alpha}=|A|^{2}\nabla^{\alpha}S,\ \ \nabla_{\alpha}j^{\alpha}=0.
\end{equation}
Here, the classical action $S$ is an equivalence class of classical trajectories, and a family of $(n-1)$D hypersurfaces $\Sigma_{\alpha}$ crossed once by the trajectories, $\dot{h}^{\alpha}d\Sigma^{\alpha}>0$. If we take $h_{n}=t$, eq. (\ref{conscurr}) rewrites
\begin{equation}
\frac{\partial\rho}{\partial t}+\partial_{a}j^{a}=0,\ \ a=1,...,n-1,
\end{equation}
where $j^{a}=\rho\dot{h}^{a}$, and describes $\rho$ as the ''distribution function for an ensemble of classical universes''.\\
The previous discussion can be easily generalized to a superposition of (\ref{cwf}), $\psi=\sum_{k}\psi_{k}=\sum_{k}A_{k}e^{iS_{k}}$. If a family of equal-time hypersurfaces can be found for all the possible $S_{k}$'s, then the total distribution function can be expressed as
\begin{equation}
\rho=\sum_{k}\rho_{k}+cross terms,
\end{equation}
where the cross terms can be shown to produce no physically-relevant interference.\\
The possibility of including small quantum subsystems, which do not modify significantly the dynamics, among the superspace variables can be taken into account. Eq. (\ref{ewdwsc}) rewrites
\begin{equation}
\left(\nabla_{0}^{2}-U_{0}-H_{q}\right)\psi=0,
\end{equation}
where the index ${_{0}}$ refers to classical variables only, and ${_{q}}$ to quantum effects only. The pertinent wave function is
\begin{equation}
\psi(h,q)=\sum_{k}\psi_{k}(h)\chi_{k}(h,q),
\end{equation}
which leads to the definition of the currents
\begin{equation}
j^{\alpha}=|\chi|^{2}|A|^{2}\nabla^{\alpha}_{0}S\equiv j^{\alpha}_{0}\rho_{\chi},
\end{equation}
\begin{equation}
j^{\nu}=-\frac{i}{2}|A|^{2}\left(\chi^{*}\nabla^{\nu}\chi-\chi\nabla^{\nu}\chi^{*}\right)\equiv\frac{1}{2}|A|^{2}j^{\nu}_{\chi},
\end{equation}
where $\alpha$ labels semiclassical variables, while $\nu$ refers to the quantum subsystem: the currents are related by the continuity conditions
\begin{equation}
\nabla_{\alpha}j^{\alpha}+\nabla_{\nu}j^{\nu}=0,\ \ \nabla_{0\alpha}j^{\alpha}_{0}=0,
\end{equation}
which lead to the probability distribution
\begin{equation}
\rho(h,q,t)=\rho_{0}(h,t)|\chi|^{2}.
\end{equation}
The normalization of probabilities can be easily checked, if one considers the volume element $d\Sigma=d\Sigma_{0} d\Sigma_{q}$, such that
\begin{equation}
\int \rho_{0}d\Sigma_{0}=1,
\end{equation}
\begin{equation}
\int|\chi|^{2}(det[g_{\mu\nu}])^{1/2}d^{m}q.
\end{equation}

\section{The idea of Third Quantization}
The quantization procedure, or, better, the attempts to quantize the WDW equation are usually referred to as ``third quantization''\cite{pel1991,gid1988,cav1994,fak90}. As a result, the prospect of creating and annihilating (interacting) universes is envisaged.\\
As a first step, the possibility of constructing a Hilbert space out of the space of the solutions $\psi[\mathcal{G}]$ must be investigated. Such a Hilbert space contains the physical states: in the coordinate representation, the state vectors are space geometries, while, in the momentum representation, space geometries with given momentum define the corresponding Fock space. The definition of this Hilbert space requires the fulfillment of three conditions, {\it i.e.},
\begin{enumerate}
	\item the hyperbolic structure of the curved metric $g_{\mu\nu}$;
	\item the stationarity of the space time, {\it i.e.}, the existence of a conformal time-like Killing vector;
	\item the time independence of  the potential term.
\end{enumerate}
In fact, these hypotheses allow one to recognize the eigenfunctions of (\ref{superham}) as energy eigenfunctions, and to split up these solutions into positive- and negative-mode solutions. Accordingly, and inner product can be defined, and a conserved current, independent of the choice of the hypersurface, can be built out of the solutions,
\begin{equation}
\Omega[\psi_{1},\psi_{2}]\equiv \prod_{x}\int D\Sigma^{ij}(x)S_{12ij}(x)[\psi_{1},\psi_{2}],
\end{equation}
where
\begin{equation}
S_{12ij}(x)=\frac{1}{2}G_{ijkl}(x)\psi_{1}[\phi]\frac{\orch\delta}{\delta h_{kl}(x)}\psi_{2}[\phi],
\end{equation}
obeys the continuity equation
\begin{equation}
\frac{\nabla}{\delta h_{ij}(x)}S_{12ij}(x)=0
\end{equation}
where the factor ordering has not been specified and the derivation is performed with respect to the Supermetric $G_{ijkl}(x)[g_{mn}]$.\\
To verify the previous hypotheses, it's easy to check that the scaled Supermetric $\tilde{G}_{ijkl}\equiv h^{1/2}G_{ijkl}$ obeys the Killing equation   
\begin{equation}
L_{t_{mn}}\tilde{G}_{ijkl}=\frac{\partial \tilde{G}_{ijkl}}{\partial h_{mn}}t_{mn}-\tilde{G}_{ijmn}\frac{\partial t_{kl}}{\partial h_{mn}}-
G_{mnkl}\frac{\partial t_{ij}}{\partial h_{mn}}=0,
\end{equation}
where $t_{mn}\propto h_{mn}$ is the conformal Killing vector, which can be normalized as $\tilde{t}_{mn}=\frac{i}{\sqrt6}h_{mn}$. Two new coordinates
can be introduced, $T$ and $\gamma_{ij}$, linked to the old ones by the relations
\begin{equation}
\tilde{t}^{mn}=\frac{\partial h_{mn}(T,\gamma_{ij})}{\partial T}, \qquad \gamma_{ij}=h^{-\frac{1}{3}}h_{ij},
\end{equation}
so that their conjugate momenta $P$ and $\tilde{\pi}^{ij}$ define the kinetic and the potential term of the rescaled super-Hamiltonian,
\begin{equation}\label{superkin}
\tilde{G}_{ijkl}\pi^{ij}\pi^{kl}=-P^{2}+\tilde{\pi}_{mn}\tilde{\pi}^{mn}
\end{equation}
and
\begin{equation}\label{superpot}
-h{}^{(3)}\!R[h_{mn}]=-e^{\sqrt{\frac{2}{3}}T}R[\gamma_{mn}]+8e^{\frac{7}{4\sqrt{6}}T}\Delta_{\gamma}e^{\frac{1}{4\sqrt{6}}T},
\end{equation}
respectively. While (\ref{superkin}) ensures that the metric is hyperbolic, (\ref{superpot}), which is the analog of the mass term in the free case, does depend on $T$ and is not positive definite.\\
To remove the last difficulty, a suitable rescaling can be taken into account, such that the rescaled potential term is time independent, while the restriction to geometries with negative curvature scalar renders it positive-definite. The impossibility of defining a global time-like Killing vector would lead, on the one hand , to the search for a local time-like Killing vector, and on the other hand, to the investigation of those (asymptotic) regions of the universe, where such a vector can be found. In the second case, however, it's worth remarking that, for a consistent description, at least two asymptotic regions (classical and quantum) should be admissible. Furthermore, the problem of singularities and topology fluctuations can be addressed if an interaction term $\propto\lambda\psi^4$ is added to the super-Hamiltonian: this way, the WDW equation would rewrite as a dynamical equation rather than as a constraint one. A more general potential term could also be added and interpreted as an effective term that labels a particular universe.

\newpage

\chapter{Loop Quantum Gravity}

\section{Holonomies and Fluxes}
We have seen in section \ref{holst} a reformulation of General Relativity, which provides us with a phase space like those of a gauge theory. This reformulation alone is not able to provide a different quantum description with respect to the Wheeler-DeWitt approach. In fact, one can quantize $A^a_i$\footnote{In what follows we will omit the superscript $^{(\gamma)}$.} and $E^i_a$, instead of $h_{ij}$ and $\pi^{ij}$, but the Stone-Von Neumann prescription insures that the two formulations are unitary equivalent.\\ 
Therefore, Loop Quantum Gravity is based on an additional technical issue, the quantization of variables adapted to the symmetries, {\it i.e.} the general covariance and the invariance under $SU(2)$ gauge transformations, one deals with in the Ashtekar-Barbero-Immirzi reformulation. At the end, the Hilbert space representation will turn to violate an hypotheses of the Von Neumann theorem, the strong continuity, such that a quantum description of the gravitational field dynamics not equivalent to the Wheeler-DeWitt comes out. The new variables are elements of the holonomy-flux algebra and they belong to the so-called ``auxiliary'' Hilbert space. Hence, constraints have to be imposed and one finds that there is one, and only one, cyclic representation which is gauge- and spatial diffeomorphisms-invariant \cite{Fl04,LOST05}. Therefore, the kinematical Hilbert space is unique.\\
The basic idea of this approach is to promote to quantum operators quantities with a geometrical meaning much more ``clear'' than connections. Inspired by a reformulation in terms of form and vector fields (which we do not propose here), it turns out that connections $A_i=A^a_i\tau_a$ are objects to be integrated on curves. Such an integration is usually performed on paths. Given a manifold, we define a path as an equivalence class of piece-wise analytic oriented curve, where two curves are identified if they differ by a re-parametrization. A loop is a closed path, while a graph is a collection of paths. A graph can also be thought of as a collection of edges, {\it i.e.} of analytic paths. It is possible to show that the space of loops with a common point can be endowed with a group structure. Therefore, one introduces the parallel transport of $A$ along a curve $x^i=\alpha^i(t)$ (holonomies), with the following definition
\begin{equation}
U_\alpha(A)=Pe^{\int_\alpha A_i(\alpha^k(t))\frac{d\alpha^i}{dt}dt}.
\end{equation}
Because of the properties of parallel transport, its transformation law under $SU(2)$ gauge transformations $\Lambda(x)$ is 
\begin{equation} 
U'=\Lambda(i)U\Lambda^{-1}(f),\label{gauge}
\end{equation}
where $i$ and $f$ are the initial and final points of $\alpha$, respectively. Hence, a gauge- and re-parametrization-invariant quantity is obtained simply by taking the Wilson loop $h_\alpha(A)$, {\it i.e.} the trace of the parallel transport along a closed path.
 
\subsection{Why a reformulation in terms of Wilson loops?}
The framework in which Wilson loops have been introduced is the path-integral formulation of QCD on a lattice, where they provide tools to study the confinement of quarks. The main point of this approach is that the potential between two static quarks can be obtained from the expectation value of Wilson loops connecting these two particles \cite{Wi74,RoLGT}. The result is that a linear potential is obtained, which produces the confinement.\\ 
Moreover, quantum states can be rewritten in terms of closed loops or of open ones, with quarks at each ending. Such states represent lines of non-Abelian electric fluxes and they turn out to be eigenstates of the Hamiltonian in the strong coupling limit, {\it i.e.} as soon as the coupling constant goes to infinity \cite{KS75}.\\ 
Almost fifteen years after their introduction, it was recognized that Wilson loops are useful in view of the canonical quantization  of a background independent model. As far as the gauge symmetry is concerned, Giles \cite{Gi81} stressed how starting from Wilson loops on the full (flat) space-time, connections can be reconstructed, modulo a gauge transformation. Therefore the knowledge of $h_\alpha(A)$ on any loop $\alpha$ gives all gauge-invariant information. This framework is not very useful in gauge theories, since no simplification occurs by passing from the space of connections to the loop space. However,  as soon as one works in a background-independent framework, it is possible to define holonomies on knot-states, {\it i.e.} on equivalence classes of loops under diffeomorphisms, so that they provide solutions to diffeomorphism and Gauss constraint and belong to the kinematical space.\\ 
However, there is a strong indication that a proper quantization of gravity is based on quantizing holonomies instead of connections in GR: the use of canonical commutation relations is not allowed for non-trivial phase space topologies. Skipping some technicalities (which can be found in \cite{Is83}), we rather prefer to give an example \cite{IsI84}. 

Let us consider a system with one positive coordinate $q$, $q>0$, and conjugate momentum $p$. After the quantization, states are defined in the Hilbert space of square integrable functions on $q$, with support on the positive real axis. By imposing canonical commutation relations, $q$ and $p$ are promoted to hermitian operators for which the following commutation relation stands 
\begin{equation}    
[\hat{q};\hat{p}]=i\hbar,
\end{equation}
which implies $\hat{p}$ to be the generator of $q$-translation, {\it i.e.} $\phi(q+\epsilon)=U_\epsilon\phi(q)=e^{i\epsilon p}\phi(q)$. However,  in this scheme, $U_\epsilon$ is no longer a unitary operator, since scalar products are no longer conserved, {\it i.e.}
\begin{equation}
\int_0^\infty(U_\epsilon\phi_1(q))^\dag U_\epsilon\phi_2(q)dq=\int_0^\infty\phi^{\dag}_1(q+\epsilon)\phi_2(q+\epsilon)dq\neq\int_0^\infty\phi^{\dag}_1(q)\phi_2(q)dq.
\end{equation}
This feature outlines that the canonical commutation relation must be replaced. In particular, a quantization based on the following commutation relation
\begin{equation}
[\hat{q};\hat{p}]=i\hbar\hat{q}
\end{equation}
does not suffer of such inconsistencies.\\ 
An analogous, but much more complicated, analysis of the Wheeler Super-space performed by Isham \cite{IsII84} leads to similar conclusions for the applicability of canonical commutation relations to General Relativity quantization. Indeed, these results stand only if configuration variables belong to a vector space. Nevertheless, an indication comes out that in Quantum Gravity elements of a non-canonical algebra have to be quantized. In this sense Rovelli and Smolin \cite{RS90} introduced the holonomy-flux algebra, even though a direct link between their work and Isham's prescriptions on GR quantization has not been establish yet. This is due to the fact that the space of holonomies is not a vector space. 

In order to introduce the physical meaning of such an algebra, its application to lattice gauge theories is presented in the following section.\\ 
For a recent framework where Wilson loops have been applied to gravity see the work \cite{HW07}, in which a positive cosmological constant is provided in a Quantum Gravity scenario.

\subsection{Lattice gauge theories}\label{latticegauge}

A lattice $k$ consists of links and plaquettes; if an orientation is chosen for them, edges $e$ and faces  $f$ are defined, respectively, as oriented links and plaquettes, but physical quantities are independent of the choice of the (arbitrary) orientation. In particular, the lattice $k$ is composed of the edges on the boundary $\partial k$ and the edges in the interior, $k^{0}$, so that
\begin{equation}
k=k^{0}\cup\partial k,
\end{equation}
and $E_{k}$ denotes the set of all the edges of $k$, $\{e\}$.\\
Connections on the lattice are applications that map the edges into elements of a (compact Lie) gauge group $G$,
\begin{equation}
g:E_{k}\rightarrow G,
\end{equation}
\begin{equation}
e\mapsto g_{e},
\end{equation}
$g_{e}$ being an element of the gauge group $G$, and the configuration space of the connections on $k$ is $A_{k}$.\\
Path integrals\footnote{Although in \cite{Co05} the whole description is developed without specifying the choice of a Minkowskian or a Euclidean background, for our purposes it will be more convenient to depict the model in a Minkowskian frame}, such as
\begin{equation}\label{w}
W[\phi]=\int \left(\prod_{e\in k}dg_{e}\right)e^{i S(g)}\phi(g),
\end{equation}
are the quantities that describe physical information. In (\ref{w}), $\left(\prod_{e\in k}dg_{e}\right)$ is the Haar measure on $G$, $S(g)$ is the action, and the function $\phi(g)\in\mathcal{L}_{0}^{2}(A_{k})$ denotes the particular physics to be described. The action 
\begin{equation}\label{faceaction}
S(g)=\sum_{f\in k^{0}}S_{f}
\end{equation}
can be written as the sum of terms referred to each face $f\in k^{0}$ (face action): each term $S_{f}$ is required to be gauge invariant and to depend on the edges of the face itself only.\\
Throughout this discussion, we will be interested in boundary-states amplitudes $\Omega[\phi]$, whose weighting functionals $\phi[\partial g]\in\mathcal{L}_{0}^{2}(A_{\partial k})$ depend on the group elements carried by boundary edges,
\begin{equation}\label{omega}
\Omega[\phi]:=\int \left(\prod_{e\in k}dg_{e}\right)e^{i S(g)}\phi^{*}(\partial g),
\end{equation}
{\it i.e.}, the wave function of a physical state, that is the probability of obtaining a physical state from vacuum.
Any lattice gauge model can be turned in a physically-equivalent description by means of the spin-foam formalism.

\paragraph{Spin-network States}
A spin network is an oriented graph, whose edges $e$ are labeled by irreducible representations of a gauge group $\rho_{e}$ (colours), and whose vertices $v$ are labeled by an intertwiner $I_{v}$ in the tensor product
\begin{equation}\label{intert}
\otimes_{i}V_{\rho_{i\quad out}}\otimes_{j}V^{*}_{\rho_{j\quad in}},
\end{equation}
where $V_{\rho}$ is the space of the irreps of $\rho$. The mathematical meaning of the intertwiners is mapping the operations on a group in the operations on another group. If one decomposes the space (\ref{intert}) in the sum of irreps, a sub-space can be found, which transforms according to the trivial representation, as it is composed of invariant vectors. In this picture, intertwiners form a map between $\otimes_{i}V_{\rho_{i\quad out}}$ and $\otimes_{j}V^{*}_{\rho_{j\quad in}}$.
A spin-network state or functional $\Psi_{S}(g)$ can be associated to a spin network, such that
\begin{equation}\label{corr}
\psi_{S}(g):g_{e}\rightarrow \rho_{g_{e}},
\end{equation}
and reads
\begin{equation}
\psi_{S}(g)=\left(\prod_{v\in k}I_{v}\right)\left(\prod_{e\in k}D_{\rho_{e}}\rho_{e}(g_{e})\right)\label{spnet}
\end{equation}
where $D_{\rho_{e}}\equiv (dimV_{\rho_{e}})^{1/2}$. The correspondence (\ref{corr}) is not one-to-one, so that spin networks can be defined equivalent if they lead to the same spin-network functional. Spin-network states define the space $\mathcal{L}_{0}^{2}(A_{k})$ of gauge-invariant functional of the connections $A_{k}$ : according to the Peter-Weyl theorem, the matrix elements of the irreps of a group form a basis for the functions of the Hilbert space of the $\mathcal{L}^{2}$ functions of the group. If an orthonormal basis $B_{k}$ for the intertwiners $I_{v}$ is chosen, {\it i.e.}

\begin{equation}
I^{a}_{\ \ b}=\frac{1}{D_{\rho_{e}}}\delta^{a}_{\ \ b},
\end{equation}
the basis for $\mathcal{L}_{0}^{2}(A_{k})$ will be orthonormal too; in this case, the spin-network functional simply rewrites 
\begin{equation}
\psi_{S}(g)=tr\left[\rho\left(\prod_{i}g_{e_{i}}\right)\right].
\end{equation}
If only the edges on the boundary $\partial k$ are taken into account, $B_{\partial k}$ is the orthonormal basis for $\mathcal{L}_{0}^{2}(A_{\partial k})$. Loops are the spin networks on the edges that surround a face (the smallest graphs possible), and induce a basis for the face action (\ref{faceaction}), so that the exponential in (\ref{w}) rewrites
\begin{equation}
e^{i S_{f}}=\sum_{S_{f}\in B_{f}}c_{S_{f}}\Psi_{S_{f}},
\end{equation}
where $c_{S_{f}}$ are suitable coefficients.\\
The loop functional is therefore the trace of the holonomy around a face in a given (irreducible) representation. 

\paragraph{Spin Foams}
Spin foams are 2-dimensional branched surfaces that carry irreps and intertwiners: the definition of analogous to that of spin networks, but one dimension has to be added. Each branched surface $F$ is composed of its unbranched components $F_{i}$, so that $F=\bigcup_{i}F_{i}$.\\
Given a spin network $\psi$, a spin foam $F$ is an application such that
\begin{equation}
\forall\psi, F:0\rightarrow\psi,
\end{equation}
and, given any two disjoint spin networks $\psi$ and $\psi'$, a spin foams $F$ is an application such that maps the former into the latter, or, equivalently,
\begin{equation}
\forall \psi, \psi', F:\psi\rightarrow\psi';\qquad F:0\rightarrow \psi^{*}\circ\psi',
\end{equation}
where $\circ$ denotes disjoint union.\\
A spin foam is non-degenerate iff each vertex is the end-point of at least one edge, each edge of at last one face, and each face carries an irrep of the group $G$.\\
Equivalence classes can be established for spin foams: spin foams are equivalent if one can be obtained from the other by affine transformation, subdivision or orientation reversal of the lattice.\\

Let's analyze in some detail how to express the path integral (\ref{omega}) in term of spin-foam amplitudes. The integration can be performed into two steps.\\
The path integrals can be integrated over $k^{0}$, as $\phi^{*}(\partial g)$ is not affected by the integration, {\it i.e.},
\begin{equation}\label{qwer}
\Omega [g]=\int_{\partial g'=g}\left(\prod_{e\in k^{0}}dg'_{e}\right)e^{i S(g')},
\end{equation}
and then inserted into (\ref{omega}), so that
\begin{equation}\label{odp}
\Omega[\phi]=\int \left(\prod_{e\in k^{0}}dg_{e}\right)\Omega(g)\phi^{*}(g),
\end{equation}
(\ref{qwer}) can be expanded into spin-network states. In fact, the exponential of the action can be expanded as
\begin{equation}
e^{i S(g)}=\prod_{f\in k^{0}}e^{i S_{f}(g)}=\prod_{f\in k^{0}}\sum _{S_{f}\in B_{f}}C_{S_{f}}\psi_{S_{f}(g)},
\end{equation}
and, when substituted in (\ref{qwer}), it brings the result
\begin{eqnarray*}
\Omega[g]=\int_{\partial g'=g}\prod_{e\in k^{0}}dg'_{e}e^{i S(g')}=\int_{\partial g'=g}\prod_{e\in k^{0}}dg'_{e}\prod_{f\in k^{0}}\sum _{S_{f}\in B_{f}}C_{S_{f}}\psi_{S_{f}}(g')=\\=\sum_{\left\{f\right\}\rightarrow\left\{S_{f}\right\}}\int_{\partial g'=g}\prod_{e\in k^{0}}dg'_{e}\prod_{f\in k^{0}}C_{S_{f}}\psi_{S_{f}(g)},
\end{eqnarray*}
where, in the last step, the sum has been drawn out of the integral, and all the possible configurations $S_{f}$ for each $f$ have been taken into account. The introduction of spin foams is suggested by the need to evaluate each term of the sum,
\begin{equation}\label{foam}
\int_{\partial g'=g}\prod_{e\in k^{0}}dg'_{e}\prod_{f\in k^{0}}C_{S_{f}}\psi_{S_{f}(g')}
\end{equation}
where spin networks are better organized into surfaces. In fact, two spin networks belong to the same unbranched surface $F_{i}$ if they share only one edge, and if this edge is not shared with any other spin network. The unbranched surfaces $F_{i}$ either are disconnected, or match other unbranched surfaces. In the latter case, the spin foam is defined as the branched surface $F=\cup_{i}F_{i}$.\\
In order to evaluate (\ref{foam}), two non-trivial cases can be distinguished, {\it i.e.}, 
\begin{enumerate}
	\item two loops match on one edge, and they carry the same label : the unbranched surface is defined as single-colored;
	\item more than two loops match on one edge, and Haar intertwiners ( a generalization of inertwiner defined formerly) have to be               introduced.
\end{enumerate}
As a result, all the elements contribute to the sum as follows
\begin{enumerate}
   \item for each vertex, a factor $A_{v}$;
   \item for each single-colored component, a factor $\prod_{i}A_{F_{i}}$, where $A_{F_{i}}\propto\prod_{f\in f_{i}}C_{f_{\rho}}$ and $C_{f_{\rho}}\propto C_{s_{f}}$, the proportionality factor being a suitable power of $dim V_{\rho}$;
   \item for each branching graph $\Gamma_{F}$, the projection properties of the Haar intertwiners have to be taken into account: as a result, for each vertex of the branching graph, one has to sum over all the possible ways to assign an intertwiners to the links of $\Gamma_{F}$.
\end{enumerate}
Collecting all the terms together, one obtains
\begin{equation}\label{omegadigi}
\Omega[g]=\sum_{F\subset k}\left(\prod_{v\in\Gamma_{F}}A_{v}\right)\left(\prod_{i}A_{F_{i}}\right)\psi_{S_{f}(g)}.
\end{equation}
The product
\begin{equation}
\prod_{f\in F_{i}}C_{f_{\rho}}
\end{equation}
in general depends on the discretization, and only in particular cases a geometrical interpretation is possible.\\
The insertion of (\ref{omegadigi}) in (\ref{odp}) gives the final expression of the path integral $\Omega[\phi]$. If one expands $\phi(g)$ in terms of the orthonormal basis of spin networks, 
\begin{equation}\label{phidigi}
\phi(g)=\sum_{S\in B(\partial k)}\phi_{S}\psi_{S}(g),
\end{equation}
(\ref{odp}) reads
\begin{equation}\label{pathin}
\Omega[\phi]=\sum_{F\subset k}\left(\prod_{v\in\Gamma_{F}}A_{v}\right)\left(\prod_{i}A_{F_{i}}\right)\phi^{*}_{S_{F}},
\end{equation}
{\it i.e.}, the only non-vanishing contributions are brought by boundary spin networks, and each spin-foam amplitude is weighted by the coefficient of the corresponding boundary state. The comparison between the path-integral formulation (\ref{omega})  and (\ref{pathin}) is eventually accomplished by noticing that the integration over connections is replaced by the sum over spin foams, and the spin-foam amplitudes wighted by the boundary functional $\phi_{S_{F}}$ play the role of the invariant measure and the exponential of the action, with a boundary weighting coefficient $\phi(g)$.    

\paragraph{Background independence}
The mismatch between the idea of background independence and the geometrical interpretation of spin-foam models can be analyzed by considering two possibilities:
\begin{enumerate}
    \item spin foams can be identified with the entire lattice, which plays the role of a discrete space-time;
    \item spin foams can be interpreted as lattice-independent geometrical objects, which live on the lattice itself. The lattice, in this case, is considered as an auxiliary field, which has to be removed in the definitive model.
\end{enumerate}
In the second case, the amplitudes described in the initial model must depend on the geometry of the spin foams only, {\it i.e.}, in the sum
\begin{equation}\label{somma}
\Omega_{k}[\phi]=\sum_{F\in k}\left(\prod_{v\in\Gamma_{F}}A_{v}\right)\left(\prod_{i}A_{F_{i}}\right)\phi^{*}_{S_{f}},
\end{equation}
each factor $A$ depends on the branching graph only. The sum in (\ref{somma}) can be extended to a background-independent sum over all the equivalence classes of spin foams $F$ on a given manifold $M$, {\it i.e.}, $\sum_{F\in k}\rightarrow\sum_{F\in M}$, so that
\begin{equation}\label{somma2}
\Omega_{k}[\phi]=\sum_{F\in M}\left(\prod_{v\in\Gamma_{F}}A_{v}\right)\left(\prod_{i}A_{F_{i}}\right)\phi^{*}_{S_{f}},
\end{equation}
where abstract (or topological) spin foams are defined by means of abstract spin-network states, the equivalence class of spin-networks states, invariant under homeomorphisms of the boundaries. The extension (\ref{somma2}) is possible only by the modification of the Hilbert space, as spin networks are not defined on the boundaries. $\partial k$. The new space of boundary states $\textsc{H}$ is defined as
\begin{equation}
\textsc{H}_{\partial M}=\left\{\sum_{i}a_{i}S_{i}:\qquad a_{i}\in C, S_{i}\subset M, n\in N\right\},
\end{equation}
{\it i.e.}, a finite combination of spin-network states, endowed with the structure of scalar product
\begin{equation}
<S,S'>=\delta_{SS'},
\end{equation}
for which the dual space $\textsc{H}^{*}_{\partial M}$ is defined, as usual, as
\begin{equation}
\textsc{H}^{*}_{\partial M}=\left\{\phi\right\}:\qquad \textsc{H}_{\partial M}\rightarrow C.
\end{equation}
The definition of such a Hilbert space is followed by the problem of overcounting, due to the homomorphisms $h$ of the manifold $M$,
\begin{equation}
h:M\rightarrow M,\qquad A(h^{*}M)=A(M), \qquad \phi_{S_{h^{*}F}}=\phi_{S-{F}},
\end{equation}
which can be gauged away \'a la Faddeev-Popov. The result is the definition of abstract or topological spin foams, and the corresponding spin-network states ar e the equivalence class of spin-network states invariant under homeomorphisms of the boundaries.

\subsection{Holonomies and fluxes in Quantum Gravity}
In this section, we demonstrate how spin networks and spin foams arise in the quantization of the gravitational field by a non-canonical algebra. 

Spin foams can be applied to covariant quantum gravity, where they play the role of the path integral \cite{hh83}, as the tool that connects different gravity states (geometries) in time. In particular, an equivalence class of 3-geometries $_{i}$ on a 3-d hypersurface $S_{i}$  is represented by a spin-network state, and the history between two different states is the spin-foam amplitude, {\it i.e.},
\begin{equation}
<h_{2}, S_{2}|h_{1}, S_{1}>=\int_{g/g(S_{1}=h_{1}), g/g(S_{2})=h_{2}}Dge^{iI_{EH}(g)}.
\end{equation}
where the measure Dg is aimed at outlining the conceptual analogy with Feynman’s
approach \cite{AJL00,FK00} rather than at defining any specific integration measure, which will be explicitly
given, when needed, throughout the calculations.\\
Of course, the composition of spin foams must be defined, such that the transition between two states is independent of the intermediate states among which the transition is decomposed, {\it i.e.}, in the sum
\begin{equation}\label{lkjh}
<h_{3}, S_{3}|h_{1}, S_{1}>=\sum_{h_{2}}<h_{3}, S_{3}|h_{2}, S_{2}><h_{2}, S_{2}|h_{1}, S_{1}>,
\end{equation}
the intermediate states $2$ must carry a trivial representation, in the sense specified in the previous paragraphs. As the probability of creating a state from the vacuum, a spin foam is defined as
\begin{equation}
|h_{1}, S_{1}>=\int_{g/g(S_{1})=h_{1}}Dge^{iI_{EH}(g)}.
\end{equation}
 
Spin-network states can be defined following a procedure which is slightly different from the previous one, in order to realize how the geometrical properties of the state fit the constraints of the ADM formulation \cite{Th01}.\\
%%%%%%%%%%%%%%%%%%%%%%%%%%%%%%%%%%%%%%%%%%%%%%%%%%%%%%%%%%%%%%%%%%%
%In particular, the Gauss, diffeomorphism and Hamiltonian constraint are encoded in the definition of the space of the physical states, annihilated by these constraints. The characterization of the physical states will be achieved by adding requests, step by step, on the properties of the general structures (such as links and vertices) introduced in the previous paragraphs.\\
%%%%%%%%%%%%%%%%%%%%%%%%%%%%%%%%%%%%%%%%%%%%%%%%%%%%%%%%%%%%%%%%%%%%%%

\paragraph{Holomy-flux algebra}
Starting from a configuration space whose elements are holonomies, other phase space variables are introduced, such that the Poisson algebra is well-defined, {\it i.e.} non-distributional. This result is achieved by smearing $E^i_a$ in two dimensions, in particular, given a surface $\sigma$, one defines $E(S)$ as
\begin{equation}
E(\sigma)=\int_\sigma\epsilon_{ijk}E^i_ad\sigma^{jk}\tau_a=\int_\sigma dudv n_iE^i_a\tau_a,
\end{equation}  
being $u$ and $v$ coordinates on the surface $\sigma$ and $n_i$ the normal vector.
The Poisson algebra is that of cylindrical functions $Cyl^{n}$, {\it i.e.} of $n$-times continuously differentiable functions of holonomies
\begin{equation}
f_\alpha(A)=f(U_{e_1},U_{e_2}\ldots)
\end{equation}
being $e_1,e_2\ldots$ edges of the graph $\alpha$.\\ 
Starting from the symplectic structure of the phase space manifold generated by $(A^a_i;E^i_a)$, we can calculate Poisson brackets between $h_\alpha(A)$ and $E_a(\sigma)$. We stress that our procedure is purely formal, since regularization has to be provided (see \cite{Th01} for details on the regularization).\\ 
Variational derivatives of $h_\alpha(A)$ and $E(\sigma)$ reads as follows 
\begin{eqnarray}
\frac{\delta h_\alpha(A)}{\delta A_i^a(x^j)}=\int_0^1\delta^3(x^j-\alpha^j(t))\frac{d\alpha^i}{dt}(U_\alpha)^t_0(A)\tau_a(U_\alpha)^1_t(A)dt\\
\frac{\delta E(\sigma)}{\delta E^i_a(x^j)}=\int_\sigma dudv n_i\delta(x^j-X^J(u;v))\tau_a,
\end{eqnarray}
$(U_\alpha)^b_a(A)$ being the parallel transport of $A$ along $\alpha=\alpha(s)$ from the point $s=a$ to $s=b$.\\
By virtue of canonical commutation relations %(\ref{canonical comm rel})
, Poisson brackets turn out to be
\begin{equation} 
\{E_a(\sigma);h_\alpha(A)\}=\int d^3x\frac{\delta E_a(\sigma)}{\delta E^i_b}\frac{h_\alpha(A)}{\delta A_i^b}.
\end{equation}
From the last relations, we see that, if $\sigma$ and $\alpha$ have no common points, we end up with expressions containing $\delta$ with different support, therefore the integration provides a vanishing result. The same result is obtained if $\alpha$ belongs to $\sigma$, since, in this case, vectors tangent to the graph are always orthogonal to $n^i$. Hence let us consider loops with a finite number of intersections with the surface $\sigma$, they can always be splitted in edges $e$, such that any edge has at most one intersection, which is the initial point. Finally, one obtains the following result
\begin{equation}
\{E_a(\sigma);h_\alpha(A)\}=8\pi G\gamma\tau_a \sum_{e\subset\alpha}h_e(A)o(e,\sigma)
\end{equation} 
$o(\alpha,\sigma)$ being $1,-1$ if orientations of $e$ and $\sigma$ agree or disagree, respectively.\\
Hence, one can define the action of $E(\sigma)$ on a generic cylindrical functional by the expression
\begin{equation}
E_a(\sigma)[f]=\{E_a(\sigma);f(h_\alpha(A))\}
\end{equation}
with $E_a(\sigma)$ acting as a vector field. The last two relations allow one to define the holonomy-flux algebra on the phase space $\it{A}$. This algebra can be equipped with a norm and the closure of $\it{A}$, $\it{\bar{A}}$, with respect to this norm is a commutative $C^*$-algebra \cite{AL95}, with the $*$ operation given by complex conjugation. The new algebra provides a distributional extension of the old one.\\ 
In what follows we will discuss the quantization in the connection representation, but the action of operators can be seen in a more intuitive way as an action on loops \cite{RS90,De97a}. This fact is based on the basic equivalence between the connections and the loop representation \cite{De97}.\\
In view of the quantization, the holonomy-flux algebra has to represented on a Hilbert space. This can be obtained by the GNS construction (see section \ref{GNS}), taking the following state $\omega$
\begin{equation}
\omega(f_\alpha(A),E_1,..,E_M)=\left\{\begin{array}{cc}\int d\mu(g_1)..d\mu(g_N)f_\alpha(g_1,..,g_N) & M=0 \\ 0 & M\neq0 \end{array}\right..
\end{equation}

This way a unique Hilbert space exists with a 3-diffeomorphisms invariant measure, which realizes a cyclic representation, with a self-adjoint flux operator. This space is that of square-integrable cylindrical function on $\it{\bar{A}}$,
\begin{equation}   
\textsc{H}_{aux}=L^2(\it{\bar{A}},d\mu),
\end{equation}
with the Ashtekar-Lewandowsky measure \cite{AL95}. A basis for this space is provided by spin networks (\ref{spnet}).

\paragraph{Gauss constraint}
Hence, we can impose the Gauss constraints: we expand functionals in $\it{\bar{A}}$ on spin networks
\begin{equation}
f_{\alpha}(A)=\sum_{S}c_{S}\psi_{S}(A)=\sum_S c_S\left(\prod_{v_S\in k}I_{v}\right)\left(\prod_{e_S\in k}D_{\rho_{e}}\rho_{e}(h_e(A))\right)
\end{equation}
and, by the Dirac prescription, we solve 
\begin{eqnarray}
\hat{G}_af_{\alpha}(A)=\bigg(\partial_i\frac{\delta}{\delta A_i^a}+\gamma\epsilon^c_{ba}\frac{\delta}{\delta A_i^c}
A_i^b\bigg)f_{\alpha}(A)=0.
\end{eqnarray}
Let us consider its action on $\psi_{S}(A)$: since it generates a gauge transformation, whose action is given by relation (\ref{gauge}), it provides us with a modification in each vertex. Therefore, $f_{\alpha}(A)$ can be made invariant under $SU(2)$ transformations by the choice of invariant intertwiners, {\it i.e.} in terms of $I_v$ components the following relation must stand 
\begin{eqnarray*}
I_{v\{j_1,\ldots,j_r,j'_1,\ldots,j'_s\}\{m_1,\ldots,m_r,m'_1,\ldots,m'_s\}\{l_1,\ldots,l_r,l'_1,\ldots,l'_s\}}U^{j_1}_{l_1n_1}\ldots U^{j_r}_{l_rn_r}U^{\dag j'_1}_{l'_1n'_1}\ldots U^{\dag j'_r}_{l'_sn'_s}=\\=I_{v\{j_1,..,j_r,j'_1,\ldots,j'_s\}\{m_1,\ldots,m_r,m'_1,\ldots,m'_s\}\{n_1,\ldots,n_r,n'_1,\ldots,n'_s\}}
\end{eqnarray*}
$r$ and $s$ labeling edges starting and ending in $v$, respectively.\\
Links that carry a representation of the SU(2) group, vertices where links intersect, and intertwiner which map the operations in the tensor product of the Hilbert spaces of the representations carried by the links define the spin-network states.\\

\paragraph{Super-momentum constraint}
The super-momentum constraint acts on $f_\alpha(A)$ as follows 
\begin{equation}
\hat{\mathcal{H}}_if_{\alpha}(A)=F^a_{ij}\frac{\delta}{\delta A_j^a}f_{\alpha}(A)=0,
\end{equation}
and it maps the curve $\alpha$ to the diffeomorphism-related one $\alpha'$. In fact, if one takes the below smeared combination of Gauss and super-momentum constraints acting on parallel transport $h_e(A)$ along an edge $e$,
\begin{equation}
\int d^3x\epsilon^i(\hat{\mathcal{H}}_i-A_i^a\hat{G}_a)h_e(A),
\end{equation}
the result obtained is the modification of $h_e(A)$ due to a 3-diffeomorphism acting on $A$ and generated by the infinitesimal vector $\epsilon$, {\it i.e.},
\begin{equation}
\int d^3x\int_0^1dth_0^t(A){\cal L}_\epsilon A^a(x)\tau_ah^1_t(A)=h_e(A+{\cal L}_\epsilon A)-h_e(A).
\end{equation}
This modification can be produced simply by operating with the inverse 3-diffeomorphisms on the edge $e$. Now the extension to a generic graph $\alpha$ is trivial.\\
Hence the operator corresponding to $\hat{H}_i$, modulo gauge transformations, annihilates those states, called s-knots, which are abstract spin-network states, defined in the Hilbert space $\textsc{H}_{diff}$. The Hilbert space $\textsc{H}_{diff}$ can be obtained from $\textsc{H}_{aux}$ by considering the invariance under diffeomorphisms. This way, s-knots are invariant under both diffeomorphisms and SU(2) gauge transformations.\\
We want to stress that ${\cal L}_\epsilon A$ is well-defined only on smooth connections, so that $\hat{H}_i$ does not act on the full algebra $\bar{A}$. The 3-diffeomorphisms invariance is restored by considering finite transformations, for which no inconsistency arises, but one has to conclude that infinitesimal 3-diffeomorphisms cannot be implemented in the LQG framework.\\  
The definition of an Hilbert space for solutions of the Gauss and of the super-momentum constraints can be given rigorously, since it can be probed \cite{ALMMT95} that such states belong to the closure of $A/G$, {\it i.e.} of the $A$ gauge invariant sector, with the Ashtekar-Lewandowsky measure. Hence, the final step is the determination of the dynamics by the solution of the super-Hamiltonian constraint.

\section{Spectrum of space-time operators}
Being the kinematical Hilbert space invariant under 3-diffeomorphisms, not under general coordinate transformations, one can define area and volume operators on it. One expects that the introduction of matter can give an invariant character under all 4-dimensional diffeomorphisms to area and volumes, so that they can be recognized as physical observables.\\
The evaluation of the spectrum of such quantities is strongly complicated because a regularization procedure has to be introduced in a manifold without any background metric. In what follows, we will give only formal proofs, having in mind that the required regularization can be provided.\\ 
Let us consider on spatial hypersurfaces a surface $\sigma$, characterized by a normal vector $n_i$ and coordinates $u,v$. The area of $S$ is given by the expression
\begin{equation}
A(\sigma)=\int_\sigma\sqrt{\delta^{ab}E^i_an_iE^i_bn_j}dudv, 
\end{equation} 
therefore the determination of its spectrum in a quantum framework is given by evaluating the action of $E^i_an_i$ on a generic parallel transport $U_e(A)$ along an edge $e$ of a spin network. In this respect, we have
\begin{equation}
[E^i_an_i](x)U_e(A)=8\pi G\gamma n_i(x)i\hbar\frac{\delta}{\delta A^a_i(x)}Pe^{\int_0^1A_j(e(t))\frac{de^j}{dt}dt} \label{area1}
\end{equation}
and, in the case only one intersection exists, coinciding with the point $t=0$, after some calculations, we obtain
\begin{equation}
=8\pi\gamma l_P^2 i o(\sigma,e)\tau_aU_e(A)\delta^2(x^i-e^i(0))
\end{equation}
$o(\sigma,e)$ being $+1$ or $-1$ if the orientations of $\sigma$ and $e$ coincide or not, respectively.
If no intersection exists or $e$ belongs to $\sigma$ the expression (\ref{area1}) vanishes.\\ 
Therefore one can easily recognize that by squaring and summing on gauge indexes, the result is as follows
\begin{equation}
\delta^{ab}E^i_an_iE^j_bn_jU_e(A)=(8\pi\gamma l_P^2)^2CU_e(A)\delta^2(x^i-e^i(0))
\end{equation}
$C$ being the Casimir of the $SU(2)$ group, $\delta^{ab}\tau_a\tau_b=j(j+1)I$.\\ 
This way, one obtains for the area operator
\begin{equation}
A(\sigma)U_e(A)=8\pi\gamma l_P^2\sqrt{j(j+1)}U_e(A),
\end{equation}
and for a generic spin network
\begin{equation}
A(\sigma)T_{\alpha\{j\}}=\sum_e8\pi\gamma l_P^2\sqrt{j_e(j_e+1)}T_{\alpha\{j\}},\label{areaspe}
\end{equation}
where the sum is on edges $e$ of the graph $\alpha$ that has one intersection with the surface $\sigma$ (as usual, in those calculations, one can always split and re-parametrize graphs, such that intersections are always one for each edge and at its initial point). It can be shown that $A(\sigma)$ is self-adjont on $\textsc{H}_{kin}$.\\
The last relation clearly shows that 
\begin{itemize}
\item spin networks are eigenstates of the area operator,
\item edges carry quanta of area;
\item the area spectrum is discrete.
\end{itemize}

Rovelli and Smolin \cite{RS95} demonstrated, by considering an arbitrary background metric, that the area operator can be expressed in terms of well-defined elements of the loop representation and that results on spectra do not depend on the metric. Moreover, Ashtekar and Lewandowski \cite{AL97} obtained same results by applying a regularization technique directly in the space of cylindrical functionals.

Let us now turn our attention to the volume operator,
\begin{equation}
V(\Omega)=\int_\Omega\sqrt{h}d^3x=\int_\Omega\sqrt{E}d^3x,\label{vol}
\end{equation}
$E$ being $\frac{1}{3!}\epsilon_{ijk}\epsilon^{abc}E_a^iE_b^jE^k_c$. Given a graph $\alpha$ and a function $h_\alpha(A)$ of parallel transports along its edges $e$, the operator $E$ acts as follows
\begin{eqnarray*}  
E(x)h_\alpha(A)=\frac{1}{3!}\epsilon_{ijk}\epsilon^{abc}E_a^i(x)E_b^j(x)E^k_c(x) h_\alpha(A)=\\=\frac{8i\pi\gamma l_P^2}{3!}\epsilon_{ijk}\epsilon^{abc}E_a^i(x)E_b^j(x)\bigg[\int_0^1dt\frac{d\alpha^k}{dt}(U_\alpha)_0^t\tau_c(U_\alpha)_t^1\delta^3(x-\alpha(t))dt\bigg]=
\end{eqnarray*}
and by acting with $E^b_j$
\begin{eqnarray*}
=\frac{(8i\pi\gamma l_P^2)^2}{3!}\epsilon_{ijk}\epsilon^{abc}E_a^i\bigg[\int_0^1dt\int_0^tds\frac{d\alpha^j}{ds}(U_\alpha)_0^s\tau_b(U_\alpha)_s^t\delta^3(x-\alpha(s))\frac{d\alpha^k}{dt}(U_\alpha)_0^t\tau_c(U_\alpha)_t^1\delta^3(x-\alpha(t))+\\+\int_0^1dt\frac{d\alpha^j}{dt}(U_\alpha)_0^t\tau_c(U_\alpha)_t^1\delta^3(x-\alpha(t))\int_t^1ds\frac{d\alpha^k}{ds}(U_\alpha)_t^s\tau_b(U_\alpha)_s^1\delta^3(x-\alpha(s))\bigg].
\end{eqnarray*}
Finally, the third $E_a^i$ operator provides us with a third integration $\int dv$ along the path $\alpha$, together with a further $\delta(x-\alpha(v))$. All these $\delta$'s restrict the integration on $x=\alpha(t)=\alpha(s)=\alpha(v)$, so we have $t=s=v$. The last condition produces a term $\epsilon_{ijk}\frac{d\alpha^i}{dt}\frac{d\alpha^j}{dt}\frac{d\alpha^k}{dt}$, which is not vanishing if and only if there exist a point in the graph $\alpha$ where tangent vectors form a set of three independent vectors. This means that a vertex must exist inside the spatial region $\Omega$. Moreover, since, after a regularization, the products of the three $\delta$-functions reduces to a single one in the vertex, the evaluation of the expression (\ref{vol}) requires simply the calculation of $E(x)$ in that point. Hence, quanta of volumes are transported by vertexes of a graph.\\
Therefore, let us consider the case in which only a vertex $\textit{v}$ with $n$ out-going edges $e_i\hspace{0.3cm}i=1,\dots,n$ is contained in $\Omega$ we obtain for the volume operator
\begin{equation}
V(\Omega)h_\alpha=(8\pi\gamma)^\frac{3}{2}l_P^3\sqrt{|q|}h_\alpha\qquad h_\alpha=\epsilon^{abc}\sum_{e,e',e''}o(\textit{v},e,e',e'')\tau^{j_e}_a\tau^{j_{e'}}_b\tau^{j_{e''}}_ch_\alpha,
\end{equation}
$o(\textit{v},e,e',e'')$ being $+1$ or $-1$ if vectors tangent to $e$, $e'$ and $e''$ in $\textit{v}$ have a positive or negative orientation with respect to $V(\Omega)$ (as already pointed out, it vanishes if tangent vectors are not independent), while the sum is performed over all edges of $\alpha$ passing through $\textit{v}$.\\ 
In general, one must sum the expression above over all vertexes contained in the region $\Omega$.\\
Although a set of eigenvectors has not been found yet, nevertheless it can be demonstrated that the spectrum is discrete.\\
The regularization procedure is not well-defined as for the area operator; in fact, an undetermined constant appears in front of the volume spectrum. Moreover, results obtained in the loop representation \cite{RS95} differ from those in the space of cylindrical functionals \cite{AL98,BR06}. However,  Giesel and Thiemann \cite{GT06} claimed, by a consistency check on the equivalence between a quantization based on triads and one based on fluxes, that the latter is the correct one and that the undetermined constant can be fixed.

Although a discrete the spectra for geometric operators is one of the most impressive
results of the LQG approach, nevertheless recently there has been a debate whether
it survives in the dynamical Hilbert space \cite{DT07,R07}.

\section{Quantum dynamics in LQG}
The implementation of the super-Hamiltonian constraint (\ref{sHam}) is up to now the major issue of the LQG program.\\ 
In this direction a useful tool is to rewrite it \cite{Th96} in terms of $A_i=A^a_i\tau_a$, $E^i=E^i_a\tau_a$ and $K_i=K^a_i\tau_a$ as follows
\begin{equation}
\mathcal{H}=\frac{1}{(2j+1)\gamma e}\bigg(Tr(F_{ij}[E^i,E^j])-\frac{\gamma^2+1}{4\gamma}[K_i,K_j][E^i,E^j]\bigg),
\end{equation}
$j$ being the dimensionality of the $SU(2)$ representation $\tau_a$. The factor $\frac{1}{e}$ can be reproduced by virtue of the volume operator, because of the relation
\begin{equation}
\frac{[E^i,E^j]}{e}(x)=\frac{1}{8\pi G}\epsilon^{ijk}2\{V(\Omega_x),A_k\},
\end{equation} 
where $V(\Omega_x)$ is an arbitrary volume containing the point $x$. It will prove to be useful the introduction of the quantity $K$,
\begin{equation}
K=\int_\Sigma K^a_iE^i_a,
\end{equation}
such that the full super-Hamiltonian constraint reads as 
\begin{equation}
\mathcal{H}=\frac{1}{(2j+1)8\pi G\gamma}\bigg(2\epsilon^{ijk}Tr(F_{ij}\{V(\Omega_x),A_k\})-\frac{\gamma^2+1}{(8\pi G)^2\gamma^4}\epsilon^{ijk}Tr([A_i,K][A_j,K][A_k,V])\bigg)
\end{equation}

The last expression allows to define the super-Hamiltonian operator on a quantum level, being the volume operator well-defined on spin networks, while $A_i$ and its curvature can be obtained by a limit procedure on holonomies along an edge or a loop with descresing length. Indeed, the limit can be taken for a small but non-vanishing length of the edges/loops involved. This feature reflects the fundamental discretness of the space manifold. 

As far as $K$ is concerned, a similar calculation can performed. A suitable regularization procedure can be defined $\cite{Th96}$ and the determination of the dynamics reduces to the study of such operators on spin-network functionals. This will involve a rather technical investigation, but, from a theoretical point of view, just a finite number of operations have to be performed. This way, the Wheeler-DeWitt equation has been reduced to a combinatorial problem.\\ 
In particular, one finds that the effect of the operator $\hat{\mathcal{H}}$ is to change graphs by removing or adding edges and to modify functionals by raising or lowering quanta of spin at edges, just like in quantum field theories field operators raise or lower the number of particles. For this reason Thiemann concluded that spin networks realize the ``non-linear Fock representation'' for Quantum Gravity.\\ 
However, a different approach consists in going over the canonical quantization and referring to explicitly covariant formulations, like those in terms of spin foams.

\paragraph{Spin foams and the Hamiltonian constraint}
In the canonical formulation, from a quantum-mechanical point of view, the Hamiltonian operator $H_{N, \vec{N}}(t)$, composed of the super-Hamiltonian and the super-momentum constraint, $H_{N, \vec{N}}(t)=\mathcal{H}[N(t)]+\vec{\mathcal{H}}[\vec{N}(t)]$, can be interpreted as the generator of quantum evolution from the initial hypersurface $\Sigma_{i}(t=0)$ to the final hypersurface $\Sigma_{f}(t=1)$, parametrized by the proper time evolution $U(T)$,
\begin{equation}
U(T)=\int_{T}dNd\vec{N}U_{N, \vec{N}}=\int_{T}dNd\vec{N}e^{-i\int_{0}^{1}dtH_{N,\vec{N}}(t)}.
\end{equation}
The evolution operator $U(T)$ encodes the dynamics of the gravitational field, and its expansion in powers of $T$ can be shown to be finite order by order. The calculation of the matrix element of such an operator between two states of the gravitational field is strictly analogous to that followed in the familiar calculation of the S-matrix elements in a gauge theory, and reads explicitly
\begin{equation}\nonumber
<s_{f}|U(T)|s_{i}>=<s_{f}|s_{i}>+(-iT)\left(\sum_{\alpha\in s_{i}}A_{\alpha}(s_{i})<s_{f}|D_{\alpha}|s_{i}>+\sum_{\alpha\in s_{f}}A_{\alpha}(s_{f})<s_{f}|D^{+}_{\alpha}|s_{i}>\right)+
\end{equation}\nonumber
\begin{equation}\label{elements}
+\frac{(-iT)^{2}}{2!}\sum{\alpha\in s_{i}}\sum_{\alpha'\in s'}A_{\alpha}(s_{i})A_{\alpha'}(s')<s_{f}|D_{\alpha'}|s'><s'|D_{\alpha}|s_{i}>+...
\end{equation}
Such a result can be obtained by splitting the calculation in several steps, {\it i.e.},
\begin{enumerate}
	\item the evolution from the initial hypersurface to the final one is expressed as a sum over intermediate hypersurfaces, as sketched in (\ref{lkjh}). The intermediate hypersurfaces differ by a small coordinate time, and the time evolution between two hypersurfaces can be written in terms of the diffeomorphism that describes the shift between them, so that $U_{N, \vec{N}}=D(g)U_{\vec{N},0}$;
	\item the expansion of $U_{\vec{N},0}$ and the insertion of the identical projector where needed leads to a sum, where, at each order $n$, the operator $D$ acts $n$ times. Its action on the states is given by the coefficients $A$, which can be evaluated in terms of the explicit form of the Hamiltonian constraint;
	\item  $U(T)$ can be worked out of $U_{\vec{N},0}$ after integrating over the lapse and the shift. The first integration follows directly, as the integrand does not depend on $N$, and the second one corresponds to the implementation of the diffeomorphism constraint.  
\end{enumerate}
As a result, the matrix elements of the operator $U$ read as (\ref{elements}), where generic spin-network states have been substituted with the corresponding s-knots states. Analyzing the geometrical meaning of the intermediate states, in which the sum has been split up, allows one to recognize (\ref{elements}) as a sum over spin foams. In fact, the time evolution of a generic surface $s_{i}$ describes a ''cylinder'', whose time slicing are spin-network states belonging to the same s-knot, unless any interaction occurs. When operating on such a state, the Hamiltonian constraint generates a new state with one new edge and two new vertices, {\it i.e.}, this structure is the elementary interaction vertex of the theory, as suggested by the comparison with ordinary gauge theories. As a generalization, the Hamiltonian constraint acts adding one dimension to the spin-network state, and such a new direction can be interpreted as time, because of the geometrical construction of (\ref{elements}), thus opening the way for the interpretation of these new states as spin foams. In fact, at the n-th order of sum, n new dimensions are added, and the sum can be written as the sum of topologically inequivalent term, where the weight of each vertex is determined by the coefficient of the Hamiltonian constraint. Furthermore, if the irreducible representation of the gauge group carried by the edge of each spin network is taken into account during the addition of the new vertices, the resulting geometrical objects fit the definition of spin foams given in the previous paragraph, {\it i.e.}, the implementation of the Hamiltonian constraint leads naturally to the sum over spin foams. Another approach to 4-d spin foams can found in \cite{BC98} and its developments, where a 4-dimensional state is constructed by the spin covering of the group $SO(4)$.

\section{Open issues in Loop Quantum Gravity}

\paragraph{Super-Hamiltonian constraint}
The main difficulties of the LQG program are connected with the implementation of the super-Hamiltonian constraint.\\
At first, some ambiguities arise in the definition of $\hat{\mathcal{H}}$ (for an example see \cite{GR01,Pe06}). Some of them are linked to the regularization procedure itself. The question is highly technical and the debate is open on the naturalness of some choices, thus on what ambiguities have a physical significance (for technical discussions about this topics see \cite{NPZ05,Th06a}). We will not discuss these issues, but we focus our attention on other points such as the implementation of the Dirac algebra of constraints, the semi-classical limit and the definition of scalar product on solutions of the full set of constraints. \\

\paragraph{Dirac algebra of constraints} The quantum implementation of the Dirac algebra of constraints (\ref{alcons1}), (\ref{alcons2}) and (\ref{alcons3}) would demonstrate the absence of anomalies, thus the consistency of the adopted quantization procedure. The first problem one faces is the impossibility to implement infinitesimal 3-diffeomorphisms in a representation of $\bar{A}$. Because of this issue, one refers to finite transformations and translates commutation relations in corresponding expressions, {\it i.e.}
\begin{eqnarray}
\hat{U}(\phi)\hat{U}(\phi')\hat{U}^{-1}(\phi)=\hat{U}(\phi\circ\phi'\circ\phi^{-1})\\
\hat{U}(\phi)\hat{\mathcal{H}}(f)\hat{U}^{-1}(\phi)=\hat{\mathcal{H}}(\phi\circ f)
\end{eqnarray}
$\phi$ being an arbitrary 3-diffeomorphism and $\hat{U}(\phi)$ its representation. The first relation stands exactly on $\textsc{H}_{aux}$, while the second is reproduced modulo 3-diffeomorphisms \cite{Th01}.\\ 
However, the main question is the implementation of the last commutation relation. Although the operator corresponding to the right-hand side of the relation (\ref{alcons3}) can be defined on $\textsc{H}_{aux}$ \cite{Th98b}, nevertheless the equivalence with the left-hand side stands only on a proper subspace (of its dual, see also \cite{LM98}). Whether the algebra of constraints has been reproduced this way is a question of debate; however this quantum implementation stands at most on-shell. The last conclusion, if true, would probe the absence of anomalies. 

\paragraph{Semi-classical limit}
The algebra of constraints must be reproduced in view of maintaining fundamental symmetries on a quantum level, but also to reproduce on semi-classical states the right transformations properties. However, no convincing semi-classical state exists up to now in LQG. Earlier attempts to develop such states were based on defining functionals with a well-defined geometric structure \cite{GR97} or which diagonalize holonomy operators  \cite{Arn99}. The issue of combining both these properties leads to the ``complexifier technique'' \cite{Th00,TW01,TW02,TW03}, which is a well-defined procedure to define coherent states for gauge theories. However, in the LQG framework the definition of semi-classical states is not unique \cite{STW01,AL01}. Moreover, since those states result to be graph-dependent, the action of the super-momentum and of the super-Hamiltonian operators takes one state into a different one.  Therefore, these constraints provide ``strong fluctuations'', such that the implementation of semi-classical conditions (small fluctuation around the expectation value) has not been accomplished yet. This means that one cannot investigate whether the quantum algebra of constraints reproduce the classical one, or if Einstein's equations are reproduced with semi-classical corrections.

The speculations above stress how the main difficulty in the development of the low-energy sector of LQG consists in combining semi-classical tools with General Covariance. In \cite{ABC05}, Ashtekar pointed out that the ``group averaging'' procedure could be a powerful tool in this direction, at least in the linearized case.

The ``group averaging'' \cite{ALMMT95,Ma00} is a procedure to find solutions of constraints. Given a constraint $\hat{C}$, acting on quantum states $\psi$, {\it i.e.},
\begin{equation}
\hat{C}\psi=0,
\end{equation} 

one can formally define a solution as
\begin{equation}
\psi_{\hat{C}}=\int d\lambda e^{i\lambda\hat{C}}\psi,
\end{equation}

being $U(\lambda)$ a one-parameter group of transformations generated by $\hat{C}$. For non-compact groups this procedure can define states which are not normalizable, thus this implies to enlarge the Hilbert space.\\
The ``group averaging'' is a formal procedure, which works well only when the group of transformations can be easily calculated from constraints. In fact, it turns to be very useful in mini-superspace models, as for instance isotropic cosmologies (see section \ref{lqc}).

As far as the determination of semi-classical states is concerned, there are example where starting from kinematical states which behave semi-classically, by a ``group averaging'' one reduces fluctuations around expectation values \cite{ABC05}. However, these results have been obtained with constraints much simpler than those of GR. Since a lot of technical problems arise in dealing with more complex cases, no definite answer whether it can be applied to LQG exists up to now.

Finally, for an example on the development of a semi-classical tetrahedron in a spin-foam model, see \cite{RS06}. 

\paragraph{Scalar products}
The introduction of a scalar product is one of the main point of any quantization procedure in presence of constraints. Here, because of the complexity of the super-Hamiltonian operator, it is a very difficult task to analyze its kernel \cite{Th98a,Th98b} and the possibility to extend the scalar product in $\textsc{H}_{diff}$ on it has not been probed.\\\\

\subsection{Master Constraint and Algebraic Quantum Gravity}

The Master constraint program proposed by Thiemann \cite{Th06} purses the resolution
of the dynamics in a canonical way. Such a procedure is based on quantizing the Master-Constriant operator $\textbf{M}$ instead of $\mathcal{H}$. $\textbf{M}$ is defined as
\begin{equation}
\textbf{M}=\int d^3x\frac{\mathcal{H}^2}{\sqrt{h}}
\end{equation}

The condition $\textbf{M}=0$ is equivalent to $\mathcal{H}=0$, thus they define the same hypersurface in the phase space. Furthermore, since $\textbf{M}$ is 3-diffeomorphisms invariant, its commutator with spatial diffeomorphisms vanishes, so that it can be represented on $H_{Diff}$.
This way, one overcomes the issue of reproducing the Dirac algebra of constraints.

As soon as $\textbf{M}$ is quantized, the full Hilbert space can be written as the direct
sum of spaces spanned by eigen-vectors of $\textbf{M}$. This way the issue of finding the physical space reduces to determine the kernel of $\textbf{M}$.

It is also possible to give a definition for $\textbf{M}$ such that it is not graph-changing
before the regolarization to take place (it implies to start with an anomalous $\mathcal{H}$).

The Master-Constraint quantization can be tested by proper semi-classical states
in the context of Algebraical Quantum Gravity (AQG) \cite{GT06a,GT06b,GT07}. This theory is
based on transporting the machinery of LQG on a fundamental algebraic graph,
from which all graphs can be derived by an embedding. In AQG one deals with
one graph only, on which normalizable coherent states can be properly defined.

Furthermore, the expectation value of $\textbf{M}$ has been evaluated, finding at the leading
order the classical expression for M. This result demostrates that AQG has GR as
semi-classical limit and makes this model very tantalizing for further developments
in Quantum Gravity (see also \cite{GT07a} for an application to a reduced phase space
quantization). The main issue with such an approach consist in in passing from
the algebraic level to the embedded one, thus in establishing a relation with results
of LQG.

\section{Loop Quantum Cosmology}\label{lqc}

The difficulties in implementating the quantization of the Hamiltonian constraint
in LQG can be solved in some minisuperspace approaches. The most remarkables
case is that of isotropic loop quantum cosmology (LQC) \cite{APS06,APS06a}, where the LQG quantization procedure is applied to a Friedmann-Robertson-Walker space-time.
The metric tensor is the following one
\begin{equation}
ds^2 = dt^2-a^2\left(\frac{1}{1-kr^2}dr^2+d\Omega^2\right) 
\end{equation}

$k$ being $1,0,-1$ for a closed, flat and open space, respectively.

The only configuration variable is the scale factor $a=a(t)$, such it can be taken as coordinates of the phase space $\{c,p\}$
\begin{equation}
|p|=a^2,\qquad c=\frac{1}{2}(k+\gamma \dot{a}).
\end{equation}

As soon as a fiducial metric ${}^0\!h_{ij}$ is introduced, with associated triads ${}^0\!e^i_a$ and connections ${}^0\!\omega^{ab}_i$, one gets
\begin{equation}
A^a_i=c\phantom1\epsilon^a_{\phantom1bc}{}^0\!\omega^{bc}_i,\qquad E^i_a=p\sqrt{{}^0\!h}{}^0\!e^i_a.
\end{equation}

Hence, holonomies can be evaluated, starting from those ones associated to straight paths $\mu{}^0\!e^i_a$, which read as
\begin{equation}
h(A)_a=\cos{\frac{\mu c}{2}}I+2\sin{\frac{\mu c}{2}}\tau_a.
\end{equation}

This way, one can take almost periodic functions $N_\mu=e^{\frac{i\mu c}{2}}$ as a basis in the configuration space. The algebra generated by $\{N_\mu,p\}$ plays the rome of the holonomy-flux algebra, such that by analogous construction of the general case the Hilbert space turns out to be $H_{kin}=L^2(\textbf{R}_{Bohr},d\mu_{Bohr})$, $\textbf{R}_{Bohr}$ being the Bohr compactification of the real line. An orthonormal basis is given by $N_\mu$ themeselves, with the measure given by 
\begin{equation}
<N_{\mu'}|N_\mu>=\delta_{\mu',\mu}.
\end{equation}

The description in terms of a countable set of basis vectors is the main achievement with repsect to Wheeler-DeWitt quantum cosmology. On a quantum level, the action of operators is given by
\begin{equation}
N_\mu\psi(c)=e^{\frac{i\mu c}{2}}\psi(c),\qquad p\psi(c)=-i\frac{8\pi\gamma l_P^2}{3}\frac{d}{dc}\psi(c),
\end{equation}

so that states $|\mu>$ can be defined for which
\begin{equation}
<c|\mu>=e^{\frac{i\mu c}{2}},\qquad p|\mu>=\frac{8\pi\gamma l_P^2}{6}\mu|\mu>.
\end{equation}

In what follows we will consider the case $k=0$, while for $k=1$ see \cite{APSV07}. 

The evaluation of the Hamiltonian constraint involves the expression of the field strength $F^a_{ij}$, which can be obtained by the limiting procedure from holonomies. The regularization procedure implies to fix a minimum value $\mu=\bar{\mu}$, just like the general case. 

Since the evaluation of the super-Hamiltonian involves the wolume operator $V$, it is useful to introduce the basis $|v>=\frac{2^{3/2}}{3^{7/4}}sign(\mu)|\mu|^{3/2}|\mu>$, such that  
\begin{equation}
V|v>=\frac{3^{7/4}}{2^{3/2}}\left(\frac{8\pi\gamma}{6}\right)^{3/2}l_P^3|v||v>.
\end{equation}

Finally, the expression of $\mathcal{H}$ reads
\begin{equation} 
\mathcal{H}^{\bar{\mu}}=sign(p)\frac{4i}{8\pi\gamma^3\bar{\mu}^3l_P^2}\sin^2{\bar{\mu}c}\left(\sin{\frac{\bar{\mu}c}{2}}V\cos{\frac{\bar{\mu}c}{2}}-\cos{\frac{\bar{\mu}c}{2}}V\sin{\frac{\bar{\mu}c}{2}}\right).\label{lqcosh}
\end{equation}

It is worth noting that the limit $\bar{\mu}\rightarrow0$ does not exists. There are different ways of fixing $\bar{\mu}$; among them a reasonable choice is \cite{CS08}
\begin{equation}
\bar{\mu}|p|=2\sqrt{3}\pi\gamma l_P^2,
\end{equation}

such that the minimum area enclosed by a loop is given by the minimum area gap predicted by LQG (\ref{areaspe}).

The expression (\ref{lqcosh}) can be made Hermitian by a proper factor ordering and a possible choice (see \cite{B02} for a different ordering) implies to rewrite it as
\begin{equation} 
\mathcal{H}^{\bar{\mu}}=sign(p)\frac{4i}{8\pi\gamma^3\bar{\mu}^3l_P^2}\sin{\bar{\mu}c}\left(\sin{\frac{\bar{\mu}c}{2}}V\cos{\frac{\bar{\mu}c}{2}}-\cos{\frac{\bar{\mu}c}{2}}V\sin{\frac{\bar{\mu}c}{2}}\right)\sin{\bar{\mu}c},
\end{equation}

which, when applied to $\psi(v)$, leads to the following difference equation
\begin{equation}
\mathcal{H}^{\bar{\mu}}\psi(v)=f_+\psi(v+4)+f_0\psi(v)+f_-\psi(v-4)
\end{equation}

with
\begin{eqnarray}
f_+(\mu)=\frac{3^{5/4}}{2^{5/2}}\sqrt{\frac{8\pi}{6}}\frac{l_P}{\gamma^3/2}|v+2|\Big||v+1|-|v+3|\Big|\\
f_-(\mu)=f_+(\mu-4)\qquad f_0(\mu)=-f_+(\mu)-f_-(\mu).
\end{eqnarray}

The replacement of the Wheeler-DeWitt differential equation with a difference one is the most impressive result of LQC. As soon as a clock-like scalar field $\phi$ is introduced the full Hamiltonian reads
\begin{equation}
\mathcal{H}_{tot}=\mathcal{H}+8\pi G\frac{p^2_\phi}{|p|^{3/2}}.
\end{equation}

such that after the quantization (by a canonical treatment for $\phi$) one ends up with the following difference equation giving the dynamics
\begin{equation}
\frac{\partial^2}{\partial\phi^2}\Psi(v,\phi)=B^{-1}(v)\mathcal{H}\Psi(v,\phi)=\Theta\Psi(v,\phi)\label{lqck}
\end{equation} 

where the inverse volume operator eigenvalues read
\begin{equation}
B(v)=\frac{1}{|p|^{3/2}}=\frac{3^{5/4}}{2^{3/2}}|v|\Big||v+1|^{1/3}-|v-1|^{1/3}\Big|^3.
\end{equation}

The boundness of the operator corresponding to $|p|^{3/2}$ has been one of the first hint towards the resolution of the cosmological singularity in LQC \cite{B01,B02}.

From the last Klein-Gordon like equation, positive frequency states can be identified and a Schr\"odinger dynamics is inferred, {\it i.e.}
\begin{equation}
i\frac{\partial^2}{\partial\phi^2}\Psi(\mu,\phi)=-\sqrt{\Theta}\psi.
\end{equation}

The physical Hilbert space can be developed from $\textbf{H}_{Kin}$ by the group averaging technique.

As far as the fate of the classical singularity is concerned, one starts from an initial semi-classical Universe, described by a state sharply picked around $\mu=\mu^\star>>\bar{\mu}$, and evolves it backward in time. What happens is that the state remains semi-classical during the evolution and a bounce occurs for $\mu\sim\bar{\mu}$, followed by an expansion phase. Therefore, the classical singularity is solved in the LQC framework.

It is very impressive to outline this scenario in terms of effective equations. In fact, it comes 
out that quantum effects provide the following modification to the Friedmann equation \cite{SVV06} at the leading order in $\bar{\mu}$
\begin{equation}
\frac{\dot{a}^2}{a^2}=\frac{8\pi G}{3}\rho\left(1-\frac{\rho}{\rho_{cr}}\right),
\end{equation}

$\rho$ being the scalar field energy density, while $\rho_{cr}=\frac{\sqrt{3}}{16\hbar G^2\pi^2\gamma^3}$. The last expression emphasizes that quantum corrections behave as a negative pressure term, responsible of the bounce. For an extension of this result to general matter fields see \cite{BHS07}, while in \cite{MS08} it is outlined how the inclusion of further orders in $\bar{\mu}$ affects significantly the Universe dynamics.

There has been increasing interest in a modification of the equation (\ref{lqck}), such that one can solve dynamical equations for any quantum states, not only semi-classical ones \cite{ACS08}. This model is known as simplified LQC (sLQC). It is expected that sLQC is a well-grounded approximation of LQC, since deviations are very small on most of the physical states. It has been demonstrated that the bounce is a proper feature of this simplified version for any state. Furthermore, the Wheeler-DeWitt dynamics has been recovered by a limiting procedure for $\bar{\mu}$, but only locally. The parameter $\bar{\mu}$ can be fixed such that the difference between observables for sLQC and the Wheeler-DeWitt theory is arbitrary small in a closed region. However, sLQC is fundamentaly discrete, since outside such a region predictions are very different from the continuous case. 

This result also points out that a fundamental scale exists at which operators must be evaluated. A possible escape from this conclusion involves the definition of an effective field theory at any scale, such that by a renormalization procedure the cut-off can be removed \cite{CVZ07}. 

%In this respect, the polymer quantization offers a scenario where such a limiting procedure can be defined (see section ...). For this reason the interplay between LQC and the polymer quantization is a subject of current investigation.  

\paragraph{}This section outlines how LQC is an arena where properties of the LQG quantization program can be tested, without most of the difficulties proper of the general framework. Indeed, LQC does not coincide with the cosmological sector of LQG, since the role of inhomogeneities is crucial on a quantum level. For instance, the avoidance of the singularity in a generic cosmological model is questionable, since the inverse volume operator is unbounded \cite{BT06}.

\section{On the physical meaning of the Immirzi parameter}
We have seen in section \ref{holst} how the Immirzi parameter, introduced as a factor in front of a topological term, plays no role classically. Nevertheless, after the quantization, it enters the spectra of observables, so it modifies physical predictions. This proves that quantizations in different $\gamma$-sectors are inequivalent.\\ 
The interpretation of this ambiguity is far from being understood. Rovelli and Thiemann \cite{RT98} analyzed the way this parameter comes out. They conclude that such an ambiguity is a consequence of two basic features:
\begin{itemize}
\item the affine structure of the configuration space, in particular the presence of two variables (the extrinsic curvature and spin connections) playing the role of connections, 
\item the quantization procedure, based on taking holonomies as fundamental variables in the Hilbert space.
\end{itemize}
In fact, as pointed out by Corichi and Krasnov \cite{CK98}, if the quantization is performed of a $U(1)$ gauge theory in the holonomy-flux representation, an ambiguity comes out, which fixes the quanta of the electric charge. For non-Abelian theories, this result does not stand, since connections cannot be rescaled without violating the gauge invariance of holonomies. However, if another connection-like variable exists, it can be combined with the old one, and the resulting theory presents a residual ambiguity.\\ 
The Immirzi ambiguity is often associated with the so-called $\theta$-sector in QCD. It consists in the CP-violating term $\theta\epsilon^{\mu\nu\rho\sigma}F^a_{\mu\nu}F^a_{\rho\sigma}$, which can be added to the Lagrangian density. Being a 4-divergence, it does not modify the classical dynamics; nevertheless, after the quantization, it can account for the observed CP violations in K-decays. However, we want to stress that, in a mathematical approach, these ambiguities are different, since the Holst modification vanishes only for the solutions of the equations of motion. Futhermore, Montesinos \cite{Mo01a} outlined that for $\gamma=\pm i$ the contribution to constraints coming out from the addition of topological terms (Euler and Pontrjagin terms) to the action is not the same as that due to the Holst modification.\\
The presence of spinors modifies this picture, therefore the introduction of fermions can give a deep insight on the physical meaning of $\gamma$ \cite{Me06,Me07,PR06}.\\ 
These speculations aim at regarding to the Immirzi parameter as a new fundamental constant, which has to be fixed from the macroscopic limit. In this direction, comparisons are made among the black hole (BH) entropy expression predicted by LQG \cite{ABCK98} and that given by BH thermodynamics (Bekenstein formula \cite{Be73}). In fact, by a counting of BH states, one gets an expression for its entropy, which is proportional to the horizon area. Reproducing the Bekenstein formula fixes a proper value for $\gamma$. A more recent estimate is based on quasi-normal modes of black holes, instead, {\it i.e.} on the investigation of damped oscillations, which arise as linear perturbations. This comparison starts from the correspondence between classical quasi-normal modes and quantum transitions \cite{Dr03,Co02}. However, the result obtained is not consistent with the previous one and till now no definite answer on the meaning of this discrepancy has been found.\\
Apart from these problems related to the assignement of a fixed $\gamma$ value, also from a theoretical point of view there are some objections againist regarding the Immirzi parameter as a fundamental quantum ambiguity. The idea is that such parameter is unphysical and it arises because variables violating some symmetry have been quantized. For instance, the covariant formulation of Alexandrov \cite{Al00,AE03} does not contain such an ambiguity, but the issue of the quantization has not been obtained yet in this framework. Similarly, some authors \cite{GMM02,GMM03} point out that the effect induced on observables by a changing of $\gamma$ can be reproduced by a boost or by a conformal scaling for the 3-metric.\\

\section{Time gauge and boost invariance}\label{timegauge}
As we point out in section \ref{holst}, the splitting procedure in the case $\gamma\neq\pm i$ is based on a partial gauge fixing, the time-gauge condition $e^0_i=0$. This way, one restricts Lorentz indexes to spatial ones and, at the same time, is able to reproduce the Gauss constraint for an $SU(2)$ gauge theory.\\ 
This standard procedure is based on the work by Barros e Sa \cite{BS01}, where he gave the proof that this gauge fixing is allowed. In fact, he considered a general Lorentz frame by taking $e^0_i=\chi_a e^a_i$, where the additional $\chi_a$-variables must be taken into account. By a Lorentz transformation $\Lambda$, such that $\chi_a=-\Lambda^0_{\phantom1a}/\Lambda^0_{\phantom10}$, one restores the time-gauge condition. This clarifies the physical meaning of $\chi_\alpha$ as velocity components of the Lorentz frame with respect to spatial hypersurface. Since, in these variables, no gauge fixing at all is performed, 3 new first-class constraints enforcing boost invariance arise. However,  the degrees of freedom counting enables one to realize that other constraints has to come out. They turn out to be second-class. The treating of a system with second-class constraints involves or solution of such constraints and the quantization of relic variables or the substitution of Poisson brackets with Dirac ones \cite{HT}. Barros e Sa \cite{BS01} adopted the first procedure and found a solution to these constraints. By resubstituting the solutions into the action, a first-class system comes out. Therefore, he concluded that boosts, being associated with first class constraints, behave as gauge transformations. Hence a canonical quantization can be performed by a gauge fixing, if no anomaly is present, since different gauge sectors are expected to be related by a unitary transformation on a quantum level.
 
In this respect it has been demonstrated \cite{CM07} that this conclusion stands in a second-order formulation. In such a framework the spatial metric reads $h_{ij}=(\delta_{ab}-\chi_a\chi_b)e^a_ie^b_j$ and the Hamiltonian formulation is given in terms of variables $(e^a_i,\chi_a)$ and conjugated momenta $(\pi^i_a,\pi^a)$.\\ 
The new set of constraints is the following one
\begin{equation}
\left\{\begin{array}{c} \mathcal{H}=\pi^i_a\pi^j_b\bigg(\frac{1}{2}E^a_iE^b_j-E^b_iE^a_j\bigg)+h{}^{3}\!R=0\\
\mathcal{H}_i=D_j(\pi^j_aE^a_i)=0\\
\Phi_{ab}=\pi^c\delta_{c[a}\chi_{b]}-\delta_{c[a}\pi^i_{b]}E^c_i=0\\
\Phi^a=\pi^a-\pi^b\chi_b\chi^a+\delta^{ab}\pi^i_b\chi_cE^c_i=0\end{array}\right.
\end{equation}
where last conditions enforce the boost invariance.\\
By fixing $\chi_a$ functions, boost constraints can be solved. Then, after the quantization of relic variables, a class of unitary operators connecting different $\chi$-sectors can be recognized. This result implies that the boost symmetry has been represented in a quantum framework, therefore observables are not affected by the $\chi_a$ functions fixing.

Alexandrov \cite{Al00,AE03} solved second-class constraints by substituting Poisson equations with Dirac ones. This way he was able to reproduce the Gauss constraints for the full Lorentz group. Then he applied the machinery of the Loop quantization program to his model and ended up with a formulation without the Immirzi ambiguity.  This result indicates that these two methods leads to inequivalent quantizations.

It is not clear yet how these two discrepant points of view can be combined together.

\section{The picture of the space-time}
The LQG quantization program provides us with a description of the quantum space-time far apart from that coming from the Wheeler-DeWitt quantization and much closer to a Quantum Field Theory for the geometry.\\ 
In the standard Wheeler-DeWitt approach, the functional space, on which one is looking for an Hilbert space structure, is that of Supermetrics. This space is an infinite one, no discrete structure arises in this context. Quantum states are linear combinations of 3-geometries. This way the link with General Relativity is easily established, but the quantization program is still lacking.\\ 
Otherwise, in LQG the phase space is similar to a gauge theory one. Configurations variables are non-local objects, {\it i.e.} holonomies, whose physical meaning is the same as Faraday lines for the electro-magnetic field. States are linear combinations of spin networks, such that a discrete structure naturally comes out. The quantization of areas and volumes makes this quantum structure manifest. The emergence of a countable set of basis vectors is a great achievement in view of a second quantization approach.

However, difficulties in performing the semi-classical limit clearly indicates that the connection between quantum states and General Relativity is missing. This lack is linked with the absence of a satisfying measurement theory for a background-independent model and it forbids a real connection with experiments, except for some mini-superspace approximations. For instance, there are hints that a non-commutative structure for space-coordinates arises \cite{ACZ98}, but there are no indications whether this point can provide testable predictions, such as violations of the Lorentz symmetry. This sort of violation would be a very useful point in view of a comparison with experiments \cite{CPSUV04}. For a link between Lorentz violations and the measure theory in Quantum Gravity see \cite{MR02}. \\ 
Recently Markopoulou and Smolin \cite{MS07} proposed that the quantization of non-local objects, proper of LQG, can provide a potential problem with non-locality also for the macroscopic metric tensor.\\ 
For these reasons the definition of semi-classical states is the basic task of the LQG quantization program.

\newpage

\end{document}